\documentstyle[preprint,aps]{revtex}
\begin{document}
\draft

\title
{The N- and 1-time Classical Descriptions of N-body Relativistic
Kinematics and the Electromagnetic Interaction}

\author{Luca Lusanna}

\address
{Sezione INFN di Firenze\\
L.go E.Fermi 2 (Arcetri)\\
50125 Firenze, Italy\\
e-mail LUSANNA@FI.INFN.IT}

\maketitle

\begin{abstract}

Given N relativistic scalar free particles described by N mass-shell first
class
constraints in their 8N-dimensional phase space, their N-time description
is obtained by means of a series of canonical transformations to a
quasi-Shanmugadhasan basis adapted to the constraints. Then the same system is
reformulated on spacelike hypersurfaces: the restriction to the family of
hyperplanes orthogonal to the total timelike momentum gives rise to a covariant
intrinsic 1-time formulation called the ``rest-frame instant form" of dynamics.
The relation between the N- and 1-time descriptions, the mass spectrum of the
system and the way how to introduce mutual interactions among the particles
are studied. Then the 1-time description of the isolated system of N charged
scalar particles plus the electromagnetic field is obtained.
The use of Grassmann
variables to describe the charges together with the determination of the
field and particle Dirac observables leads to a formulation without infinite
self-energies and with mutual Coulomb interactions extracted from classical
electromagnetic field theory. A comparison with the Feshbach-Villars
Hamiltonian
formulation of the Klein-Gordon equation is made. Finally a 1-time covariant
formulation of relativistic statistical mechanics is found.

\vskip 1truecm
\noindent November 1995
\vskip 1truecm
\noindent This work has been partially supported by the network ``Constrained
Dynamical Systems" of the E.U. Programme ``Human Capital and Mobility".

\end{abstract}
\vfill\eject

\section
{Introduction}

Relevant physical models are described by singular Lagrangians, so that
their Hamiltonian formulation is based on the Dirac theory of
constraints\cite{diraca}. The qualitative aspects of this theory are now
understood\cite{lusa}, in particular the role of the Shanmugadhasan
canonical transformation\cite{shan} in the determination of a canonical basis
of Dirac's observables allowing the elimination of gauge degrees of
freedom from the classical description of physical systems\cite{lusb}. This
programme was initiated by Dirac\cite{diracb} for the electromagnetic field
with charged fermions. More recently, Dirac's observables for Yang-Mills theory
with fermions (whose typical application is QCD) have been found\cite{lusc} in
suitable function spaces in which the Gribov ambiguity is absent. Also the ones
for the Abelian Higgs model are known\cite{lusd} and those for the $SU(2)
\times U(1)$ electroweak theory with fermions can be found\cite{luse}
with the same method that works for the Abelian case. The main task along these
lines will now be the search of Dirac's observables for tetrad gravity in the
case of asymptotically flat 3-manifolds.

The price for having only physical degrees of freedom is the nonlocal (and in
general nonpolynomial) nature of the physical Hamiltonians and
Lagrangians, as already known from Dirac's work on the electromagnetic
field\cite{diracb}, for which the origin of the difficulties is
the Coulomb self-interaction of the fermion fields.
Two obstacles appear immediately: 1) the lack of manifest Lorentz covariance
of the Hamiltonian formalism, which requires the choice of a 3+1 splitting of
Minkowski spacetime; 2) the inapplicability of the standard methods of
regularization and renormalization due to the nonlocality (and
nonpolynomiality) of the interactions and the failure of the power
counting rule.

In Ref.\cite{lush}, I reviewed these problems and I pointed out
that the Lorentz
covariance problem can be solved by reformulating the theory on spacelike
hypersurfaces\cite{diraca} (see also Ref.\cite{lusc}) and then reducing it to
the hyperplanes orthogonal to the total momentum $p^{\mu}$ (for the
configurations in which it is timelike, $p^2 > 0$; see Ref.\cite{ku} for a
general study of the embeddings of spacelike hypersurfaces in a given
Riemannian spacetime). In this way, the breaking of Lorentz
covariance is reduced to a minimum: only the
three degrees of freedom that  describe
the canonical centre of mass 3-position of the overall isolated system are
not covariant. In Ref.\cite{lush}, it is also stressed that only one tool has
until now emerged for attacking the difficult task of quantizing a nonlocal
(and
generically nonpolynomial) theory. Namely, the problem of the center-of-mass
extended relativistic systems in irreducible representations of the Poincar\'e
group with $P^2 > 0$, $W^2=-P^2{\vec {\bar S}}^2\not= 0$ (they are dense in
the set of all allowed field configurations) identifies a finite worldtube of
noncovariance of the canonical center of mass, whose radius $\rho =\sqrt{-W^2}
/P^2=|\, {\vec {\bar S}}\, |\, /\sqrt{P^2}$ represents a classical intrinsic
unit of length that can be used as a ultraviolet cutoff at the quantum
level in the spirit of Dirac and Yukawa. As noted in Ref.[6], the distances
corresponding to the interior of the worldtube are connected with problems
coming from both quantum theory and general relativity: 1) pair production
occurs if an attempt is made
to localize particles at these distances; 2) relativistic extended
bodies with a material radius less than $\rho$ cannot have the classical
energy density positive definite everywhere in every reference frame, and the
peripheral rotation velocity may be higher than the velocity of light.
Therefore, the worldtube is the flat remnant of the energy  conditions of
general relativity; in this theory,
the radius $\rho$ is defined in terms of the asymptotic
Poincar\'e group that exists in the case of asymptotically flat 3-manifolds.

In this way, one has perceives the possibility of formulating a
``rest-frame field theory" on these special hyperplanes\cite{lush},
in which the ultraviolet cutoff just identified
could be used in a constructive way. The asymptotic states of this theory
should be connected with the description of free particles on these
hyperplanes. But this implies that the phase space description of the particles
involves only 6 degrees of freedom per particle, that there is an 1-time
description (``rest-frame instant form" of the dynamics
in the language of Ref.\cite{diracc}, the time
parameter being the rest frame time, which describes the foliation of
Minkowski spacetime with these hyperplanes), and that there is no room left
for relative times and energies, which are precisely the variables connected
with the spurious solutions of the Bethe-Salpeter equation\cite{saz}
for relativistic bound states
in ordinary quantum field theory. Actually, the standard Fock space describes
asymptotic states defined as tensor products of free one-particle states
without any restriction on their mutual temporal ordering (an in-particle
may be in the absolute future of another in-particle): the classical background
is an N-time description of N free particles, each one of which has 8 degrees
of
freedom and an associated mass-shell first class constraint (connected with
the inverse of the standard propagators).

Therefore, in Ref.\cite{lush}
there was a review of the problems of relativistic
particle mechanics, especially the two main ones: i) the No-Interaction
Theorem\cite{nit,nita} (see Ref.\cite{chel} for a review),
which turns out to be also present in Galilean mechanics\cite{pons} and to be
connected with the use of an N-time description; ii) the many definitions of
relativistic center-of-mass position (see Ref.\cite{com}
for a review), since the Lorentz signature of Minkowski spacetime precludes
the existence of an object with all the properties of the nonrelativistic
center-of-mass.

Behind all these descriptions there is the theory of the irreducible
representations of the canonical realizations of the Poincar\'e group\cite
{pp} in given phase spaces.

In this paper, I will consider N free relativistic scalar particles, and I will
study their description both in the N-time approach (following jointly Ref.
\cite{lusf} and the complete analysis of the N=2 case in Ref.\cite{longhi})
and in the 1-time approach, which was developed in Ref.\cite{karp}
for completely different  reasons. While the N-time theory identifies a set of
variables that makes it possible to disentangle the
relative times and energies from the relevant variables, the 1-time theory
allows the identification of the branches of the mass spectrum of the
N-body isolated system. Then the two descriptions are compared, and it
is found that the final physically relevant variables are the same in the two
approaches and that they also seem to be the best relativistic kinematical
variables (carrying the knowledge of the geometry of the timelike Poincar\'e
orbits) for the study of interactions.

In Section II, there is a study of the canonical transformations needed to
find a quasi-Shanmugadhasan canonical basis adapted to the N-1 combinations of
the original first class constraints that define the vanishing of the
relevant N-1 relative energies. Due to the nonlinearity of the final
canonical transformation, it is not known how to find the mass spectrum and the
inverse of the transformation.

In Section III, the system is reformulated on spacelike hypersurfaces, on
which each particle is now identified only by three coordinates. After the
identification of the first class constraints, the reduction
to the special intrinsic family of spacelike hyperplanes orthogonal to the
total four-momentum, when it is timelike, is studied. The 1-time covariant
rest-frame instant form of dynamics is identified.

In Section IV, the N- and 1- time descriptions are compared, and the mass
spectrum is found.

In Section V, there is a study of the mass spectrum in the N-time theory. Then
action-at-a-distance interactions are introduced among the particles in the
1-time theory with a comment on their separability.

In Section VI, the isolated system of N charged scalar particles plus the
electromagnetic field is formulated on spacelike hypersurfaces and then
reduced to the hyperplanes orthogonal to the total timelike four-momentum.
The charges of the particles are described in a pseudoclassical way by means of
Grassmann variables. The Dirac observables of the particles and of the
electromagnetic field with respect to electromagnetic gauge transformations are
found. The final resulting four first class constraints contain the
interparticle Coulomb potential (extracted covariantly from the classical
electromagnetic field theory) but not the classical electromagnetic
self-energy due to the vanishing of the square of the Grassmann charges: the
underlying hypothesis of charge quantization generates a regularization of the
pseudoclassical description.

In Section VII, the 1-time theory of one charged scalar particle plus the
electromagnetic field is compared with the Feshbach-Villars
Hamiltonian formulation of the Klein-Gordon equation in an external
electromagnetic field.

InSection VIII, after a review of the main problems of covariant relativistic
statistical mechanics, its reformulation in the 1-time theory is given.

The Conclusions IX contain some final remarks and the outline of future
research.

In Appendix A, there is a review of the properties of the standard Wigner boost
for timelike Poincar\'e orbits. In Appendix B, the formulas for the case N=2
are reviewed. In Appendix C, some notation in the description with
spacelike hypersurfaces is given.

\vfill\eject

\section
{The N-time Theory}

A system of N free scalar relativistic particles is usually described in
phase space by 8N variables $x^{\mu}_i,\, p^{\mu}_i,$ i=1,..,N, $\lbrace
x^{\mu}_i,p^{\nu}_j\rbrace =-\delta_{ij}\eta^{\mu\nu}$ [$\eta^{\mu\nu}=
(1,-1,-1,-1)$] restricted by N first class constraints [to start with, we
assume
arbitrary masses $m_i$ for the particles]

\begin{equation}
\phi_i =p^2_i-m^2_i \approx 0,\quad\quad
\lbrace \phi_i,\phi_j\rbrace =0.
\label{1}
\end{equation}

The evolution in a scalar parameter $\tau$ is described by the Dirac
Hamiltonian

\begin{equation}
H_D=\sum_{i=1}^N\lambda_i(\tau )\, \phi_i,
\label{2}
\end{equation}

\noindent where $\lambda_i(\tau )$ are Dirac's multipliers. This description
is obtained by starting from the Lagrangian

\begin{equation}
{\cal L}=- \sum_{i=1}^N m_i\sqrt{ {\dot x}^2_i(\tau ) },
\label{3}
\end{equation}

\noindent whose associated canonical momenta are

\begin{equation}
p^{\mu}_i=- { {\partial {\cal L}}\over {\partial {\dot x}_{i\mu}} }=
m_i\, { {{\dot x}^{\mu}_i}\over {\sqrt{{\dot x}^2_i}} }.
\label{4}
\end{equation}

Therefore, the configuration position variables, describing the worldlines
of the particles, are $q^{\mu}_i(\tau )=x^{\mu}_i(\tau )$. An alternative
N-time Hamiltonian description is the multi-time one \cite{lusb},
in which the N
scalar time parameters $\tau_i$ are defined by $d\tau_i=\lambda_i(\tau )
d\tau$. In this description one has $q^{\mu}_i(\tau_i)=x^{\mu}_i(\tau_i)$,
$\lbrace x^{\mu}_i(\tau_i),p^{\nu}_j(\tau_j)\rbrace
=-\delta_{ij}\eta^{\mu\nu}$,
and the first class constraints $\phi_i\approx 0$ are the Hamiltonians for
the evolution in the $\tau_i$'s [${ {\partial A(x_k,p_k)}\over {\partial
\tau_i} }{\buildrel \circ \over =}\lbrace A,\phi_i\rbrace$ are the many-time
Hamilton  equations and $\lbrace \phi_i,\phi_j\rbrace =0$ are
their integrability
conditions; ${\buildrel \circ \over =}$ means evaluated on the solutions of
the equations of motion].
The Lagrangian description derives from the
action $S=\sum_{i=1}^N\int d\tau_i\, {\cal L}_i$ with ${\cal L}_i=-
m_i\, \sqrt{{\dot x}^2_i(\tau_i)}$.

Let us remark that in presence of interactions, the No-Interaction-Theorem
\cite{nit,nita} implies $q^{\mu}_i(\tau_i)\not= x^{\mu}_i(\tau_1,..,\tau_N)$.
As
shown in Ref.\cite{pons} (see also Ref.\cite{lush}),
this result is also present at the
nonrelativistic level: it is not connected with the Lorentz signature but with
the many-time description. If the particles are charged, the minimal coupling
to an external electromagnetic field requires the canonical covariant
coordinates $x^{\mu}_i$'s and not the configuration  $q^{\mu}_i$'s, which
describe the particle worldlines: therefore the electric charges cannot be
localized on the worldlines in the interacting case.

In Ref.\cite{lusf}, a study of the kinematics of N relativistic free scalar
particles was started with a double aim. Firstly, the spurious
solutions of the Bethe-Salpeter equation\cite{saz} stimulated a search for
nonlinear canonical transformation replacing N-1
linear combinations of the first class constraints $\phi_i\approx 0$ with
suitable``relative energy variables". Secondly, it was
investigated how to introduce action-at-a-distance interactions
while preserving the
first class nature of the constraints (only nonseparable interactions were
found).

Subsequently, for the case N=2 with $p^2 > 0$ [$p^{\mu}=\sum_{i=1}^Np^{\mu}_i$
is the total momentum] suitable canonical variables adapted to the timelike
orbits of the Poincar\'e group were found. Then, in Ref.\cite{seplon}, a study
of the N=3 case was initiated, with the aim of understanding how to describe
separable action-at-a-distance interactions.

Collecting all these results and simplifying the notation, we start with
the canonical tranformation from the basis $x^{\mu}_i, p^{\mu}_i$, i=1,..,N,
to a first set of linear center-of-mass and relative variables [a=1,..,N-1]

\begin{equation}
\begin{minipage}[t]{1cm}
\begin{tabular}{|l|} \hline
$x^{\mu}_i$ \\
$p^{\mu}_i$ \\ \hline
\end{tabular}
\end{minipage} \ {\longrightarrow \hspace{.2cm}} \
\begin{minipage}[t]{2 cm}
\begin{tabular}{|ll|} \hline
$x^{\mu}$   & $p^{\mu}$   \\ \hline
$R^{\mu}_a$ & $Q^{\mu}_a$ \\ \hline
\end{tabular}
\end{minipage}
\label{5}
\end{equation}

\noindent defined by

\begin{eqnarray}
&x^{\mu}={1\over \sqrt{N}}\sum_{i=1}^N{\hat \gamma}_ix^{\mu}_i,\quad\quad
&\lbrace x^{\mu},p^{\nu}\rbrace =-\eta^{\mu\nu}\nonumber \\
&p^{\mu}=\sum_{i=1}^Np^{\mu}_i,\quad\quad\quad\quad &{}\nonumber \\
&R^{\mu}_a=\sqrt{N} \sum_{i=1}^N{\hat \gamma}_{ai}x^{\mu}_i,\quad\quad
&\lbrace R^{\mu}_a,Q^{\nu}_b\rbrace =-\delta_{ab}\eta^{\mu\nu}\nonumber \\
&Q^{\mu}_a={1\over \sqrt{N}}\sum_{i=1}^N{\hat \gamma}_{ai}p^{\mu}_i,
\quad\quad &{},
\label{6}
\end{eqnarray}

\noindent where ${\hat {\vec \gamma}}$, ${\hat {\vec \gamma}}_a$ is a basis
of orthogonal unit vectors satisfying

\begin{eqnarray}
&&\sum_{i=1}^N{\hat \gamma}_i=\sqrt{N},\quad\quad\quad \sum_{i=1}^N{\hat
\gamma}_{ai}=0,\nonumber \\
&&\sum_{i=1}^N{\hat \gamma}_i{\hat \gamma}_{ai}=0,\quad\quad\quad \sum_{i=1}^N
{\hat \gamma}_{ai}{\hat \gamma}_{bi}=\delta_{ab},\nonumber \\
&&{\hat \gamma}_i{\hat \gamma}_j+\sum_{a=1}^{N-1}{\hat \gamma}_{ai}{\hat
\gamma}
_{aj}=\delta_{ij}.
\label{7}
\end{eqnarray}

{}From now on, we shall choose ${\hat \gamma}_i=1/\sqrt{N}$, i=1,..,N, so that
we have

\begin{equation}
x^{\mu}={1\over N}\sum_{i=1}^Nx^{\mu}_i,\quad\quad\quad \sum_{a=1}^{N-1}
{\hat \gamma}_{ai}{\hat \gamma}_{aj}=\delta_{ij}-{1\over N}.
\label{8}
\end{equation}

Here
$x^{\mu}$ is a canonical covariant center-of-mass coordinate, which however
does not have free motion, since ${\dot x}^{\mu}=\lbrace x^{\mu},H_D\rbrace =-
{2\over N}\sum_{i=1}^N\lambda_i(\tau )p^{\mu}_i/\sqrt{p^2_i}\not=
\lambda (\tau )p^{\mu}/\eta \sqrt{p^2}=\lambda (\tau )u^{\mu}(p)$ [$\lambda
(\tau )=\sqrt{{\dot x}^2}$; $\eta =sign\, p^o$]; it moves with a ``classical
zitterbewegung" that depends on the choice of the arbitrary Dirac multipliers
$\lambda_i(\tau )$ (it is a gauge effect).

The inverse canonical transformation is

\begin{eqnarray}
&&x^{\mu}_i=x^{\mu}+{1\over {\sqrt{N}} }\sum_{a=1}^{N-1}{\hat \gamma}_{ai}
R^{\mu}_a\nonumber \\
&&p^{\mu}_i={1\over N}p^{\mu}+\sqrt{N}\sum_{a=1}^{N-1}{\hat \gamma}_{ai}Q^{\mu}
_a.
\label{9}
\end{eqnarray}

The conserved Poincar\'e generators are

\begin{eqnarray}
&&p^{\mu}=\sum_{i=1}^N p^{\mu}_i\nonumber \\
&&J^{\mu\nu}=\sum_{i=1}^N(x^{\mu}_ip^{\nu}_i-x^{\nu}_ip^{\mu}_i)=L^{\mu\nu}+
S^{\mu\nu}\nonumber \\
&&{}\nonumber \\
&&L^{\mu\nu}=x^{\mu}p^{\nu}-x^{\nu}p^{\mu}\nonumber \\
&&S^{\mu\nu}=\sum_{a=1}^{N-1}(R^{\mu}_aQ^{\nu}_a-R^{\nu}_aQ^{\mu}_a).
\label{10}
\end{eqnarray}

The isolated system of N particles is assumed to belong to an irreducible
timelike representation of the Poincar\'e group with Casimirs $p^2 > 0$ and
$W^2=-p^2{\vec {\bar S}}^2$ [$W^{\mu}={1\over 2}\epsilon^{\mu\nu\alpha\beta}
p_{\nu}S_{\alpha\beta}$ is the Pauli-Lubanski fourvector ($\epsilon_{0123}
=1$, $\epsilon^{ijk}=\epsilon_{ijk}$);
${\vec {\bar S}}$ is the rest-frame Thomas spin (see later on)].

Note that $L^{\mu\nu}$ and $S^{\mu\nu}$ are not separately constants
of the motion due to the ``classical zitterbewegung of $x^{\mu}$" (which is an
analogue of the zitterbewegung associated with the Dirac position operator
for the electron).

Extending the results of Ref.\cite{longhi} for N=2 to arbitrary N
(see Appendix A for the notation), one defines
the following further canonical transformation

\begin{equation}
\begin{minipage}[t]{2cm}
\begin{tabular}{|ll|} \hline
$x^{\mu}$    &  $p^{\mu}$  \\ \hline
$R^{\mu}_a$  & $Q^{\mu}_a{}$ \\ \hline
\end{tabular}
\end{minipage} \ {\longrightarrow \hspace{.2cm}} \
\begin{minipage}[t]{3cm}
\begin{tabular}{|ll|l|} \hline
\multicolumn{2}{|c}{${\tilde x}^{\mu}$} & { $p^{\mu}$} \\ \hline
{}  &  {}  & $T_{Ra}$  \\
${\vec \rho}_a$  &  ${\vec \pi}_a$  &  {}  \\
{}  &  {}  &  $\epsilon_{Ra}$  \\ \hline
\end{tabular}
\end{minipage}
\label{11}
\end{equation}

\begin{equation}
\lbrace {\tilde x}^{\mu},p^{\nu}\rbrace =-\eta^{\mu\nu},\quad
\lbrace T_{Ra},\epsilon_{Rb}\rbrace =-\delta_{ab},\quad
\lbrace \rho^i_a,\pi^j_b\rbrace =\delta_{ab}\delta^{ij},
\label{12}
\end{equation}

\noindent with (one has $p_{\mu}{\tilde x}^{\mu}=p_{\mu}x^{\mu}$)

\begin{eqnarray}
&&{\tilde x}^{\mu}=x^{\mu}+{1\over 2}\, \epsilon^A_{\nu}(u(p))\eta_{AB}
{ {\partial \epsilon^B_{\rho}(u(p))}\over {\partial p_{\mu}} }\, S^{\nu\rho}
=\nonumber \\
&&=x^{\mu}-{ 1\over {\eta \sqrt{p^2}(p^o+\eta \sqrt{p^2})} }\, [p_{\nu}
S^{\nu\mu}+\eta \sqrt{p^2} (S^{o\mu}-S^{o\nu}{ {p_{\nu}p^{\mu}}\over {p^2} })]
\nonumber \\
&&{}\nonumber \\
&&p^{\mu}=p^{\mu}\nonumber \\
&&{}\nonumber \\
&&T_{Ra}=\epsilon^o_{\mu}(u(p))R^{\mu}_a={ {p\cdot R_a}\over {\eta \sqrt{p^2}}
}
\nonumber \\
&&\epsilon_{Ra}=\epsilon^o_{\mu}(u(p))Q^{\mu}_a=
{ {p\cdot Q_a}\over {\eta \sqrt{p^2}} }\nonumber \\
&&\rho^r_a=\epsilon^r_{\mu}(u(p))R^{\mu}_a=
R^r_a-{ {p^r}\over {\eta \sqrt{p^2}} }
(R^o_a-{ {{\vec p}\cdot {{\vec R}_a} }\over {p^o+\eta \sqrt{p^2}} })
\nonumber \\
&&\pi^r_a=\epsilon^r_{\mu}(u(p))Q^{\mu}_a=Q^r_a-{ {p^r}\over {\eta \sqrt{p^2}}
}
(Q^o_a-{ {{\vec p}\cdot {{\vec Q}_a} }\over {p^o+\eta \sqrt{p^2}} }),
\label{13}
\end{eqnarray}

\noindent where Eqs.(\ref{A10}) and (\ref{A11}) have been used to evaluate
${\tilde x}^{\mu}$.

As shown in Ref.\cite{longhi} and in Appendix A, the canonical transformation
of Eqs.(\ref{13}) is defined by boosting at rest the relative variables
$R^{\mu}_a$, $Q^{\mu}_a$ with the standard Wigner boost $L^A{}_{\mu}
(\stackrel{\circ}{p},p)=\epsilon^A_{\mu}(u(p))$ of Eqs.(\ref{A2}),
(\ref{A3}) for timelike orbits $p^2 > 0$. In Eq.(\ref{11}),
$T_{Ra}=\epsilon^{\bar o}_{\mu}(u(p))R^{\mu}_a$ and $\epsilon_{Ra}=
\epsilon^{\bar o}_{\mu}(u(p))Q^{\mu}_a$ are Poincar\'e-scalar relative times
and energies, respectively; $\rho^r_a=\epsilon^r_{\mu}(u(p))R^{\mu}_a$ and
$\pi^r_a=\epsilon^r_{\mu}(u(p))Q^{\mu}_a$ are, respectively,
relative three-coordinates and three-momenta, which transform as Wigner
spin-1 3-vectors. It turns out that the
new canonical center-of-mass coordinate ${\tilde x}^{\mu}$ is not
a fourvector under Lorentz boosts (it has only Euclidean covariance),
so that it cannot define a geometrical center-of-mass worldline, but only
a frame-dependent line (pseudo-worldline).

The Lorentz generators become

\begin{eqnarray}
&&J^{\mu\nu}={\tilde L}^{\mu\nu}+{\tilde S}^{\mu\nu}\nonumber \\
&&{}\nonumber \\
&&{\tilde L}^{\mu\nu}={\tilde x}^{\mu}p^{\nu}-{\tilde x}^{\nu}p^{\mu}
\nonumber \\
&&{\tilde S}^{\mu\nu}=S^{\mu\nu}-{1\over 2}\epsilon^A_{\rho}(u(p))\eta_{AB}
({ {\partial \epsilon^B_{\sigma}(u(p))}\over {\partial p_{\mu}} }\, p^{\nu}-
{ {\partial \epsilon^B_{\sigma}(u(p))}\over {\partial p_{\nu}} }\, p^{\mu})
S^{\rho\sigma}=\nonumber \\
&&=S^{\mu\nu}+{1\over {\eta \sqrt{p^2}(p^o+\eta \sqrt{p^2})} } [p_{\beta}
(S^{\beta\mu}p^{\nu}-S^{\beta\nu}p^{\mu})+\eta \sqrt{p^2}(S^{o\mu}p^{\nu}-
S^{o\nu}p^{\mu})]\nonumber \\
&&{}\nonumber \\
&&{\tilde S}^{oi}=-{1\over {\eta \sqrt{p^2}} }[(p^o-\eta \sqrt{p^2})S^{oi}+
{ {p^k(S^{ko}p^i-S^{ki}p^o)}\over {p^o+\eta \sqrt{p^2}} }]\nonumber \\
&&{\tilde S}^{ij}=S^{ij}+{1\over {\eta \sqrt{p^2}} }(S^{oi}p^j-S^{oj}p^i)-
{ {p^k(S^{ki}p^j-S^{kj}p^i)}\over {\eta \sqrt{p^2}(p^o+\eta \sqrt{p^2})} }.
\label{14}
\end{eqnarray}

If we introduce the rest-frame spin tensor (the last line gives the Thomas
spin)

\begin{eqnarray}
&&{\bar S}^{AB}=\epsilon^A_{\mu}(u(p))\epsilon^B_{\nu}(u(p))S^{\mu\nu}
\nonumber \\
&&{}\nonumber \\
&&{\bar S}^{\bar or}=\sum_{a=1}^{N-1}(T_{Ra}\pi^r_a-\rho^r_a\epsilon_{Ra})
\nonumber \\
&&{\bar S}^{rs}=\sum_{a=1}^{N-1}(\rho^r_a\pi^s_a-\rho^s_a\pi^r_a)\nonumber \\
&&{}\nonumber \\
&&{\bar S}^r={1\over 2}\epsilon^{rst}{\bar S}^{st}=\sum_{a=1}^{N-1}{\bar
S}^r_a,
\quad\quad {\vec {\bar S}}_a={\vec \rho}_a\times {\vec \pi}_a
\label{15}
\end{eqnarray}

\noindent we find the following form of the Poincar\'e generators

\begin{eqnarray}
&&J^{ij}={\tilde x}^ip^j-{\tilde x}^jp^i+\delta^{ir}
\delta^{js}{\bar S}^{rs}\nonumber \\
&&J^{oi}={\tilde x}^op^i-{\tilde x}^ip^o- { {\delta^{ir}{\bar S}^{rs}p^s}\over
{p^o+\eta \sqrt{p^2}} }.
\label{16}
\end{eqnarray}

\noindent and the following form of ${\tilde x}^{\mu}$

\begin{eqnarray}
{\tilde x}^{\mu}&=&x^{\mu}-{1\over {\eta \sqrt{p^2}} }[\eta^{\mu}_A({\bar S}
^{\bar oA}-{ {{\bar S}^{Ar}p^r}\over {p^o+\eta \sqrt{p^2}} })+
{ {p^{\mu}+2\eta \sqrt{p^2}\eta^{\mu o}}\over {\eta \sqrt{p^2}(p^o+\eta
\sqrt{p^2})} } {\bar S}^{\bar or}p^r]\nonumber \\
{}&{}&{}\nonumber \\
&{\tilde x}^o&=x^o-{1\over {p^2} }\sum_{a=1}^{N-1}(T_{Ra}\vec p\cdot {\vec \pi}
_a-\epsilon_{Ra}\vec p\cdot {\vec \rho}_a)\nonumber \\
&{\tilde x}^i&=x^i-{1\over {\eta \sqrt{p^2}} }\sum_{a=1}^{N-1}[T_{Ra}\pi^i_a-
\epsilon_{Ra}\rho^i_a-\nonumber \\
&-&{ {\rho^i_a\vec p\cdot {\vec \pi}_a-\pi^i_a\vec p\cdot{\vec \rho}_a}\over
{p^o+\eta \sqrt{p^2}} }+{ {p^i(T_{Ra}\vec p\cdot {\vec \pi}_a-\epsilon_{Ra}
\vec p\cdot {\vec \rho}_a)}\over {\eta \sqrt{p^2}(p^o+\eta
\sqrt{p^2})} }].
\label{17}
\end{eqnarray}

Therefore, we have

\begin{eqnarray}
&&{\tilde S}^{ij}=\delta^{ir}\delta^{js}{\bar S}^{rs}\nonumber \\
&&{\tilde S}^{oi}=-{ {\delta^{ir}{\bar S}^{rs}p^s}
\over {p^o+\eta \sqrt{p^2}} }\nonumber \\
&&{}\nonumber \\
&&{\tilde {\bar S}}^{AB}=\epsilon^A_{\mu}(u(p))\epsilon^B_{\nu}(u(p)){\tilde
S}^{\mu\nu}=\nonumber \\
&&={\bar S}^{AB}-{1\over 2}\eta \sqrt{p^2}(\epsilon^A_{\mu}(u(p))\eta^B_o-
\epsilon^B_{\mu}(u(p))\eta^A_o){ {\partial \epsilon^E_{\sigma}(u(p))}\over
{\partial p_{\mu}} }\epsilon^{\sigma}_D(u(p))\eta_{EC} {\bar S}^{CD}=
\nonumber \\
&&={\bar S}^{AB}.
\label{18}
\end{eqnarray}

The inverse canonical transformation is

\begin{eqnarray}
x^o&=&{\tilde x}^o+{1\over {p^2}}\sum_{a=1}^{N-1}(T_{Ra}\vec p\cdot {\vec \pi}
_a-\epsilon_{Ra}\vec p\cdot {\vec \rho}_a)\nonumber \\
\vec x&=&{\vec {\tilde x}}+{1\over {\eta \sqrt{p^2}} }\sum_{a=1}^{N-1}(T_{Ra}
{\vec \pi}_a-\epsilon_{Ra}{\vec \rho}_a)+\nonumber \\
&+&{1\over {\eta \sqrt{p^2}(p^o+\eta \sqrt{p^2})} }\sum_{a=1}^{N-1}[\vec p
\cdot {\vec \rho}_a {\vec \pi}_a-\vec p\cdot {\vec \pi}_a{\vec \rho}_a+
{ {\vec p}\over {\eta \sqrt{p^2}} }(T_{Ra}\vec p\cdot {\vec \pi}_a-
\epsilon_{Ra}\vec p\cdot {\vec \rho}_a)]\nonumber \\
R^o_a&=&{1\over {\eta \sqrt{p^2}} }(T_{Ra}p^o+\vec p\cdot {\vec \rho}_a)
\nonumber \\
{\vec R}_a&=&{\vec \rho}_a+{ {\vec p}\over {\eta \sqrt{p^2}} }(T_{Ra}+
{ {\vec p\cdot {\vec \rho}_a}\over {p^o+\eta \sqrt{p^2}} })\nonumber \\
Q^o_a&=&{1\over {\eta \sqrt{p^2}} }(\epsilon_{Ra}p^o+\vec p\cdot {\vec \pi}_a)
\nonumber \\
{\vec Q}_a&=&{\vec \pi}_a+{ {\vec p}\over {\eta \sqrt{p^2}} }(\epsilon_{Ra}+
{ {\vec p\cdot {\vec \pi}_a}\over {p^o+\eta \sqrt{p^2}} })
\label{19}
\end{eqnarray}

\noindent and the original variables $x^{\mu}_i$, $p^{\mu}_i$ have the
following
form in this new canonical basis:

\begin{eqnarray}
x^o_i&=&{\tilde x}^o+{1\over {\eta \sqrt{p^2}}}\sum_{a=1}^{N-1}[{1\over
{\sqrt{N}} }{\hat \gamma}_{ai}(T_{Ra}p^o+\vec p\cdot {\vec \rho}_a)+
{1\over {\eta \sqrt{p^2}}}(T_{Ra}\vec p\cdot {\vec \pi}_a-\epsilon_{Ra}
\vec p\cdot {\vec \rho}_a)]\nonumber \\
{\vec x}_i&=&{\vec {\tilde x}}+{1\over {\eta \sqrt{p^2}}}\sum_{a=1}^{N-1}
[T_{Ra}{\vec \pi}_a-\epsilon_{Ra}{\vec \rho}_a)+{ {\vec p\cdot {\vec \rho}_a
{\vec \pi}_a-\vec p\cdot {\vec \pi}_a{\vec \rho}_a}\over {p^o+\eta
\sqrt{p^2}} }+\nonumber \\
&+& {{\vec p}\over {\eta \sqrt{p^2}} }{ {T_{Ra}\vec p\cdot {\vec \pi}_a-
\epsilon_{Ra}\vec p\cdot {\vec \rho}_a}\over {p^o+\eta \sqrt{p^2}} }]+
{1\over {\sqrt{N}} }\sum_{a=1}^{N-1}{\hat \gamma}_{ai}[{\vec \rho}_a+{{\vec p}
\over {\eta \sqrt{p^2}} }(T_{Ra}+{ {\vec p\cdot {\vec \rho}_a}\over {p^o+
\eta \sqrt{p^2}} })]\nonumber \\
p^o_i&=&{{p^o}\over N}+{{\sqrt{N}}\over {\eta \sqrt{p^2}} }\sum_{a=1}^{N-1}
{\hat \gamma}_{ai}(p^o\epsilon_{Ra}+\vec p\cdot {\vec \pi}_a)\nonumber \\
{\vec p}_i&=&{{\vec p}\over N}+\sqrt{N}\sum_{a=1}^{N-1}{\hat \gamma}_{ai}
[{\vec \pi}_a+{{\vec p}\over {\eta \sqrt{p^2}} }(\epsilon_{Ra}+ { {\vec p\cdot
{\vec \pi}_a}\over {p^o+\eta \sqrt{p^2}} })].
\label{20}
\end{eqnarray}

As shown in Ref.\cite{longhi}, it is then convenient to perform the following
canonical transformation

\begin{equation}
\begin{minipage}[t]{3cm}
\begin{tabular}{|ll|l|} \hline
\multicolumn{2}{|c}{${\tilde x}^{\mu}$} & {$p^{\mu}$} \\ \hline
{}  &  {}  &  $T_{Ra}$  \\
${\vec \rho}_a$  &  ${\vec \pi}_a$  &  {}  \\
{}  &  {}  &  $\epsilon_{Ra}$  \\ \hline
\end{tabular}
\end{minipage} \ {\longrightarrow \hspace{.2cm}} \
\begin{minipage}[t]{3cm}
\begin{tabular}{|ll|l|} \hline
{}  &  {}  &  T  \\
$\vec z$  &  $\vec k$  & {}  \\
{}  &  {}  &  $\epsilon$  \\ \hline
{}  &  {}  &  $T_{Ra}$  \\
${\vec \rho}_a$  &  ${\vec \pi}_a$  &  {}  \\
{}  &  {}  &  $\epsilon_{Ra}$  \\ \hline
\end{tabular}
\end{minipage}
\label{21}
\end{equation}

\noindent defined by

\begin{eqnarray}
&&\epsilon =\eta \sqrt{p^2}  \nonumber \\
&&T={{p\cdot {\tilde x}}\over {\eta \sqrt{p^2}} }={{p\cdot x}\over {\eta
\sqrt{p^2}} }=u(p)\cdot x  \nonumber \\
&&\vec k=\vec u(p)={{\vec p}\over {\eta \sqrt{p^2}} },\quad k^o=u^o(p)=\sqrt{1+
{\vec k}^2}\nonumber \\
&&\vec z={\eta \sqrt{p^2}}({\vec {\tilde x}}-{{\vec p}\over {p^o}}{\tilde x}^o)
\nonumber \\
&&{}  \nonumber \\
&&\lbrace z^i,k^j\rbrace =\delta^{ij}, \quad\quad \lbrace T,\epsilon \rbrace
=-1.
\label{22}
\end{eqnarray}

\noindent Its inverse is

\begin{eqnarray}
&&{\tilde x}^o=\sqrt{1+{\vec k}^2}(T+{{\vec k\cdot \vec z}\over {\epsilon}})
\nonumber \\
&&{\vec {\tilde x}}={{\vec z}\over {\epsilon}}+(T+{{\vec k\cdot \vec z}\over
{\epsilon}})\, \vec k\nonumber \\
&&p^o=\epsilon \sqrt{1+{\vec k}^2}\nonumber \\
&&\vec p=\epsilon \vec k,
\label{23}
\end{eqnarray}

\noindent and the form of the Poincar\'e generators is now

\begin{eqnarray}
&&p^{\mu}=\epsilon k^{\mu}\nonumber \\
&&J^{ij}=z^ik^j-z^jk^i+\delta^{ir}\delta^{js}{\bar S}^{rs}\nonumber \\
&&J^{oi}=-z^ik^o+{{\delta^{ir}{\bar S}^{rs}p^s}\over {\epsilon (1+\sqrt{1+
{\vec k}^2})} },\quad\quad J^{io}=-J^{oi}.
\label{24}
\end{eqnarray}

Here
T is the Lorentz-scalar rest-frame time and $\epsilon$ is the invariant mass
of the isolated system of N particles. Actually, Eqs.(\ref{21}) define as many
canonical transformations as there are  disjoint branches of the mass spectrum
(half with $\eta =sign\, p^0 > 0$ and half with $\eta < 0$). All the
noncovariance of ${\tilde x}^{\mu}$ has been shifted to the 3-vector
$\vec z$ [it has the Euclidean covariance (see Appendix A) of the
kinematical stability subgroup of the Poincar\'e group of timelike orbits]. The
noncovariance of ${\tilde x}^{\mu}$ and $\vec z$ is
a consequence of, also in the
free case, the Lorentz signature of Minkowski spacetime and of the
geometry of timelike Poincar\'e orbits, when one uses the associated instant
form of dynamics according to Dirac\cite{diracc}. In contrast to what is often
said, it is not due to the
relativistic version of the No-Interaction-Theorem: until it was recognized
that
this theorem is independent of the Lorentz
signature\cite{pons}, it was difficult to separate this phenomenon from the
complications of relativistic kinematics. In the canonical subspace $\vec z,
\vec k$ (treated independently from the canonical pair $T, \epsilon$, whose
presence is due to the covariant description with first class constraints)
one has $\vec z=\eta \sqrt{p^2} \vec \kappa$, where the canonical
noncovariant three-coordinate $\vec \kappa =-\vec K/p^o+{\vec {\bar S}}
\times \vec p/p^o(p^o+\eta \sqrt{p^2})$ [from Eq.(\ref{24}) with $K^i=J^{oi}$]
can be shown to depend only on the Poincar\'e generators (see the theory
of the canonical realizations of the Poincar\'e group\cite{pp} as a special
case of the canonical realizations of Lie groups\cite{ppp}), to be the
classical basis of the Newton-Wigner three-position operator\cite{nw}, and to
have free motion without classical zitterbewegung\cite{pauri} (the problem
of classical zitterbewegung will be clarified in Ref.\cite{luc}).
While these results
were obtained in the standard instant form of dynamics $x^o=const.$, now they
are recovered (in terms of $\vec z$) in a covariant version of the instant
form, the ``rest-frame instant form $T=const.$".
The vector $\vec z$ describes the Jacobi
data for the center-of-mass three-variable in this rest-frame instant form
and $J^{\mu\nu}$ in Eqs.(\ref{24}) is split in two terms
that are separately constants of
the motion; in this respect, ${\tilde x}^{\mu}$ has all the properties, except
covariance, of the Foldy-Wouthuysen mean position for the electron\cite{fw}
(see also Ref.\cite{pauri},
 where a covariant canonical zitterbewegung-free ${\check
x}^{\mu}$ is shown to exist for certain pole-dipole systems, though no
statement
is made about extended systems). As shown in Ref.\cite{pauri,lusc,lush}, there
are other two notions of center of mass that can be build solely in terms of
the
Poincar\'e generators, coincide with $\vec \kappa$ in the rest frame and have
the same three-velocity of $\vec \kappa$. In the instant form $x^o=const.$,
they are the Fokker center of inertia\cite{fok} $\vec Y$ and the Moeller
center of energy\cite{moll} $\vec R$; in the rest-frame instant form $T=
const.$ they become $\eta \sqrt{p^2}\vec Y$ and $\eta \sqrt{p^2}\vec R$,
respectively. The vector
$\vec Y$ is covariant but not canonical ($\lbrace Y^i,Y^j\rbrace
\not= 0$), while $\vec R$ is neither covariant nor canonical. Since in the rest
frame $\vec \kappa =\vec Y=\vec R$, in every frame the Fokker center of inertia
describes by construction the worldline obtained by applying to the rest-frame
$\vec \kappa$ the Lorentz transformation appropriate for going from the rest
frame to the given frame. In contrast, $\vec R$ is defined by replacing
the masses $m_i$ in the Newtonian center of mass by the energies $p^o_i$.
Both $\vec \kappa$ and $\vec R$ define in every frame a pseudo-worldline;
if we draw all these pseudo-worldlines in a given reference frame, we
get\cite{pauri,moll} a worldtube around the worldline of the Fokker center
of inertia, whose scalar intrinsic radius
$\rho$\cite{moll} is determined by the
Poincar\'e Casimirs for timelike orbits with spin

\begin{equation}
\rho ={{\sqrt{-W^2}}\over {p^2}}={{|\, {\vec {\bar S}}\, |}\over {\sqrt{p^2}}
}.
\label{25}
\end{equation}

\noindent See Refs.\cite{lusc,lush} for the remarkable properties of the
worldtube and for the proposal to use $\rho$ as a ultraviolet cutoff.

It should be remarked that the definitions of $\vec \kappa , \vec Y, \vec R$
are
naturally given in phase space. Even in the free case ($q^{\mu}_i=x^{\mu}_i$),
their definitions in configuration space in terms of the $q^{\mu}_i$'s are
very involved, being velocity-dependent. This is the reason why the
Newtonian definition $\vec x=\sum_{i=1}^Nm_i{\vec q}_i/\sum_{i=1}^Nm_i$ in
configuration space does not extend trivially to the relativistic case.

However the canonical variables of Eqs.(\ref{21}), (\ref{22}) are not the
final ones adapted to the first class constraints of Eq.(\ref{1}). This can
be seen by replacing those constraints with the following linear combinations

\begin{eqnarray}
\chi &=&N\sum_{i=1}^N\phi_i=p^2-N\sum_{i=1}^N[m^2_i-N\sum_{b,c=1}^{N-1}{\hat
\gamma}_{bi}{\hat \gamma}_{ci}Q_c\cdot Q_c]=\nonumber \\
&=&p^2-N(\sum_{i=1}^N\, m^2_i-N\sum_{b=1}^{N-1}Q_b^2)=\epsilon^2-N[\sum_{i=1}^N
\, m^2_i-N\sum_{b=1}^{N-1}(\epsilon^2_{Rb}-{\vec \pi}_b^2)]\approx 0
\nonumber \\
\chi_a&=&{{\sqrt{N}}\over 2}\sum_{i=1}^N{\hat \gamma}_{ai}\phi_i=p\cdot Q_a-
{{\sqrt{N}}\over 2}\sum_{i=1}^N{\hat \gamma}_{ai}[m^2_i-N\sum_{b,c=1}^{N-1}
{\hat \gamma}_{bi}{\hat \gamma}_{ci}Q_c\cdot Q_c]=\nonumber \\
&=&p_{\mu}[Q^{\mu}_a-{{\sqrt{N}}\over 2}{{p^{\mu}}\over {p^2}}\sum_{i=1}^N
{\hat \gamma}_{ai}(m^2_i-N\sum_{b,c=1}^{N-1}
{\hat \gamma}_{bi}{\hat \gamma}_{ci}Q_c\cdot Q_c)]=\nonumber \\
&=&\epsilon \epsilon_{Ra}-{{\sqrt{N}}\over 2}\sum_{i=1}^N{\hat \gamma}_{ai}
[m^2_i-N\sum_{b,c=1}^{N-1}{\hat \gamma}_{bi}{\hat \gamma}_{ci}(\epsilon_{Rb}
\epsilon_{Rc}-{\vec \pi}_b\cdot {\vec \pi}_c)]=\epsilon {\hat \epsilon}_{Ra}
\approx 0\nonumber \\
&{}&\nonumber \\
\phi_i&=&{1\over {N^2}}\chi +{2\over {\sqrt{N}} }\sum_{a=1}^{N-1}{\hat \gamma}
_{ai}\chi_a ={1\over {N^2}}\chi +{{2\epsilon}\over {\sqrt{N}} }\sum_{a=1}^{N-1}
{\hat \gamma}_{ai}{\hat \epsilon}_{Ra}\approx 0,
\label{26}
\end{eqnarray}

\noindent where ${\hat \epsilon}_{Ra}$ will be explicitly defined in Eqs.
(\ref{31}) later on.

We see that the N-1 constraints $\chi_a\approx 0$ do not determine the
$\epsilon_{Ra}$ directly, but they form a system of N-1 quadratic algebraic
equations for them. With the constraint $\chi \approx 0$ we obtain
a system of N equations for $\epsilon$ and the $\epsilon_{Ra}$'s.
By substitution we could arrive at a single ($\epsilon$-dependent) equation for
each $\epsilon_{Ra}$ of order $2^{N-1}$; therefore, in general there will be
$2^{N-1}$ sets of different solutions for the $\epsilon_{Ra}$'s, a=1,..,N-1.
Each one of these sets of solutions,
substituted into $\chi \approx 0$, should give
the equation for the spectrum of the invariant mass $\epsilon$ of the isolated
system of N particles in terms of $m^2_i$ and ${\vec \pi}_b\cdot {\vec \pi}_c$.
Since each free particle has two disjoint branches of its mass spectrum
($\eta_i\sqrt{p^2_i}\approx \pm m_i$), the whole system will have $2^N$
disjoint branches for $\epsilon$ ($2^{N-1}$ for $p^2=\epsilon^2$); $2^N$ is
a topological number, namely, the dimension of the zeroth homoty group of the
mass-spectrum hypersurface in phase space. Even in the free case, these
branches
are not known because of the difficulty of solving the nonlinear equations for
the $\epsilon_{Ra}$'s.

By defining $H_D=\lambda (\tau )\chi +\sum_{a=1}^{N-1}\lambda_a(\tau )\chi_a$,
one sees that the splittings $J^{\mu\nu}=L^{\mu\nu}+S^{\mu\nu}={\tilde L}
^{\mu\nu}+{\tilde S}^{\mu\nu}$ do not define separate constants of the motion
$L^{\mu\nu}, S^{\mu\nu}$ or ${\tilde L}^{\mu\nu}, {\tilde S}^{\mu\nu}$; only
Eqs.(\ref{24}) split $J^{\mu\nu}$ into a pair of constants of the motion.

In any case, Eqs.(\ref{26}) suggest the following nonlinear (point in the
momenta) canonical transformation\cite{lusf} of the canonical variables of
Eqs.(\ref{5})

\begin{equation}
\begin{minipage}[t]{2cm}
\begin{tabular}{|ll|} \hline
$x^{\mu}$    &  $p^{\mu}$  \\ \hline
$R^{\mu}_a$  & $Q^{\mu}_a{}$ \\ \hline
\end{tabular}
\end{minipage} \ {\longrightarrow \hspace{.2cm}} \
\begin{minipage}[t]{2cm}
\begin{tabular}{|ll|} \hline
${\hat x}^{\mu}$    &  $p^{\mu}$  \\ \hline
${\hat R}^{\mu}_a$  & ${\hat Q}^{\mu}_a{}$ \\ \hline
\end{tabular}
\end{minipage}
\label{27}
\end{equation}

\begin{eqnarray}
{\hat x}^{\mu}&=&x^{\mu}+{N\over {2p^2}}(\eta^{\mu\nu}-{{2p^{\mu}p^{\nu}}\over
{p^2}})\sum_{i=1}^N(m^2_i-N\sum_{u,v=1}^{N-1}{\hat \gamma}_{ui}{\hat \gamma}
_{vi}Q_u\cdot Q_v)\times \nonumber \\
&\times&[{1\over {\sqrt{N}}}\sum_{a=1}^{N-1}{\hat \gamma}_{ai}
R_{a\nu}-\nonumber \\
&-&\sum_{b,c=1}^{N-1}p\cdot R_bQ_{c\nu}\sum_{j=1}^N{ {(N\delta_{ij}-1){\hat
\gamma}_{cj}}\over {p^2+N\sqrt{N}\sum_{d=1}^{N-1}{\hat \gamma}_{dj}p\cdot
Q_d} }\times \nonumber \\
&\times&({\hat \gamma}_{bj}+
{  {\sum_{k=1}^N
   { {{\hat \gamma}_{bk}\sum_{u=1}^{N-1}{\hat \gamma}_{uk}p\cdot Q_u}\over
     {p^2+N\sqrt{N}\sum_{v=1}^{N-1}{\hat \gamma}_{vk}p\cdot Q_v} }
                                                                    }\over
   {\sum_{h=1}^N
   { {\sum_{u=1}^{N-1}{\hat \gamma}_{uh}p\cdot Q_u}\over
     {p^2+N\sqrt{N}\sum_{v=1}^{N-1}{\hat \gamma}_{vh}p\cdot Q_v} } }  })]=
\nonumber \\
&=&x^{\mu}+{N\over {2\epsilon^2}}\sum_{i=1}^N[m^2_i-N\sum_{u,v=1}^{N-1}{\hat
\gamma}_{ui}{\hat \gamma}_{vi}(\epsilon_{Ru}\epsilon_{Rv}-{\vec \pi}_u\cdot
{\vec \pi}_v)]\times \nonumber \\
&\times&[{1\over {\sqrt{N}}}\sum_{a=1}^{N-1}{\hat \gamma}_{ai}
(\epsilon^{\mu}_r(u(p))\rho^r_a-u^{\mu}(p)T_{Ra})-\nonumber \\
&-&\sum_{b,c=1}^{N-1}T_{Rb}(\epsilon^{\mu}_r(u(p))\pi^r_a-u^{\mu}(p)\epsilon
_{Ra})\sum_{j=1}^N { {(N\delta_{ij}-1){\hat \gamma}_{cj}}\over
{\epsilon +N\sqrt{N}\sum_{d=1}^{N-1}{\hat \gamma}_{dj}\epsilon_{Rd}} }\times
\nonumber \\
&\times& ({\hat \gamma}_{bj}+
{  {\sum_{k=1}^N
   { {{\hat \gamma}_{bk}\sum_{u=1}^{N-1}{\hat \gamma}_{uk}\epsilon_{Ru}}\over
     {\epsilon +N\sqrt{N}\sum_{v=1}^{N-1}{\hat \gamma}_{vk}\epsilon_{Rv}} }
                                                                    }\over
   {\sum_{h=1}^N
   { {\sum_{u=1}^{N-1}{\hat \gamma}_{uh}\epsilon_{Ru}}\over
     {\epsilon +N\sqrt{N}\sum_{v=1}^{N-1}{\hat \gamma}_{vh}\epsilon_{Rv}} } }
}
)]\nonumber \\
p^{\mu}&=&p^{\mu}\nonumber \\
{\hat R}^{\mu}_a&=&R^{\mu}_a-n\sqrt{N}\sum_{b,c=1}^{N-1}p\cdot R_bQ^{\mu}_c
\sum_{i=1}^N{ {{\hat \gamma}_{ai}{\hat \gamma}_{ci}}\over {p^2+N\sqrt{N}
\sum_{d=1}^{N-1}{\hat \gamma}_{di}p\cdot Q_d} }\times \nonumber \\
&\times& ({\hat \gamma}_{bi}+
{  {\sum_{k=1}^N
   { {{\hat \gamma}_{bk}\sum_{u=1}^{N-1}{\hat \gamma}_{uk}p\cdot Q_u}\over
     {p^2+N\sqrt{N}\sum_{v=1}^{N-1}{\hat \gamma}_{vk}p\cdot Q_v} }
                                                                    }\over
   {\sum_{h=1}^N
   { {\sum_{u=1}^{N-1}{\hat \gamma}_{uh}p\cdot Q_u}\over
     {p^2+N\sqrt{N}\sum_{v=1}^{N-1}{\hat \gamma}_{vh}p\cdot Q_v} } }  })]=
\nonumber \\
&=&R^{\mu}_a-n\sqrt{N}\sum_{b,c=1}^{N-1}T_{Rb}Q^{\mu}_c\sum_{i=1}^N
{ {{\hat \gamma}_{ai}{\hat \gamma}_{ci}}\over {\epsilon +N\sqrt{N}\sum_{d=1}
^{N-1}{\hat \gamma}_{di}\epsilon_{Rd}} }\times \nonumber \\
&\times&({\hat \gamma}_{bi}+
{  {\sum_{k=1}^N
   { {{\hat \gamma}_{bk}\sum_{u=1}^{N-1}{\hat \gamma}_{uk}\epsilon_{Ru}}\over
     {\epsilon +N\sqrt{N}\sum_{v=1}^{N-1}{\hat \gamma}_{vk}\epsilon_{Rv}} }
                                                                    }\over
   {\sum_{h=1}^N
   { {\sum_{u=1}^{N-1}{\hat \gamma}_{uh}\epsilon_{Ru}}\over
     {\epsilon +N\sqrt{N}\sum_{v=1}^{N-1}{\hat \gamma}_{vh}\epsilon_{Rv}} } }
}
)]\nonumber \\
{\hat Q}^{\mu}_a&=&Q^{\mu}_a-{{\sqrt{N}}\over 2}{{p^{\mu}}\over
{p^2}}\sum_{i=1}
^N{\hat \gamma}_{ai}(m^2_i-N\sum_{b,c=1}^{N-1}{\hat \gamma}_{bi}{\hat \gamma}
_{ci}Q_b\cdot Q_c)=\nonumber \\
&=&Q^{\mu}_a-{{\sqrt{N}}\over 2}{{p^{\mu}}\over {p^2}}\sum_{i=1}
^N{\hat \gamma}_{ai}[m^2_i-N\sum_{b,c=1}^{N-1}{\hat \gamma}_{bi}{\hat \gamma}
_{ci}(\epsilon_{Rb}\epsilon_{Rc}-{\vec \pi}_b\cdot {\vec \pi}_c)].
\label{28}
\end{eqnarray}

We now  have [it can be checked that eqs.(\ref{10}) are reproduced by inserting
${\hat x}^{\mu}$, ${\hat R}^{\mu}_a$, ${\hat Q}^{\mu}_a$ from Eqs.(\ref{28})]

\begin{eqnarray}
J^{\mu\nu}&=&{\hat L}^{\mu\nu}+{\hat S}^{\mu\nu}\nonumber \\
&&{\hat L}^{\mu\nu}={\hat x}^{\mu}p^{\nu}-{\hat x}^{\nu}p^{\mu}\nonumber \\
&&{\hat S}^{\mu\nu}=\sum_{a=1}^{N-1}({\hat R}^{\mu}_a{\hat Q}^{\nu}_a-{\hat R}
^{\nu}_a{\hat Q}^{\mu}_a).
\label{29}
\end{eqnarray}

The subsequent canonical transformation

\begin{equation}
\begin{minipage}[t]{2cm}
\begin{tabular}{|ll|} \hline
${\hat x}^{\mu}$    &  $p^{\mu}$  \\ \hline
${\hat R}^{\mu}_a$  & ${\hat Q}^{\mu}_a{}$ \\ \hline
\end{tabular}
\end{minipage} \ {\longrightarrow \hspace{.2cm}} \
\begin{minipage}[t]{3cm}
\begin{tabular}{|ll|l|} \hline
\multicolumn{2}{|c}{${\hat {\tilde x}}^{\mu}$} & { $p^{\mu}$} \\ \hline
{}  &  {}  & ${\hat T}_{Ra}$  \\
${\hat {\vec \rho}}_a$  &  ${\hat {\vec \pi}}_a$  &  {}  \\
{}  &  {}  &  ${\hat \epsilon}_{Ra}$  \\ \hline
\end{tabular}
\end{minipage}
\label{30}
\end{equation}

\noindent defines the following analogues of the variables of Eqs.(\ref{13})

\begin{eqnarray}
{\hat {\tilde x}}^{\mu}&=&{\hat x}^{\mu}+{1\over 2}\, \epsilon^A_{\nu}(u(p))
\eta_{AB}{ {\partial \epsilon^B_{\rho}(u(p))}\over {\partial p_{\mu}} }\,
{\hat S}^{\nu\rho}=\nonumber \\
&=&{\hat x}^{\mu}-{ 1\over {\eta \sqrt{p^2}(p^o+\eta \sqrt{p^2})} }\, [p_{\nu}
{\hat S}^{\nu\mu}+\eta \sqrt{p^2} ({\hat S}^{o\mu}-{\hat S}^{o\nu}
{ {p_{\nu}p^{\mu}}\over {p^2} })]\nonumber \\
p^{\mu}&=&p^{\mu}\nonumber \\
{\hat T}_{Ra}&=&{ {p\cdot {\hat R}_a}\over {\eta \sqrt{p^2}} }=T_{Ra}-N\sqrt{N}
\epsilon \sum_{b,c=1}^{N-1}T_{Rb}\epsilon_{Rc}\times \nonumber \\
&\times& \sum_{i=1}^N
{ {{\hat \gamma}_{ai}{\hat \gamma}_{ci}}\over {\epsilon +N\sqrt{N}\sum_{d=1}
^{N-1}{\hat \gamma}_{di}\epsilon_{Rd}} }
({\hat \gamma}_{bi}+
{  {\sum_{k=1}^N
   { {{\hat \gamma}_{bk}\sum_{u=1}^{N-1}{\hat \gamma}_{uk}\epsilon_{Ru}}\over
     {\epsilon +N\sqrt{N}\sum_{v=1}^{N-1}{\hat \gamma}_{vk}\epsilon_{Rv}} }
                                                                    }\over
   {\sum_{h=1}^N
   { {\sum_{u=1}^{N-1}{\hat \gamma}_{uh}\epsilon_{Ru}}\over
     {\epsilon +N\sqrt{N}\sum_{v=1}^{N-1}{\hat \gamma}_{vh}\epsilon_{Rv}} } }
}
)]\nonumber \\
{\hat \epsilon}_{Ra}&=&{{p\cdot {\hat Q}_a}\over {\eta \sqrt{p^2}}}={1\over
{\epsilon}}\chi_a=\nonumber \\
&=&{1\over {\epsilon}}[\epsilon \epsilon_{Ra}-{{\sqrt{N}}\over 2}\sum_{i=1}^N
{\hat \gamma}_{ai}(m^2_i-N\sum_{b,c=1}^{N-1}{\hat \gamma}_{bi}
{\hat \gamma}_{ci}[\epsilon_{Rb}\epsilon_{Rc}-{\vec \pi}_b\cdot {\vec \pi}_c])]
\approx 0\nonumber \\
{\hat {\vec \rho}}_a&=&{\vec \rho}_a-N\sqrt{N}\sum_{b,c=1}^{N-1}T_{Rb}{\vec
\pi}
_c\times \nonumber \\
&\times& \sum_{i=1}^N
{ {{\hat \gamma}_{ai}{\hat \gamma}_{ci}}\over {\epsilon +N\sqrt{N}\sum_{d=1}
^{N-1}{\hat \gamma}_{di}\epsilon_{Rd}} }
({\hat \gamma}_{bi}+
{  {\sum_{k=1}^N
   { {{\hat \gamma}_{bk}\sum_{u=1}^{N-1}{\hat \gamma}_{uk}\epsilon_{Ru}}\over
     {\epsilon +N\sqrt{N}\sum_{v=1}^{N-1}{\hat \gamma}_{vk}\epsilon_{Rv}} }
                                                                    }\over
   {\sum_{h=1}^N
   { {\sum_{u=1}^{N-1}{\hat \gamma}_{uh}\epsilon_{Ru}}\over
     {\epsilon +N\sqrt{N}\sum_{v=1}^{N-1}{\hat \gamma}_{vh}\epsilon_{Rv}} } }
}
)]\nonumber \\
{\hat {\vec \pi}}_a&=&{\vec \pi}_a.
\label{31}
\end{eqnarray}

We now  have

\begin{eqnarray}
J^{\mu\nu}&=&{\hat {\tilde L}}^{\mu\nu}+{\hat {\tilde S}}^{\mu\nu}\nonumber \\
&&{\hat {\tilde L}}^{\mu\nu}={\hat {\tilde x}}^{\mu}p^{\nu}-{\hat {\tilde x}}
^{\nu}p^{\mu}\nonumber \\
&&{\hat {\tilde S}}^{\mu\nu}={\hat S}^{\mu\nu}-{1\over
2}\epsilon^A_{\rho}(u(p))
\eta_{AB}({ {\partial \epsilon^B_{\sigma}(u(p))}\over {\partial p_{\mu}} }\,
p^{\nu}-{ {\partial \epsilon^B_{\sigma}(u(p))}\over {\partial p_{\nu}} }\,
p^{\mu}){\hat S}^{\rho\sigma}=\nonumber \\
&&={\hat S}^{\mu\nu}+{1\over {\eta \sqrt{p^2}(p^o+\eta \sqrt{p^2})} }
[p_{\beta}
({\hat S}^{\beta\mu}p^{\nu}-{\hat S}^{\beta\nu}p^{\mu})+\eta \sqrt{p^2}
({\hat S}^{o\mu}p^{\nu}-{\hat S}^{o\nu}p^{\mu})]\nonumber \\
&&{}\nonumber \\
&&{\hat {\tilde S}}^{oi}=-{1\over {\eta \sqrt{p^2}} }[(p^o-\eta \sqrt{p^2})
{\hat S}^{oi}+{ {p^k({\hat S}^{ko}p^i-{\hat S}^{ki}p^o)}\over
{p^o+\eta \sqrt{p^2}} }]\nonumber \\
&&{\hat {\tilde S}}^{ij}={\hat S}^{ij}+{1\over {\eta \sqrt{p^2}} }({\hat S}
^{oi}p^j-{\hat S}^{oj}p^i)-{ {p^k({\hat S}^{ki}p^j-{\hat S}^{kj}p^i)}\over
{\eta \sqrt{p^2}(p^o+\eta \sqrt{p^2})} }\nonumber \\
&&{}\nonumber \\
&&{\hat {\bar S}}^{AB}=\epsilon^A_{\mu}(u(p))\epsilon^B_{\nu}(u(p)){\hat S}
^{\mu\nu}\nonumber \\
&&{}\nonumber \\
&&{\hat {\bar S}}^{\bar or}=\sum_{a=1}^{N-1}({\hat T}_{Ra}\pi^r_a-
{\hat \rho}^r_a{\hat \epsilon}_{Ra})\nonumber \\
&&{\hat {\bar S}}^{rs}=\sum_{a=1}^{N-1}({\hat \rho}^r_a\pi^s_a-
{\hat \rho}^s_a\pi^r_a)\nonumber \\
&&{}\nonumber \\
&&{\hat {\bar S}}^r={1\over 2}\epsilon^{rst}{\hat {\bar S}}^{st}=\sum_{a=1}
^{N-1}{\hat {\bar S}}^r_a,
\quad\quad {\hat {\vec {\bar S}}}_a={\hat {\vec \rho}}_a\times {\vec \pi}_a
\nonumber \\
&&{}\nonumber \\
&&{\hat {\tilde x}}^{\mu}={\hat x}^{\mu}-{1\over {\eta \sqrt{p^2}} }[\eta^{\mu}
_A({\hat {\bar S}}^{\bar oA}-{ {{\hat {\bar S}}^{Ar}p^r}\over {p^o+\eta
\sqrt{p^2}} })+{ {p^{\mu}+2\eta \sqrt{p^2}\eta^{\mu o}}\over {\eta \sqrt{p^2}
(p^o+\eta \sqrt{p^2})} } {\hat {\bar S}}^{\bar or}p^r].
\label{32}
\end{eqnarray}

Note that if we put $T_{Ra}=0$, a=1,..,N-1, then we get

\begin{eqnarray}
&&{\hat T}_{Ra}=0,\nonumber \\
&&{\hat {\vec \rho}}_a={\vec \rho}_a,\nonumber \\
&&{\hat {\bar S}}^i={\bar S}^i,\nonumber \\
&&{\hat {\bar S}}^{\bar or}=-\sum_{a=1}^{N-1}\rho^r_a{\hat \epsilon}_{Ra}
\approx 0,\nonumber \\
&&{\hat {\tilde x}}^{\mu}\approx {\hat x}^{\mu}+{ {\eta^{\mu}_s{\hat {\bar S}}
^{sr}p^r}\over {\eta \sqrt{p^2}(p^o+\eta \sqrt{p^2})} },\nonumber \\
&&p\cdot {\hat {\tilde x}}=p\cdot {\hat x}=p\cdot x.
\label{33}
\end{eqnarray}

The final canonical transformation, the analogue of Eq.(\ref{21}), is

\begin{equation}
\begin{minipage}[t]{3cm}
\begin{tabular}{|ll|l|} \hline
\multicolumn{2}{|c}{${\hat {\tilde x}}^{\mu}$} & {$p^{\mu}$} \\ \hline
{}  &  {}  &  ${\hat T}_{Ra}$  \\
${\hat {\vec \rho}}_a$  &  ${\vec \pi}_a$  &  {}  \\
{}  &  {}  &  ${\hat \epsilon}_{Ra}$  \\ \hline
\end{tabular}
\end{minipage} \ {\longrightarrow \hspace{.2cm}} \
\begin{minipage}[t]{3cm}
\begin{tabular}{|ll|l|} \hline
{}  &  {}  &  ${\hat T}$  \\
${\hat {\vec z}}$  &  $\vec k$  & {}  \\
{}  &  {}  &  $\epsilon$  \\ \hline
{}  &  {}  &  ${\hat T}_{Ra}$  \\
${\hat {\vec \rho}}_a$  &  ${\vec \pi}_a$  &  {}  \\
{}  &  {}  &  ${\hat \epsilon}_{Ra}$  \\ \hline
\end{tabular}
\end{minipage}
\label{34}
\end{equation}

\noindent with

\begin{eqnarray}
&&\hat T={{p\cdot {\hat {\tilde x}}}\over {\eta \sqrt{p^2}}}={{p\cdot {\hat x}}
\over {\eta \sqrt{p^2}}}\nonumber \\
&&\epsilon =\eta \sqrt{p^2}\nonumber \\
&&{\hat {\vec z}}=\eta \sqrt{p^2}({\hat {\vec {\tilde x}}}-{{\vec p}\over
{p^o}}{\hat {\tilde x}}^o)\nonumber \\
&&\vec k={{\vec p}\over {\eta \sqrt{p^2}}}.
\label{35}
\end{eqnarray}

The canonical basis of Eqs.(\ref{34}) is a quasi-Shanmugadhasan basis because
the N-1 relative energy variables vanish if the constraints $\chi_a
\approx 0$ are used;
therefore, it seems suitable for the description of interactions, which
yield relativistic bound states without relative energy excitations at the
quantum level. Only the mass-spectrum constraint is not contained in the basis,
because it describes $2^N$ disjoint branches

\begin{equation}
\epsilon -\lambda_A(m_i, {\vec
\pi}_a\cdot {\vec \pi}_b)\approx 0,\quad\quad A=1,..,2^N.
\label{36}
\end{equation}

Therefore, in the free case one could define $2^N$ Shanmugadhasan canonical
transformations containing also the variables $\epsilon -\lambda_A$; this is
also possible for all Liouville integrable interactions. Moreover,
Eqs.(\ref{34})
are adapted to the Poincar\'e Casimir $p^2=\epsilon^2$, single out a rest-frame
time $\hat T$ as the natural time parameter, and show that the relative
times ${\hat T}_{Ra}$ are gauge variables: namely, it is a freedom of the
observer whether to describe the N particles with a 1-time theory (for
instance by adding the gauge-fixings ${\hat T}_{Ra}=0$) or with
an N-time theory [or even with an F-time one with $1 < F < N$].

The only problem with the canonical transformation (\ref{31}) is that
to get its explicit inversion one has to solve the system of equations

\begin{eqnarray}
{{N\sqrt{N}}\over 2}\sum_{b,c=1}^{N-1}(\sum_{i=1}^N{\hat \gamma}_{ai}
{\hat \gamma}_{bi}{\hat \gamma}_{ci}&)&\epsilon_{Rb}\epsilon_{Rc}+
\epsilon \epsilon_{Ra}-\epsilon {\hat \epsilon}_{Ra}-\nonumber \\
&-&{{\sqrt{N}}\over 2}\sum_{i=1}^N{\hat \gamma}_{ai}[m^2_i+N\sum_{b,c=1}
^{N-1}{\hat \gamma}_{bi}{\hat \gamma}_{ci}{\vec \pi}_b\cdot {\vec \pi}_c]=0,
\label{37}
\end{eqnarray}

\noindent in order
to obtain the old variables $\epsilon_{Ra}$ in terms of the new
${\hat \epsilon}_{Ra}$ and  the kinetic terms ${\vec \pi}_b\cdot {\vec \pi}
_c$. For ${\hat \epsilon}_{Ra}$, this is the same set of equations discussed
previously, whose solution is needed also to determine the branches of the
mass spectrum. So far, the inversion is known only for N=2\cite{longhi,lusf},
as can be checked in Appendix B. Actually, it turns out that one should solve
Eqs.(\ref{37}) together with the first of Eqs.(\ref{26}),

$\chi=\epsilon^2-N[\sum_{i=1}^N
\, m^2_i-N\sum_{b=1}^{N-1}(\epsilon^2_{Rb}-{\vec \pi}_b^2)]\approx 0,$

\noindent which determines the mass specturm, as a system of N equations
in the N variables $\epsilon$, $\epsilon_{Ra}$, a=1,..,N.

Equations (\ref{37}) are still untractable even if we restrict ourselves to a
rest-frame
1-time theory by adding the gauge-fixings $T_{Ra}\approx 0$, which imply
${\hat T}_{Ra}\approx 0$ and all the results of Eqs.(\ref{33}),
so that one can eliminate the N-1 pairs
of second class constraints ${\hat T}_{Ra}\approx 0$, ${\hat \epsilon}_{Ra}
\approx 0$ by going to Dirac brackets.

We shall see in the next Section that there is an approach to the 1-time
theory that allows us to find the $2^N$ branches of the mass spectrum and the
$2^{N-1}$ solutions of Eqs.(\ref{37}).

\vfill\eject

\section{The 1-time Theory}

Since in the case of a charged scalar particle interacting with the
electromagnetic field it is not clear how, in the absence of a covariant notion
of equal times, to evaluate the Poisson bracket
of the particle mass-spectrum constraint ${(p(\tau )-eA(x(\tau )))}^2-m^2
\approx 0$ and of the field primary constraint $\pi^o(\vec z,z^o)\approx 0$,
this system was reformulated in Ref.\cite{karp}
on spacelike hypersurfaces following Dirac's approach to
parametrized field theory\cite{diraca}. To describe the intersection of a
particle worldline with $\Sigma (\tau )$, only three coordinates $\vec
\sigma =\vec \eta (\tau )$ are needed and not four, since the time variable
is the parameter $\tau$ labelling the hypersurfaces of the family. But this
implies a covariant solution of the constraint $p^2-m^2\approx 0$ and a
choice of the sign $\eta =sign\, p^o$. For N free particles, one obtains a
1-time description for every choice of $\eta_i=sign\, p_i^o$ with
particle coordinates ${\vec \eta}_i(\tau )$ and with $x^{\mu}_i(\tau )=
z^{\mu}(\tau ,{\vec \eta}_i(\tau ))$ describing the i-th particle worldline
[so that ${\dot x}^{\mu}_i(\tau )=z^{\mu}_{\tau}(\tau , {\vec \eta}_i(\tau ))+
z^{\mu}_{\check r}(\tau ,{\vec \eta}_i(\tau )){\dot \eta}_i^{\check r}(\tau
)$];
moreover, $sign\, {\dot x}^o_i(\tau )=\eta_i$. See Appendix C for the
notations.

As shown in Ref.\cite{karp}, the system is described by the action

\begin{eqnarray}
S&=& \int d\tau d^3\sigma \, {\cal L}(\tau ,\vec
\sigma )=\int d\tau L(\tau )\nonumber \\
&&{\cal L}(\tau ,\vec \sigma )=-\sum_{i=1}^N\delta^3(\vec \sigma -{\vec \eta}_i
(\tau ))\eta_im_i\sqrt{ g_{\tau\tau}(\tau ,\vec \sigma )+2g_{\tau {\check r}}
(\tau ,\vec \sigma ){\dot \eta}^{\check r}_i(\tau )+g_{{\check r}{\check s}}
(\tau ,\vec \sigma ){\dot \eta}_i^{\check r}(\tau ){\dot \eta}_i^{\check s}
(\tau )  }\nonumber \\
&&L(\tau )=-\sum_{i=1}^N\eta_im_i\sqrt{ g_{\tau\tau}(\tau ,{\vec \eta}_i
(\tau ))+2g_{\tau {\check r}}(\tau ,{\vec \eta}_i(\tau )){\dot \eta}^{\check r}
_i(\tau )+g_{{\check r}{\check s}}(\tau ,{\vec \eta}_i(\tau )){\dot \eta}_i
^{\check r}(\tau ){\dot \eta}_i^{\check s}(\tau )  },
\label{38}
\end{eqnarray}

\noindent where the configuration variables are $z^{\mu}(\tau ,\vec \sigma )$
and ${\vec \eta}_i(\tau )$, i=1,..,N. The action is invariant under separate
$\tau$- and $\vec \sigma$-reparametrizations.

The canonical momenta are

\begin{eqnarray}
\rho_{\mu}(\tau ,\vec \sigma )&=&-{ {\partial {\cal L}(\tau ,\vec \sigma )}
\over {\partial z^{\mu}_{\tau}(\tau ,\vec \sigma )} }=\sum_{i=1}^N\delta^3
(\vec \sigma -{\vec \eta}_i(\tau ))\eta_im_i\nonumber \\
&&{ {z_{\tau\mu}(\tau ,\vec \sigma )+z_{{\check r}\mu}(\tau ,\vec \sigma )
{\dot \eta}_i^{\check r}(\tau )}\over {\sqrt{g_{\tau\tau}(\tau ,\vec \sigma )+
2g_{\tau {\check r}}(\tau ,\vec \sigma ){\dot \eta}_i^{\check r}(\tau )+
g_{{\check r}{\check s}}(\tau ,\vec \sigma ){\dot \eta}_i^{\check r}(\tau
){\dot
\eta}_i^{\check s}(\tau ) }} }=\nonumber \\
&=&[(\rho_{\nu}l^{\nu})l_{\mu}+(\rho_{\nu}z^{\nu}_{\check r})\gamma^{{\check r}
{\check s}}z_{{\check s}\mu}](\tau ,\vec \sigma )\nonumber \\
&&{}\nonumber \\
\kappa_{i{\check r}}(\tau )&=&-{ {\partial L(\tau )}\over {\partial {\dot
\eta}_i^{\check r}(\tau )} }=\nonumber \\
&=&\eta_im_i{ {g_{\tau {\check r}}(\tau ,{\vec \eta}_i(\tau ))+g_{{\check r}
{\check s}}(\tau ,{\vec \eta}_i(\tau )){\dot \eta}_i^{\check s}(\tau )}\over
{ \sqrt{g_{\tau\tau}(\tau ,{\vec \eta}_i(\tau ))+
2g_{\tau {\check r}}(\tau ,{\vec \eta}_i(\tau )){\dot \eta}_i^{\check r}(\tau
)+
g_{{\check r}{\check s}}(\tau ,{\vec \eta}_i(\tau )){\dot \eta}_i^{\check r}
(\tau ){\dot \eta}_i^{\check s}(\tau ) }} },
\label{39}
\end{eqnarray}

\noindent and the following Poisson brackets are assumed

\begin{eqnarray}
&&\lbrace z^{\mu}(\tau ,\vec \sigma ),\rho_{\nu}(\tau ,{\vec \sigma}^{'}\rbrace
=-\eta^{\mu}_{\nu}\delta^3(\vec \sigma -{\vec \sigma}^{'})\nonumber \\
&&\lbrace \eta^{\check r}_i(\tau ),\kappa_j^{\check s}(\tau )\rbrace =
\delta_{ij}\delta^{{\check r}{\check s}}.
\label{40}
\end{eqnarray}

{}From Eqs.(\ref{39}) one obtains the following four primary constraints by
using
Eqs.(\ref{C4}), (\ref{C5})

\begin{eqnarray}
{\cal H}_{\mu}(\tau ,\vec \sigma )&=& \rho_{\mu}(\tau ,\vec \sigma )-l_{\mu}
(\tau ,\vec \sigma )\sum_{i=1}^N\delta^3(\vec \sigma -{\vec \eta}_i(\tau ))
\eta_i\sqrt{ m^2_i-\gamma^{{\check r}{\check s}}(\tau ,\vec \sigma )
\kappa_{i{\check r}}(\tau )\kappa_{i{\check s}}(\tau ) }-\nonumber \\
&-&z_{{\check r}\mu}
(\tau ,\vec \sigma )\gamma^{{\check r}{\check s}}(\tau ,\vec \sigma )
\sum_{i=1}^N\delta^3(\vec \sigma -{\vec \eta}_i(\tau ))\kappa_{i{\check s}}
\approx 0,
\label{41}
\end{eqnarray}

\noindent which satisfy

\begin{equation}
\lbrace {\cal H}_{\mu}(\tau ,\vec \sigma ),{\cal H}_{\nu}(\tau ,{\vec \sigma}
^{'})\rbrace =0.
\label{42}
\end{equation}

Since the canonical Hamiltonian vanishes, one has the Dirac Hamiltonian
[$\lambda^{\mu}(\tau ,\vec \sigma )$ are Dirac's multipliers]

\begin{equation}
H_D=\int d^3\sigma \lambda^{\mu}(\tau ,\vec \sigma ){\cal H}_{\mu}(\tau ,\vec
\sigma ),
\label{43}
\end{equation}

\noindent and one finds that $\lbrace {\cal H}_{\mu}(\tau ,\vec \sigma ),H_D
\rbrace =0$. Therefore, there are only the four first class constraints of
Eq.(\ref{41}). The constraints ${\cal H}_{\mu}(\tau ,\vec \sigma )\approx 0$
describe the arbitrariness of the
foliation: physical results do not depend on its choice.

The conserved Poincar\'e generators are (the suffix ``s" denotes the
hypersurface $\Sigma (\tau )$)

\begin{eqnarray}
&&p^{\mu}_s=\int d^3\sigma \rho^{\mu}(\tau ,\vec \sigma )\nonumber \\
&&J_s^{\mu\nu}=\int d^3\sigma [z^{\mu}(\tau ,\vec \sigma )\rho^{\nu}(\tau ,
\vec \sigma )-z^{\nu}(\tau ,\vec \sigma )\rho^{\mu}(\tau ,\vec \sigma )],
\label{44}
\end{eqnarray}

\noindent and one has

\begin{equation}
\lbrace z^{\mu}(\tau ,\vec \sigma ),p^{\nu}_s\rbrace =-\eta^{\mu}_{\nu}
\label{45}
\end{equation}

\begin{eqnarray}
\int d^3\sigma {\cal H}_{\mu}(\tau ,\vec \sigma )&=& p_s^{\mu}-\sum_{i=1}^N
l^{\mu}(\tau ,{\vec \eta}_i(\tau ))\eta_i \sqrt{ m^2_i-\gamma^{{\check
r}{\check
s}}(\tau ,{\vec \eta}_i(\tau ))\kappa_{i{\check r}}(\tau )\kappa_{i{\check s}}
(\tau ) }-\nonumber \\
&-&\sum_{i=1}^Nz^{\mu}_{\check r}(\tau ,{\vec \eta}_i(\tau ))\gamma^{{\check
r}{\check s}}(\tau ,{\vec \eta}_i(\tau ))\kappa_{i{\check s}}(\tau )
\approx 0.
\label{46}
\end{eqnarray}

Let us now restrict ourselves to spacelike hyperplanes $\Sigma_H(\tau )$ by
imposing the gauge-fixings

\begin{eqnarray}
\zeta^{\mu}(\tau ,\vec \sigma )&=&z^{\mu}(\tau ,\vec \sigma )-x_s^{\mu}(\tau )-
b^{\mu}_{\check r}(\tau )\sigma^{\check r}\approx 0\nonumber \\
&&{}\nonumber \\
&&\lbrace \zeta^{\mu}(\tau ,\vec \sigma ),{\cal H}_{\nu}(\tau ,{\vec \sigma}
^{'})\rbrace =-\eta^{\mu}_{\nu}\delta^3(\vec \sigma -{\vec \sigma}^{'}),
\label{47}
\end{eqnarray}

\noindent where $b^{\mu}_{\check r}(\tau )$, ${\check r}=1,2,3$, are three
orthonormal vectors such that the constant (future pointing) normal to the
hyperplane is

\begin{equation}
l^{\mu}(\tau ,\vec \sigma )\approx l^{\mu}=b^{\mu}_{\tau}=\epsilon^{\mu}
{}_{\alpha\beta\gamma}b^{\alpha}_{\check 1}(\tau )b^{\beta}_{\check 2}(\tau )
b^{\gamma}_{\check 3}(\tau ).
\label{48}
\end{equation}

Therefore, we get

\begin{eqnarray}
&&z^{\mu}_{\check r}(\tau ,\vec \sigma )\approx b^{\mu}_{\check r}(\tau )
\nonumber \\
&&z^{\mu}_{\tau}(\tau ,\vec \sigma )\approx {\dot x}^{\mu}_s(\tau )+{\dot
b}^{\mu}_{\check r}(\tau )\sigma^{\check r}\nonumber \\
&&g_{{\check r}{\check s}}(\tau ,\vec \sigma )\approx -\delta_{{\check
r}{\check s}},\quad\quad \gamma^{{\check r}{\check s}}(\tau ,\vec \sigma )
\approx -\delta^{{\check r}{\check s}},\quad\quad \gamma (\tau ,\vec \sigma )
\approx 1.
\label{49}
\end{eqnarray}

By introducing the Dirac brackets for the resulting second class constraints

\begin{equation}
\lbrace A,B\rbrace {}^{*}=\lbrace A,B\rbrace -\int d^3\sigma [\lbrace A,\zeta
^{\mu}(\tau ,\vec \sigma )\rbrace \lbrace {\cal H}_{\mu}(\tau ,\vec \sigma ),
B\rbrace -\lbrace A,{\cal H}_{\mu}(\tau ,\vec \sigma )\rbrace \lbrace \zeta
^{\mu}(\tau ,\vec \sigma ),B\rbrace ],
\label{50}
\end{equation}

\noindent one finds by using Eq.(\ref{45})

\begin{equation}
\lbrace x_s^{\mu}(\tau ),p^{\nu}_s(\tau )\rbrace {}^{*}=-\eta^{\mu\nu}.
\label{51}
\end{equation}

The ten degrees of freedom describing the hyperplane are $x^{\mu}_s(\tau )$
with conjugate momentum $p^{\mu}_s$ and six variables $\phi_{\lambda}(\tau )$,
$\lambda =1,..,6$, which parametrize the orthonormal tetrad $b^{\mu}_{\check A}
(\tau )$, with their conjugate momenta $T_{\lambda}(\tau )$.

The preservation in time of the gauge-fixings $\zeta^{\mu}(\tau ,\vec \sigma )
\approx 0$ implies

\begin{equation}
{d\over {d\tau}}\zeta^{\mu}(\tau ,\vec \sigma )=\lbrace \zeta^{\mu}(\tau ,\vec
\sigma ),H_D\rbrace =-\lambda^{\mu}(\tau ,\vec \sigma )-{\dot x}^{\mu}_s(\tau
)-{\dot b}^{\mu}_{\check r}(\tau )\sigma^{\check r}\approx 0,
\label{52}
\end{equation}

\noindent so that one has [by using ${\dot b}^{\mu}_{\tau}=0$ and ${\dot b}
^{\mu}_{\check r}(\tau )b^{\nu}_{\check r}(\tau )=-b^{\mu}_{\check r}(\tau )
{\dot b}^{\nu}_{\check r}(\tau )$]

\begin{eqnarray}
\lambda^{\mu}(\tau ,\vec \sigma )&\approx&{\tilde \lambda}^{\mu}(\tau )+
{\tilde \lambda}^{\mu}{}_{\nu}(\tau )b^{\nu}_{\check r}(\tau )
\sigma^{\check r}\nonumber \\
&&{\tilde \lambda}^{\mu}(\tau )=-{\dot x}^{\mu}_s(\tau )\nonumber \\
&&{\tilde \lambda}^{\mu\nu}(\tau )=-{\tilde \lambda}^{\nu\mu}(\tau )={1\over 2}
[{\dot b}^{\mu}_{\check r}(\tau )b^{\nu}_{\check r}(\tau )-b^{\mu}_{\check r}
(\tau ){\dot b}^{\nu}_{\check r}(\tau )].
\label{53}
\end{eqnarray}

Therefore the Dirac Hamiltonian becomes

\begin{equation}
H_D={\tilde \lambda}^{\mu}(\tau ){\tilde {\cal H}}_{\mu}(\tau )-{1\over 2}
{\tilde \lambda}^{\mu\nu}(\tau ){\tilde {\cal H}}_{\mu\nu}(\tau ),
\label{54}
\end{equation}

\noindent and only the following ten first class constraints are left

\begin{eqnarray}
{\tilde {\cal H}}^{\mu}(\tau )&=&\int d^3\sigma {\cal H}^{\mu}(\tau ,\vec
\sigma )=\, p^{\mu}_s-l^{\mu}\sum_{i=1}^N\eta_i\sqrt{m^2_i+{\vec \kappa}^2_i
(\tau )}+b^{\mu}_{\check r}(\tau )\sum_{i=1}^N\kappa_{i{\check r}}(\tau )
\approx 0,\nonumber \\
{\tilde {\cal H}}^{\mu\nu}(\tau )&=&b^{\mu}_{\check r}(\tau )\int d^3\sigma
\sigma^{\check r}\, {\cal H}^{\nu}(\tau ,\vec \sigma )-b^{\nu}_{\check r}(\tau
)
\int d^3\sigma \sigma^{\check r}\, {\cal H}^{\mu}(\tau ,\vec \sigma )=
\nonumber \\
&=&S_s^{\mu\nu}(\tau )-[b^{\mu}_{\check r}(\tau )b^{\nu}_{\tau}-b^{\nu}_{\check
r}(\tau )b^{\mu}_{\tau}]\sum_{i=1}^N\eta_i^{\check r}(\tau )\eta_i
\sqrt{m^2_i+{\vec \kappa}^2_i(\tau )}-\nonumber \\
&-&[b^{\mu}_{\check r}(\tau )b^{\nu}_{\check s}(\tau )-b^{\nu}_{\check r}(\tau
)
b^{\mu}_{\check s}(\tau )]\sum_{i=1}^N\eta_i^{\check r}(\tau )\kappa_i^{\check
s}(\tau )\approx 0.
\label{55}
\end{eqnarray}

Here $S^{\mu\nu}_s$ is the spin part of the Lorentz generators

\begin{eqnarray}
J^{\mu\nu}_s&=&x^{\mu}_sp^{\nu}_s-x^{\nu}_sp^{\mu}_s+S^{\mu\nu}_s\nonumber \\
&&S^{\mu\nu}_s=b^{\mu}_{\check r}(\tau )\int d^3\sigma \sigma^{\check r}
\rho^{\nu}(\tau ,\vec \sigma )-b^{\nu}_{\check r}(\tau )\int d^3\sigma
\sigma^{\check r}\rho^{\mu}(\tau ,\vec \sigma ).
\label{56}
\end{eqnarray}

As shown in Ref.\cite{hans} (see also the Appendix of Ref.\cite{bard}),
instead of finding
$\phi_{\lambda}(\tau ), T_{\lambda}(\tau )$, one can use the redundant
variables $b^{\mu}_{\check A}(\tau ), S_s^{\mu\nu}(\tau )$, with the
following Dirac brackets assuring the validity of the orthonormality
condition $\eta^{\mu\nu}-b^{\mu}_{\check A}\eta^{{\check A}{\check b}}b^{\nu}
_{\check B}=0$ [$C^{\mu\nu\alpha\beta}_{\gamma\delta}=\eta^{\nu}_{\gamma}
\eta^{\alpha}_{\delta}\eta^{\mu\beta}+\eta^{\mu}_{\gamma}\eta^{\beta}_{\delta}
\eta^{\nu\alpha}-\eta^{\nu}_{\gamma}\eta^{\beta}_{\delta}\eta^{\mu\alpha}-
\eta^{\mu}_{\gamma}\eta^{\alpha}_{\delta}\eta^{\nu\beta}$
are the structure constants of the Lorentz group]

\begin{eqnarray}
&&\lbrace S_s^{\mu\nu},b^{\rho}_{\check A}\rbrace {}^{*}=\eta^{\rho\nu}
b^{\mu}_{\check A}-\eta^{\rho\mu}b^{\nu}_{\check A}\nonumber \\
&&\lbrace S^{\mu\nu}_s,S_s^{\alpha\beta}\rbrace {}^{*}=C^{\mu\nu\alpha\beta}
_{\gamma\delta}S_s^{\gamma\delta},
\label{57}
\end{eqnarray}

\noindent so that while ${\tilde {\cal H}}^{\mu}(\tau )\approx 0$ has zero
Dirac bracket with itself and with ${\tilde {\cal H}}^{\mu\nu}(\tau )
\approx 0$ these last six constraints have the Dirac brackets

\begin{equation}
\lbrace {\tilde {\cal H}}^{\mu\nu}(\tau ),{\tilde {\cal H}}^{\alpha\beta}
(\tau )\rbrace {}^{*}=C^{\mu\nu\alpha\beta}_{\gamma\delta}{\tilde {\cal H}}
^{\gamma\delta}(\tau )\approx 0.
\label{58}
\end{equation}

Let us now restrict ourselves to configurations with $p_s^2 > 0$ and let us
boost at rest with the Wigner boost $L^{\mu}{}_{\nu}(\stackrel{\circ}{p_s},
p_s)$ the variables $b^{\mu}_{\check A}$, $S_s^{\mu\nu}$ of the non-Darboux
basis

${}$

$x^{\mu}_s, p^{\mu}_s, b^{\mu}_{\check A}, S_s^{\mu\nu}, \eta_i^{\check r},
\kappa_i^{\check r}$

${}$

\noindent of the
Dirac brackets $\lbrace .,.\rbrace {}^{*}$ [in Refs.\cite{lusc} this step is
missing; the final results in those papers are not changed, but the reduction
associated with the following Eqs.(\ref{62}) is not sufficient to get them
as erroneously stated in Refs.\cite{lusc}].
The following new non-Darboux
basis is obtained (${\tilde x}^{\mu}_s$ is no more a fourvector)

\begin{eqnarray}
&&{\tilde x}^{\mu}_s=x_s^{\mu}+{1\over 2}\, \epsilon^A_{\nu}(u(p_s))\eta_{AB}
{ {\partial \epsilon^B_{\rho}(u(p_s))}\over {\partial p_{s\mu}} }\,
S^{\nu\rho}_s=\nonumber \\
&&=x_s^{\mu}-{ 1\over {\eta \sqrt{p_s^2}(p_s^o+\eta \sqrt{p_s^2})} }\,
[p_{s\nu}
S^{\nu\mu}_s+\eta \sqrt{p_s^2} (S_s^{o\mu}-S_s^{o\nu}{ {p_{s\nu}p_s^{\mu}}\over
{p_s^2} })]=\nonumber \\
&&=x_s^{\mu}-{1\over {\eta_s \sqrt{p_s^2}} }[\eta^{\mu}_A({\bar S}_s^{\bar oA}
-{ {{\bar S}_s^{Ar}p_s^r}\over {p_s^o+\eta_s \sqrt{p_s^2}} })+{ {p_s^{\mu}+2
\eta_s \sqrt{p_s^2}\eta^{\mu o}}\over {\eta_s \sqrt{p_s^2}(p_s^o+\eta_s
\sqrt{p_s^2})} }{\bar S}_s^{\bar or}p_s^r]\nonumber \\
&&{}\nonumber \\
&&p^{\mu}_s=p_s^{\mu}\nonumber \\
&&{}\nonumber \\
&&\eta_i^{\check r}=\eta_i^{\check r}\nonumber \\
&&\kappa_i^{\check r}=\kappa_i^{\check r}\nonumber \\
&&{}\nonumber \\
&&b^A_{\check r}=\epsilon^A_{\mu}(u(p_s))b^{\mu}_{\check r}\nonumber \\
&&{\tilde S}_s^{\mu\nu}=S^{\mu\nu}_s-{1\over 2}\epsilon^A_{\rho}(u(p_s))\eta
_{AB}({ {\partial \epsilon^B_{\sigma}(u(p_s))}\over {\partial p_{s\mu}} }\,
p^{\nu}_s-{ {\partial \epsilon^B_{\sigma}(u(p_s))}\over {\partial p_{s\nu}} }\,
 p_s^{\mu})S^{\rho\sigma}_s=\nonumber \\
&&=S^{\mu\nu}_s+{1\over {\eta \sqrt{p_s^2}(p_s^o+\eta \sqrt{p_s^2})} }
[p_{s\beta}(S^{\beta\mu}_sp_s^{\nu}-S_s^{\beta\nu}p_s^{\mu})+\eta \sqrt{p_s^2}
(S_s^{o\mu}p_s^{\nu}-S^{o\nu}_sp_s^{\mu})]\nonumber \\
&&{}\nonumber \\
&&J^{\mu\nu}_s={\tilde x}_s^{\mu}p_s^{\nu}-{\tilde x}^{\nu}_sp_s^{\mu}+{\tilde
S}_s^{\mu\nu},
\label{59}
\end{eqnarray}

We have [cf. Eq.(\ref{A7})]

\begin{eqnarray}
&&\lbrace {\tilde x}^{\mu}_s,p^{\nu}_s\rbrace {}^{*}=0\nonumber \\
&&\lbrace {\tilde S}^{oi}_s,b^r_{\check A}\rbrace {}^{*}={ {\delta^{is}(p^r_s
b^s_{\check A}-p^s_sb^r_{\check A})}\over {p^o_s+\eta_s\sqrt{p^2_s}} }
\nonumber \\
&&\lbrace {\tilde S}_s^{ij},b^r_{\check A}\rbrace {}^{*}=(\delta^{ir}\delta
^{js}-\delta^{is}\delta^{jr})b^s_{\check A}\nonumber \\
&&\lbrace {\tilde S}_s^{\mu\nu},{\tilde S}_s^{\alpha\beta}\rbrace {}^{*}=
C^{\mu\nu\alpha\beta}_{\gamma\delta}{\tilde S}_s^{\gamma\delta},
\label{60}
\end{eqnarray}

\noindent and we can define

\begin{eqnarray}
{\bar S}_s^{AB}&=&\epsilon^A_{\mu}(u(p_s))\epsilon^B_{\nu}(u(p_s))S_s^{\mu\nu}
\approx [b^A_{\check r}(\tau )b^B_{\tau}-b^B_{\check r}(\tau )b^A_{\tau}]
\sum_{i=1}^N\eta_i^{\check r}(\tau )\eta_i\sqrt{m^2_i+{\vec \kappa}_i^2(\tau
)}+
\nonumber \\
&+&[b^A_{\check r}(\tau )b^B_{\check s}(\tau )-b^B_{\check r}(\tau )b^A
_{\check s}(\tau )]\sum_{i=1}^N\eta_i^{\check r}(\tau )\kappa_i^{\check s}
(\tau ).
\label{61}
\end{eqnarray}

Let us now add six more gauge-fixings by selecting the special family of
spacelike hyperplanes orthogonal to $p^{\mu}_s$ (this is possible for
$p^2_s > 0$), which can be called the `Wigner foliation' of Minkowski
spacetime. This can be done by requiring (only six conditions are
independent)

\begin{eqnarray}
T^{\mu}_{\check A}(\tau )&=&b^{\mu}_{\check A}(\tau )-\epsilon^{\mu}
_{A={\check A}}(u(p_s))\approx 0\nonumber \\
&&\Rightarrow \quad b^A_{\check A}(\tau )=\epsilon^A_{\mu}(u(p_s))b^{\mu}
_{\check A}(\tau )\approx \eta^A_{\check A}.
\label{62}
\end{eqnarray}

Now the tetrad $b^{\mu}_{\check A}$ has become $\epsilon^{\mu}_A(u(p_s))$ and
the indices `${\check r}$' are forced to coincide with the Wigner spin-1
indices
`r', while $\bar o=\tau$ is a Lorentz-scalar index. One has

\begin{eqnarray}
{\bar S}_s^{AB}&\approx& (\eta^A_{\check r}\eta^B_{\tau}-\eta^B_{\check r}
\eta^A_{\tau})\sum_{i=1}^N\eta^{\check r}_i(\tau )\eta_i\sqrt{m_i^2+{\vec
\kappa}_i^2(\tau )}+\nonumber \\
&+&(\eta^A_{\check r}\eta^B_{\check s}-\eta^B_{\check r}\eta^A_{\check s})
\sum_{i=1}^N\eta_i^{\check r}(\tau )\kappa_i^{\check s}(\tau )\nonumber \\
&&{}\nonumber \\
{\bar S}_s^{rs}&\approx& \sum_{i=1}^N(\eta^r_i\kappa_i^s-\eta_i^s\kappa_i^r)
\nonumber \\
{\bar S}_s^{\bar or}&\approx& -{\bar S}_s^{r\bar o}=-\sum_{i=1}^N\eta^r_i\eta_i
\sqrt{m_i^2+{\vec \kappa}_i^2}\nonumber \\
&&{}\nonumber \\
J_s^{ij}&\approx& {\tilde x}^i_sp_s^j-{\tilde x}_s^jp^i_s+\delta^{ir}\delta
^{js}{\bar S}_s^{rs}\nonumber \\
J_s^{oi}&\approx&{\tilde x}^op^i_s-{\tilde x}_s^ip^o_s-{ {\delta^{ir}{\bar
S}_s^{rs}p^s_s}\over {p^o_s+\eta_s\sqrt{p^2_s}} }.
\label{63}
\end{eqnarray}

The time constancy of $T^{\mu}_{\check A}\approx 0$ with respect to the Dirac
Hamiltonian of Eq.(\ref{54}) gives

\begin{eqnarray}
{d\over {d\tau}}[b^{\mu}_{\check r}(\tau )-\epsilon^{\mu}_r(u(p_s))]&=&
\lbrace b^{\mu}_{\check r}(\tau )-\epsilon^{\mu}_r(u(p_s)),H_D\rbrace {}^{*}=
\nonumber \\
&=&{1\over 2}{\tilde \lambda}^{\alpha\beta}(\tau )\lbrace b^{\mu}_{\check r}
(\tau ),S_{s\alpha\beta}(\tau )\rbrace {}^{*}={\tilde \lambda}^{\mu\alpha}
(\tau )b_{\check r\alpha}(\tau )\approx 0\nonumber \\
&\Rightarrow& {\tilde \lambda}^{\mu\nu}(\tau )\approx 0,
\label{64}
\end{eqnarray}

\noindent so that the independent gauge-fixings contained in Eqs.(\ref{62}) and
the constraints ${\tilde {\cal H}}^{\mu\nu}(\tau )\approx 0$ form six pairs of
second class constraints.

Besides Eqs.(\ref{49}), now we have [remember that ${\dot x}_s^{\mu}(\tau )=
-{\tilde \lambda}^{\mu}(\tau )$]

\begin{eqnarray}
&&l^{\mu}=b^{\mu}_{\tau}=u^{\mu}(p_s)\nonumber \\
&&z^{\mu}_{\tau}(\tau )={\dot x}^{\mu}_s(\tau )=\sqrt{g(\tau )}u^{\mu}(p_s)-
{\dot x}_{s\nu}(\tau )\epsilon^{\mu}_r(u(p_s))\epsilon^{\nu}_r(u(p_s))
\nonumber \\
&&g(\tau )={[{\dot x}_{s\mu}(\tau )u^{\mu}(p_s))]}^2\nonumber \\
&&g_{\tau\tau}={\dot x}^2_s,\quad\quad g^{\tau\tau}={1\over g},\quad\quad
g^{\tau{\check r}}={1\over g}{\dot x}_{s\mu}\delta^{{\check r}s}\epsilon^{\mu}
_s(u(p_s))\nonumber \\
&&g_{\tau{\check r}}={\dot x}_{s\mu}\delta_{{\check r}s}\epsilon_s^{\mu}
(u(p_s)),\quad\quad g^{{\check r}{\check s}}=-\delta^{{\check r}{\check s}}+
{ {\delta^{{\check r}u}\delta^{{\check s}v}}\over {g(\tau )}}
{\dot x}_{s\mu}\epsilon^{\mu}_u(u(p_s))
{\dot x}_{s\nu}\epsilon^{\nu}_v(u(p_s)).
\label{65}
\end{eqnarray}

On the hyperplane $\Sigma_W(\tau )$ all the degrees of freedom $z^{\mu}(\tau ,
\vec \sigma )$ are reduced to the four degrees of freedom ${\tilde x}^{\mu}_s
(\tau )$, which replace $x^{\mu}_s$. The Dirac
Hamiltonian is now $H_D={\tilde \lambda}^{\mu}(\tau ){\tilde {\cal H}}_{\mu}
(\tau )$ with

\begin{equation}
{\tilde {\cal H}}^{\mu}(\tau )=p_s^{\mu}-u^{\mu}(p_s)\sum_{i=1}^N\eta_i
\sqrt{m^2_i+{\vec \kappa}^2_i(\tau )}-\epsilon^{\mu}_r(u(p_s))\sum_{i=1}^N
\kappa^r_i(\tau )\approx 0.
\label{66}
\end{equation}

To find the new Dirac brackets, one needs to evaluate the matrix of the
old Dirac brackets of the second class constraints (without extracting the
independent ones)

\begin{eqnarray}
C=\left(
\begin{array}{cccc}
\lbrace {\tilde {\cal H}}^{\alpha\beta},{\tilde {\cal H}}^{\gamma\delta}
\rbrace {}^{*}\approx 0 & \lbrace {\tilde {\cal H}}^{\alpha\beta},T^{\sigma}
_{\check B}\rbrace {}^{*}=\\
{} & =\delta_{{\check B}B}[\eta^{\sigma\beta}\epsilon
^{\alpha}_B(u(p_s))-\eta^{\sigma\alpha}\epsilon_B^{\beta}(u(p_s))] \\
\lbrace T^{\rho}_{\check A},{\tilde {\cal H}}^{\gamma\delta}\rbrace {}^{*}
= & \lbrace T^{\rho}_{\check A},T^{\sigma}_{\check B}\rbrace {}^{*}=0\\
=\delta_{{\check A}A}[\eta^{\rho\gamma}\epsilon^{\delta}_A(u(p_s))-\eta
^{\rho\delta}\epsilon^{\gamma}_A(u(p_s))] & {}.
\end{array} \right)
\label{67}
\end{eqnarray}

Since the constraints are redundant, this matrix has the following left and
right null eigenvectors: $\left( \begin{array}{c}  a_{\alpha\beta}=a_{\beta
\alpha} \\ 0 \end{array} \right)$  [$a_{\alpha\beta}$ arbitrary],
$\left( \begin{array}{c}   0 \\ \epsilon^B_{\sigma}(u(p_s))
\end{array} \right)$. Therefore, according to Ref.\cite{bclu}, one has to find
a left and right quasi-inverse
$\bar C$, $\bar CC=C\bar C=D$, such that $\bar C$ and D have the same left and
right null eigenvectors. One finds

\begin{eqnarray}
\bar C&=&\left( \begin{array}{cc}
0_{\gamma\delta\mu\nu} & {1\over
4}[\eta_{\gamma\tau}\epsilon^D_{\delta}(u(p_s))
-\eta_{\delta\tau}\epsilon^D_{\gamma}(u(p_s))] \\ {1\over 4}[\eta_{\sigma\nu}
\epsilon^B_{\mu}(u(p_s))-\eta_{\sigma\mu}\epsilon^B_{\nu}(u(p_s))] &
0^{BD}_{\sigma\tau} \end{array} \right) \nonumber \\
&&{}\nonumber \\
\bar CC=C\bar C=D&=&\left( \begin{array}{cc}
{1\over 2}(\eta^{\alpha}_{\mu}\eta^{\beta}_{\nu}-\eta^{\alpha}_{\nu}\eta_{\mu}
^{\beta}) & 0^{\alpha\beta D}_{\tau} \\
0^{\rho}_{A\mu\nu} & {1\over 2}(\eta^{\rho}_{\tau}\eta^D_A-\epsilon^{D\rho}
(u(p_s))\epsilon_{A\tau}(u(p_s)) \end{array} \right)
\label{68}
\end{eqnarray}

\noindent and the new Dirac brackets are

\begin{eqnarray}
\lbrace A,B\rbrace {}^{**}&=&\lbrace A,B\rbrace {}^{*}-{1\over 4}[\lbrace
A,{\tilde {\cal H}}^{\gamma\delta}\rbrace {}^{*}[\eta_{\gamma\tau}\epsilon^D
_{\delta}(u(p_s))-\eta_{\delta\tau}\epsilon^D_{\gamma}(u(p_s))]\lbrace
T^{\tau}_D,B\rbrace {}^{*}+\nonumber \\
&+&\lbrace A,T^{\sigma}_B\rbrace {}^{*}[\eta_{\sigma\nu}\epsilon^B_{\mu}(u(p
_s))-\eta_{\sigma\mu}\epsilon^B_{\nu}(u(p_s))]\lbrace {\tilde {\cal
H}}^{\mu\nu}
,B\rbrace {}^{*}].
\label{69}
\end{eqnarray}

\noindent While the check of $\lbrace {\tilde {\cal H}}^{\alpha\beta},B\rbrace
{}^{**}=0$ is immediate, we must use the relation $b_{{\check A}\mu}T^{\mu}
_D\epsilon^{D\rho}=-T^{\rho}_{\check A}$ [at this level we have
$T^{\mu}_{\check
A}=T^{\mu}_A$] to check $\lbrace T^{\rho}_A,B\rbrace {}^{**}=0$.

Then, we find the following brackets for the remaining variables ${\tilde
x}^{\mu}_s, p_s^{\mu}, \eta^r_i, \kappa_i^r$

\begin{eqnarray}
&&\lbrace {\tilde x}^{\mu}_s,p^{\nu}_s\rbrace {}^{**}=-\eta^{\mu\nu}\nonumber
\\
&&\lbrace \eta_i^r,\kappa^s_j\rbrace {}^{**}=\delta_{ij}\delta^{rs},
\label{70}
\end{eqnarray}

\noindent and the following form of the Poincar\'e generators

\begin{eqnarray}
p^{\mu}_s&&{}\nonumber \\
J^{\mu\nu}_s&=&{\tilde x}^{\mu}_sp^{\nu}_s-{\tilde x}^{\nu}_sp_s^{\mu}+{\tilde
S}_s^{\mu\nu}\nonumber \\
&&{\tilde S}_s^{oi}=-{ {\delta^{ir}{\bar S}_s^{rs}p_s^s}\over {p^o_s+\eta_s
\sqrt{p_s^2}} }\nonumber \\
&&{\tilde S}_s^{ij}=\delta^{ir}\delta^{js}{\bar S}_s^{rs}.
\label{71}
\end{eqnarray}

\noindent Therefore, ${\tilde x}_s^{\mu}$ is not a fourvector and ${\vec \eta}
_i, {\vec \kappa}i$ transform as Wigner spin-1 3-vectors.

Let us come back to the four first class constraints ${\tilde {\cal H}}^{\mu}
(\tau )\approx 0$, $\lbrace {\tilde {\cal H}}^{\mu},{\tilde {\cal H}}^{\nu}
\rbrace {}^{**}=0$, of Eq.(\ref{66}). They can be rewritten in the following
form

\begin{eqnarray}
{\cal H}(\tau )&=&u^{\mu}(p_s){\tilde {\cal H}}_{\mu}(\tau )=
\eta_s\sqrt{p^2_s}-\sum_{i=1}^N\eta_i\sqrt{m_i^2+{\vec \kappa}^2_i
(\tau )}=\epsilon_s-\sum_{i=1}^N\eta_i\sqrt{m_i^2+{\vec \kappa}^2_i
(\tau )}\approx 0\nonumber \\
{\vec {\cal H}}_p(\tau )&=&{\vec \kappa}_{+}=\sum_{i=1}^N{\vec \kappa}_i(\tau )
\approx 0.
\label{72}
\end{eqnarray}

\noindent The first one gives the mass spectrum of the isolated system, while
the other three say that the total 3-momentum of the N particles on the
hyperplane $\Sigma_W(\tau )$ vanishes. The Dirac Hamiltonian is now $H_D=
\lambda (\tau ){\cal H}(\tau )-\vec \lambda (\tau )\cdot {\vec {\cal H}}_p
(\tau )$ and we have ${\dot {\tilde x}}_s^{\mu}=\lbrace {\tilde x}_s^{\mu},
H_D\rbrace {}^{**}=-\lambda (\tau )u^{\mu}(p_s)$. Therefore, while the old
$x^{\mu}_s$ had a velocity ${\dot x}_s^{\mu}$ not parallel to the normal
$l^{\mu}=u^{\mu}(p_s)$ to the hyperplane as shown by Eqs.(\ref{65}), the new
${\tilde x}_s^{\mu}$ has ${\dot {\tilde x}}^{\mu}_s \| l^{\mu}$ and
no classical zitterbewegung. Moreover, we have that $T_s=l\cdot {\tilde x}_s=
l\cdot x_s$ is the Lorentz-invariant rest frame time.

Let us do the following canonical transformation [with the ${\hat \gamma}
_{ai}$ of Eqs.(\ref{6}), (\ref{7})]

\begin{equation}
\begin{minipage}[t]{2cm}
\begin{tabular}{|ll|} \hline
${\tilde x}_s^{\mu}$    &  $p_s^{\mu}$  \\ \hline
${\vec \eta}_i$  & ${\vec \kappa}_i$ \\ \hline
\end{tabular}
\end{minipage} \ {\longrightarrow \hspace{.2cm}} \
\begin{minipage}[t]{3cm}
\begin{tabular}{|ll|l|} \hline
{} & {} & $T_s$  \\
${\vec z}_s$ & ${\vec k}_s$ & { }  \\
{} & {} & $\epsilon_s$  \\ \hline
${\vec \rho}^{'}_a$  &  ${\vec \pi}^{'}_a$  & {}  \\ \hline
${\vec \eta}_{+}$  &  ${\vec \kappa}_{+}$  &  {}  \\ \hline
\end{tabular}
\end{minipage}
\label{73}
\end{equation}

\begin{eqnarray}
&&T_s={ {p_s\cdot {\tilde x}_s}\over {\eta_s\sqrt{p^2_s}}}=
{ {p_s\cdot x_s}\over {\eta_s\sqrt{p^2_s}}}\nonumber \\
&&\epsilon_s=\eta_s\sqrt{p_s^2}\nonumber \\
&&{\vec z}_s=\eta_s\sqrt{p_s^2}({\vec {\tilde x}}_s-{ {{\vec p}_s}\over
{p^o_s}}{\tilde x}^o_s)\nonumber \\
&&{\vec k}_s={{{\vec p}_s}\over {\eta_s\sqrt{p_s^2}}}\nonumber \\
&&{\vec \eta}_{+}={1\over N}\sum_{i=1}^N{\vec \eta}_i\nonumber \\
&&{\vec \kappa}_{+}=\sum_{i=1}^N{\vec \kappa}_i\nonumber \\
&&{\vec \rho}^{'}_a=\sqrt{N}\sum_{i=1}^N{\hat \gamma}_{ai}{\vec
\eta}_i\nonumber
\\
&&{\vec \pi}^{'}_a={1\over {\sqrt{N}}}\sum_{i=1}^N{\hat \gamma}_{ai}{\vec
\kappa}_i,
\label{74}
\end{eqnarray}

\noindent whose inverse is

\begin{eqnarray}
&&{\tilde x}^o_s=\sqrt{1+{\vec k}^2_s}(T_s+{ { {\vec k}_s\cdot {\vec z}_s}
\over {\epsilon_s} })\nonumber \\
&&{\vec {\tilde x}}_s={ { {\vec z}_s}\over {\epsilon_s}}+(T_s+{ { {\vec k}_s
\cdot {\vec z}_s}\over {\epsilon_s} }){\vec k}_s\nonumber \\
&&p^o_s=\epsilon_s\sqrt{1+{\vec k}^2_s}\nonumber \\
&&{\vec p}_s=\epsilon_s{\vec k}_s\nonumber \\
&&{\vec \eta}_i={\vec \eta}_{+}+{1\over {\sqrt{N}}}\sum_{a=1}^{N-1}{\hat
\gamma}_{ai}{\vec \rho}^{'}_a\nonumber \\
&&{\vec \kappa}_i={1\over N}{\vec \kappa}_{+}+\sqrt{N}\sum_{a=1}^{N-1}
{\hat \gamma}_{ai}{\vec \pi}^{'}_a.
\label{75}
\end{eqnarray}

By using the constraints ${\vec \kappa}_{+}\approx 0$, we get

\begin{eqnarray}
&&{\vec \kappa}^2_i\approx N{(\sum_{a=1}^{N-1}{\hat \gamma}_{ai}{\vec \pi}
^{'}_a)}^2\nonumber \\
&&{\bar S}_s^{rs}\approx \sum_{a=1}^{N-1}(\rho^{{'}r}_a\pi^{{'}s}_a-
\rho^{{'}s}_a\pi^{{'}r}_a)\nonumber \\
&&{\bar S}_s^{\bar or}\approx -\eta^r_{+}\sum_{i=1}^N\eta_i\sqrt{m^2_i+N
{(\sum_{a=1}^{N-1}{\hat \gamma}_{ai}{\vec \pi}
^{'}_a)}^2}-{1\over {\sqrt{N}}}\sum_{a=1}^{N-1}\rho^{{'}r}_a\sum_{i=1}^N
{\hat \gamma}_{ai}\eta_i\sqrt{m^2_i+N
{(\sum_{a=1}^{N-1}{\hat \gamma}_{ai}{\vec \pi}
^{'}_a)}^2}.
\label{76}
\end{eqnarray}

Note that we have N+1 3-vectors ${\vec z}_s, {\vec \eta}_{+}, {\vec
\rho}^{'}_a$ for the N particles, but that the constraints
${\vec \kappa}_{+}\approx 0$ imply that the
${\vec \eta}_{+}$'s are gauge variables. We can decouple these 3 canonical
pairs from the others: the simplest way is to add the gauge-fixings ${\vec
\eta}_{+}\approx 0$ [let us remark that $z^{\mu}(\tau ,\vec 0)=x^{\mu}_s
(\tau )$, i.e. this is the origin of the coordinate system on the
hyperplane from the point of view of Minkowski spacetime];
so we remain with the correct number of N-1 relative
variables ${\vec \rho}^{'}_a, {\vec \pi}^{'}_a$ plus the center-of-mass ones
${\tilde x}^{\mu}_s, p_s^{\mu}$ and with [compare with ${\bar S}^{\bar or}=-
\sum_{a=1}^{N-1}\rho^r_a\epsilon_{Ra}$ at $T_{Ra}=0$, eq.(\ref{15})]

\begin{equation}
{\bar S}_s^{\bar or}=-{1\over {\sqrt{N}}}\sum_{a=1}^{N-1}\rho^{{'}r}_a\sum
_{i=1}^N{\hat \gamma}_{ai}\eta_i\sqrt{m^2_i+N
{(\sum_{a=1}^{N-1}{\hat \gamma}_{ai}{\vec \pi}
^{'}_a)}^2}=-\sum_{a=1}^{N-1}\rho^{{'}r}_a\epsilon_{Ra},
\label{77}
\end{equation}

\noindent where we anticipated a result of next Section.

In this way we obtain the rest-frame instant
form $T_s=const.$ of the reduced problem for the N-body system with the
center-of-mass motion separated out. This constant motion is taken into
account by the observer which looks at the hyperplane $\Sigma_W(\tau )$.
The evolution in $T_s$ of the reduced system is governed by the remaining
constraint

\begin{equation}
{\hat {\cal H}}(\tau )
=\epsilon_s-\sum_{i=1}^N\eta_i\sqrt{m^2_i+N{(\sum_{a=1}^{N-1}
{\hat \gamma}_{ai}{\vec \pi}^{'}_a)}^2}=\epsilon_s -H^{(T_s)}_R\approx 0.
\label{78}
\end{equation}

If we would add the gauge-fixing $T_s-\tau \approx 0$ and would go to Dirac
brackets, we would obtain a Hamilton-Jacobi description in terms of Jacobi
data. If we want to reintroduce an evolution for the relative variables,
the natural time parameter is $T_s$ and the associated Hamiltonian is $\epsilon
_s\equiv H^{(T_s)}_R$; the Jacobi data would then become the Cauchy data for
this evolution. In the nonrelativistic limit $H^{(T_s)}_R$ would go (modulo
$m_ic^2$ terms) in the Hamiltonian $H_R$ of the reduced [N-1 relative
variables]
problem: $H={{{\vec p}^2}\over {2\sum_{i=1}^Nm_i}}+H_R$ would be the total
Newtonian Hamiltonian.

The vectors
${\vec z}_s, {\vec k}_s$ describe the frame-dependent position of the canonical
center of mass; in the rest frame, $\stackrel{\circ}{p}{}^{\mu}_s=\epsilon_s
(1;\vec 0), \stackrel{\circ}{{\vec z}_s}$ defines a point in the hyperplane
$\Sigma_W(\tau )$ orthogonal to $\stackrel{\circ}{p}{}_s^{\mu}$, which can be
used to build the covariant noncanonical Fokker center of inertia. Therefore,
the worldtube defined by the frame-dependent canonical center-of-mass
positions
${\vec \zeta}_s={\vec z}_s/\eta_s\sqrt{p^2_s}$ arises naturally also in this
1-time descritpion (the rest-frame instant form $T_s=const.$). Let us note
that, while $\epsilon_s=\eta_s\sqrt{p_s^2}$ is determined by the invariant
mass of the physical system on the hypersurface [see Eq.(\ref{78})], ${\vec k}
_s$ describes the orientation of the hyperplane with normal $l^{\mu}=u^{\mu}
(p_s)$ embedded in Minkowski spacetime with respect to an arbitrary
observer: ${\vec k}_s$
is the collective velocity of the physical system as seen from
outside the hyperplane and not the three-momentum, determined by the
energy-momentum tensor, on the hypersurface, which is the vanishing ${\vec
\kappa}_{+}=0$, since the hyperplane corresponds to the rest frame.

For generic masses $m_i$, Eq.(\ref{78})
describes all the $2^N$ branches of the mass
spectrum by considering all the possible combinations of $\eta_i=sign\, p^o_i
=\pm$, i.e. the solutions in $\epsilon$ of $\chi \approx 0$ of Eq.(\ref{26})
when ${\hat \epsilon}_{Ra}\approx 0$. In the 1-time theory there is a different
action (\ref{38}) associated with each branch of the mass spectrum of the
N-time theory.

\vfill\eject

\section{The connection between the 1- and N-time descriptions}

We have arrived at two descriptions of N scalar free particles: i) the
1-time theory with variables ${\tilde x}_s^{\mu}, p_s^{\mu}, {\vec \rho}_a
^{'}, {\vec \pi}_a^{'}, {\vec \eta}_{+}, {\vec \kappa}_{+}$ and with the
first class constraints (\ref{72}), i.e. ${\vec \kappa}_{+}\approx 0$ and
${\hat {\cal H}}=\epsilon_s-\sum_{i=1}^N\eta_i\sqrt{m_i^2+N{(\sum_{a=1}^{N-1}
{\hat \gamma}_{ai}{\vec \pi}^{'}_a)}^2}\approx 0$; ii) the N-time theory
with variables ${\hat {\tilde x}}^{\mu}, p^{\mu}, {\hat {\vec \rho}}_a,
{\vec \pi}_a, {\hat T}_{Ra}$ and with the first class constraints ${\hat
\epsilon}_{Ra}\approx 0$ plus a mass-spectrum constraint $\chi \approx 0$,
Eq.(\ref{26}). The last line of Eqs.(\ref{21}), giving the original constraints
$\phi_i
\approx 0$, shows that it is the mass-spectrum constraint $\chi \approx 0$
that governs the dynamics in both descriptions, since ${\hat {\cal H}}\approx
0$ identifies one of the $2^N$ branches of $\chi \approx 0$.

To compare the two descriptions, we have to identify two isomorphic phase
spaces: i) in the 1-time theory, we add the gauge-fixings ${\vec \eta}_{+}
\approx 0$ [natural from the comparison of Eqs.(\ref{15})
and (\ref{77})], so that
the first phase space is spanned by ${\tilde x}_s^{\mu}, p_s^{\mu}, {\vec
\rho}_a^{'}, {\vec \pi}_a^{'},$ with the constraint ${\hat {\cal H}}\approx
0$; ii) in the N-time theory we add the gauge-fixings $T_{Ra}\approx 0$
implying Eqs.(\ref{33}) [so that the final variables of the second phase
space are ${\hat {\tilde x}}^{\mu},
p^{\mu}, {\hat {\vec \rho}}_a,{\vec \pi}_a$], and we choose among the
$2^N$ branches of the mass spectrum the one corresponding to the given
choice of the signs $\eta_i$ of the 1-time theory.

Since we do not know the expression of the original variables $x^{\mu}_i,
p_i^{\mu}$ in terms of the variables of Eqs.(\ref{31}), to find the relation
among the variables of the two phase spaces we have initially to go back to
Eqs.(\ref{9})

\begin{eqnarray}
x^{\mu}_i&=&x^{\mu}+{1\over {\sqrt{N}}}\sum_{a=1}^{N-1}{\hat \gamma}_{ai}
R^{\mu}_a= x^{\mu}+{1\over {\sqrt{N}}}\sum_{a=1}^{N-1}{\hat \gamma}_{ai}
\epsilon^{\mu}_A(u(p))\epsilon^A_{\nu}(u(p))R^{\nu}_a=\nonumber \\
&=&x^{\mu}+{1\over {\sqrt{N}}}\sum_{a=1}^{N-1}{\hat \gamma}_{ai}({{p^{\mu}}
\over {\epsilon}}T_{Ra}+\epsilon^{\mu}_r(u(p))\rho^r_a)\rightarrow
\nonumber \\
&&{\rightarrow}_{T_{Ra}={\hat \epsilon}_{Ra}=0} x^{\mu}{|}_{T_{Ra}={\hat
\epsilon}_{Ra}=0}+{1\over {\sqrt{N}}}\epsilon^{\mu}_r(u(p))\sum_{a=1}^{N-1}
{\hat \gamma}_{ai}\rho^r_a\nonumber \\
&&{}\nonumber \\
p^{\mu}_i&=&{1\over N}p^{\mu}+\sqrt{N}\sum_{a=1}^{N-1}{\hat \gamma}_{ai}Q_a
^{\mu}=\nonumber \\
&=&({{\epsilon}\over N}+\sqrt{N}\sum_{a=1}^{N-1}{\hat \gamma}_{ai}\epsilon
_{Ra})u^{\mu}(p)+\sqrt{N}\epsilon^{\mu}_r(u(p))\sum_{a=1}^{N-1}{\hat \gamma}
_{ai}\pi^r_a\rightarrow \nonumber \\
&&{\rightarrow}_{T_{Ra}={\hat \epsilon}_{Ra}=0} ?
\label{79}
\end{eqnarray}

On the other hand, Eqs.(\ref{39}), (\ref{62}) and (\ref{66}) imply the
following identification with the variables on the hyperplane orthogonal
to $p_s^{\mu}$

\begin{eqnarray}
x^{\mu}_i{|}_{T_{Ra}={\hat \epsilon}_{Ra}=0}&=&z^{\mu}(\tau ,{\vec \eta}_i
(\tau ))\approx x^{\mu}_s+\epsilon^{\mu}_{\check r}(u(p_s))\eta^{\check r}_i
(\tau )\approx {|}_{{\vec \eta}_{+}=0}\nonumber \\
&\approx&
{\tilde x}^{\mu}_s-{1\over 2}\, \epsilon^A_{\nu}(u(p_s))\eta_{AB}
{ {\partial \epsilon^B_{\rho}(u(p_s))}\over {\partial p_{s\mu}} }\,
S^{\nu\rho}_s+{1\over {\sqrt{N}}}\epsilon^{\mu}_r(u(p_s))\sum_{a=1}^{N-1}
{\hat \gamma}_{ai}\rho^{{'}r}_a=\nonumber \\
&=&{\tilde x}_s^{\mu}+
{ 1\over {\eta \sqrt{p_s^2}(p_s^o+\eta \sqrt{p_s^2})} }\, [p_{s\nu}
S^{\nu\mu}_s+\eta \sqrt{p_s^2} (S_s^{o\mu}-S_s^{o\nu}{ {p_{s\nu}p^{s\mu}}\over
{p_s^2} })]+\nonumber \\
&+&{1\over {\sqrt{N}}}\epsilon^{\mu}_r(u(p_s))\sum_{a=1}^{N-1}
{\hat \gamma}_{ai}\rho^{{'}r}_a=\nonumber \\
&=&{\tilde x}_s^{\mu}+{1\over {\eta_s \sqrt{p_s^2}} }[\eta^{\mu}_A({\bar S}_s
^{\bar oA}-{ {{\bar S}_s^{Ar}p_s^r}\over {p_s^o+\eta_s \sqrt{p_s^2}} })+
{ {p_s^{\mu}+2\eta_s \sqrt{p_s^2}\eta^{\mu o}}\over {\eta_s \sqrt{p_s^2}(p_s^o
+\eta_s \sqrt{p_s^2})} }{\bar S}_s^{\bar or}p_s^r]+\nonumber \\
&+&{1\over {\sqrt{N}}}\epsilon^{\mu}_r(u(p_s))\sum_{a=1}^{N-1}
{\hat \gamma}_{ai}\rho^{{'}r}_a\nonumber \\
&&{}\nonumber \\
p^{\mu}_i{|}_{T_{Ra}={\hat \epsilon}_{Ra}=0}&=&\eta_i\sqrt{m_i^2+{\vec \kappa}
^2_i(\tau )}u^{\mu}(p_s)+\epsilon^{\mu}_{\check r}(u(p_s))\kappa^{\check r}_i
(\tau )=\nonumber \\
&=&\eta_i\sqrt{m^2_i+{\vec \kappa}_i^2(\tau )}u^{\mu}(p_s)+\epsilon^{\mu}
_{\check r}(u(p_s))[{1\over N}\kappa_{+}^{\check r}+\sqrt{N}\sum_{a=1}^{N-1}
{\hat \gamma}_{ai}\pi^{{'}{\check r}}_a]
\approx {|}_{{\vec \kappa}_{+}=0}\nonumber \\
&\approx&\eta_i\sqrt{m^2_i+N{(\sum_{a=1}^{N-1}{\hat \gamma}_{ai}{\vec \pi}^{'}
_a(\tau ))}^2}u^{\mu}(p_s)+\sqrt{N}\epsilon^{\mu}_r(u(p_s))\sum_{a=1}^{N-1}
{\hat \gamma}_{ai}\pi^{{'}r}_a.
\label{80}
\end{eqnarray}

We see that $p^{\mu}_i{|}_{T_{Ra}={\hat \epsilon}_{Ra}=0}$ satisfies
automatically $p_i^2=m_i^2$ and that

\begin{equation}
p^{\mu}=\sum_{i=1}^Np^{\mu}_i\approx {|}_{T_{Ra}={\hat \epsilon}_{Ra}=0}
\sum_{i=1}^N\eta_i\sqrt{m^2_i+N{(\sum_{a=1}^{N-1}{\hat \gamma}_{ai}{\vec \pi}
^{'}_a(\tau ))}^2} {{p^{\mu}_s}\over {\epsilon_s}}+\epsilon^{\mu}_r(u(p_s))
\kappa^r_{+}\approx p_s^{\mu}
\label{81}
\end{equation}

\noindent if we have

\begin{equation}
\epsilon \approx \epsilon_s\approx \sum_{i=1}^N\eta_i\sqrt{m^2_i+N{(\sum_{a=1}
^{N-1}{\hat \gamma}_{ai}{\vec \pi}^{'}_a(\tau ))}^2},
\label{82}
\end{equation}

\noindent by using Eqs.(\ref{59}). Moreover, we have

\begin{eqnarray}
Q^{\mu}_a{|}_{T_{Ra}={\hat \epsilon}_{Ra}=0}&=&{1\over {\sqrt{N}}}\sum_{i=1}
^N{\hat \gamma}_{ai}p^{\mu}_i{|}_{T_{Ra}={\hat \epsilon}_{Ra}=0}=\nonumber \\
&=&{1\over {\sqrt{N}}}\sum_{i=1}^N{\hat \gamma}_{ai}\eta_i\sqrt{m_i^2+{\vec
\kappa}^2_i}u^{\mu}(p_s)+\epsilon^{\mu}_r(u(p_s))\kappa^r_{+}\approx
\nonumber \\
&\approx& {1\over {\sqrt{N}}}\sum_{i=1}^N{\hat \gamma}_{ai}\eta_i
\sqrt{m^2_i+N{(\sum_{a=1}^{N-1}{\hat \gamma}_{ai}{\vec \pi}
^{'}_a(\tau ))}^2}u^{\mu}(p),
\label{83}
\end{eqnarray}

\noindent so that for ${\hat T}_{Ra}={\hat \epsilon}_{Ra}={\vec \kappa}_{+}=0$
we find

\begin{equation}
\epsilon_{Ra}{|}_{T_{Ra}={\hat \epsilon}_{Ra}=0}\approx {1\over {\sqrt{N}}}
\sum_{i=1}^N{\hat \gamma}_{ai}\eta_i
\sqrt{m^2_i+N{(\sum_{a=1}^{N-1}{\hat \gamma}_{ai}{\vec \pi}
^{'}_a(\tau ))}^2}.
\label{84}
\end{equation}

\noindent Equation (\ref{84}),
with Eq.(\ref{8}), implies that the second one of Eqs.
(\ref{79}) becomes the second one of Eq.(\ref{80}) with ${\vec \pi}^{'}_a=
{\vec \pi}_a$. Equation (\ref{84}) gives all the looked for solutions of
Eq.(\ref{26}) [or of Eqs.({37})]
for ${\hat T}_{Ra}={\hat \epsilon}_{Ra}=0$ according to which branch of
$\epsilon \approx \epsilon_s$ has been chosen in Eq.(\ref{82}), as can be
explicitly checked.

Moreover, for each branch the mass spectrum, given by Eq.(\ref{82}) for a given
choice of the signs $\eta_i$, we have the following solution of Eqs.(\ref{37})
without the restriction ${\hat T}_{Ra}={\hat \epsilon}_{Ra}=0$ [remember
that ${\vec \pi}_a={\hat {\vec \pi}}_a={\vec \pi}_a^{'}$]

\begin{equation}
\epsilon_{Ra}={1\over {\sqrt{N}}}\sum_{i=1}^N{\hat \gamma}_{ai}\eta_i
\sqrt{m^2_i+N{(\sum_{a=1}^{N-1}{\hat \gamma}_{ai}{\vec \pi}
^{'}_a(\tau ))}^2+{{2\epsilon}\over {\sqrt{N}}}\sum_{b=1}^{N-1}{\hat \gamma}
_{bi}{\hat \epsilon}_{Rb} }
\label{85}
\end{equation}

\noindent if Eq.(\ref{82}) is rewritten in the weakly equivalent form

\begin{equation}
\epsilon \approx \epsilon_s\approx \sum_{i=1}^N\eta_i\sqrt{m^2_i+N{(\sum_{a=1}
^{N-1}{\hat \gamma}_{ai}{\vec \pi}^{'}_a(\tau ))}^2+{{2\epsilon}\over
{\sqrt{N}}}\sum_{b=1}^{N-1}{\hat \gamma}_{bi}{\hat \epsilon}_{Rb} }.
\label{86}
\end{equation}

On the branch defined by Eq.(\ref{86}), Eqs.(\ref{85}) imply the following
form for $p^{\mu}_i$ of Eqs.(\ref{20})

\begin{eqnarray}
&&p_i^o\approx {1\over {\epsilon}}[p^o\eta_i\sqrt{m^2_i+N{(\sum_{a=1}
^{N-1}{\hat \gamma}_{ai}{\vec \pi}^{'}_a(\tau ))}^2+{{2\epsilon}\over
{\sqrt{N}}}\sum_{b=1}^{N-1}{\hat \gamma}_{bi}{\hat \epsilon}_{Rb} } +
\sqrt{N}\sum_{a=1}^{N-1}{\hat \gamma}_{ai}\vec p\cdot {\vec \pi}_a]
\nonumber \\
&&{\vec p}_i\approx \sqrt{N}[\sum_{a=1}^{N-1}{\hat \gamma}_{ai}{\vec \pi}_a+
{{\vec p}\over {\epsilon}}({{\sum_{a=1}^{N-1}{\hat \gamma}_{ai}\vec p\cdot
{\vec \pi}_a}\over {p^o+\epsilon}}+\nonumber \\
&&+{1\over {\sqrt{N}}}\eta_i\sqrt{m^2_i+N{(\sum_{a=1}
^{N-1}{\hat \gamma}_{ai}{\vec \pi}^{'}_a(\tau ))}^2+{{2\epsilon}\over
{\sqrt{N}}}\sum_{b=1}^{N-1}{\hat \gamma}_{bi}{\hat \epsilon}_{Rb} }],
\label{87}
\end{eqnarray}

\noindent and Eqs.(\ref{26}) give

\begin{equation}
\phi_i\approx {{2\epsilon}\over {\sqrt{N}}}\sum_{a=1}^{N-1}{\hat \gamma}_{ai}
{\hat \epsilon}_{Ra}\approx 0,
\label{88}
\end{equation}

\noindent consistently with $\chi \approx 0$ due to Eq.(\ref{86}). Therefore,
one
has [$\prod_{\eta_i}$ means a product over all the $2^N$ choices of the signs
$\eta_i$]

\begin{eqnarray}
\chi &=& \epsilon^2-N[\sum_{i=1}^Nm_i^2-N\sum_{a=1}^{N-1}(\epsilon_{Ra}^2-
{\vec \pi}_a^2)]=\nonumber \\
&=& {[\epsilon]}^{-2^{N-1}}\, \prod_{\eta_i}[\epsilon -\sum_{i=1}^N
\eta_i\sqrt{m^2_i+N{(\sum_{a=1}
^{N-1}{\hat \gamma}_{ai}{\vec \pi}^{'}_a(\tau ))}^2+{{2\epsilon}\over
{\sqrt{N}}}\sum_{b=1}^{N-1}{\hat \gamma}_{bi}{\hat \epsilon}_{Rb} }]
\approx 0.
\label{89}
\end{eqnarray}

Even with Equations
(\ref{85}), (\ref{86}), we cannot get the explicit inverse
canonical transformation giving $x^{\mu}_i, p^{\mu}_i$ of Eqs.(\ref{20}) in
terms of ${\hat {\tilde x}}^{\mu}, p^{\mu}, {\hat T}_{Ra}, {\hat \epsilon}
_{Ra}, {\hat {\vec \rho}}_a, {\hat {\vec \pi}}_a={\vec \pi}_a$ of
Eqs.(\ref{31}) independently from the choice of the branch of the mass
spectrum,
 since Eqs.(\ref{85}) are a branch-dependent solution of Eqs.(\ref{37}).

While Equations (\ref{87}) give the answer for the second half of
Eqs.(\ref{79}), the discussion of the first half is more involved.

By using Eqs.(\ref{31}), (\ref{33}), (\ref{28}), (\ref{15}) evaluated at
${\hat T}_{Ra}={\hat \epsilon}_{Ra}=0$, we get

\begin{eqnarray}
&&{\hat {\tilde x}}^{\mu}\approx {\hat x}^{\mu}+{ {\eta^{\mu}_s{\hat {\bar
S}}^{sr}p^r}\over {\epsilon (p^o+\epsilon )} }\approx {\hat x}^{\mu}+
{ {\eta^{\mu}_s{\bar S}^{sr}p^r}\over {\epsilon (p^o+\epsilon )} }\nonumber \\
&&{\hat x}^{\mu}\approx x^{\mu}+{1\over {\epsilon}}\epsilon^{\mu}_r(u(p))
\sum_{a=1}^{N-1}\rho^r_a\epsilon_{Ra}=x^{\mu}-{1\over {\epsilon}}\epsilon^{\mu}
_r(u(p)){\bar S}^{\bar or}\nonumber \\
&&{}\nonumber \\
&&\Rightarrow {\hat {\tilde x}}^{\mu}\approx x^{\mu}-{1\over {\epsilon}}
\epsilon^{\mu}_r(u(p)){\bar S}^{\bar or}+
{ {\eta^{\mu}_s{\bar S}^{sr}p^r}\over {\epsilon (p^o+\epsilon )} },
\label{90}
\end{eqnarray}

so that Eqs.(\ref{79}) and (\ref{80}) imply at ${\hat T}_{Ra}={\hat \epsilon}
_{Ra}=0$

\begin{eqnarray}
x^{\mu}_i&\approx &x^{\mu}+{1\over {\sqrt{N}}}\epsilon^{\mu}_r(u(p))
\sum_{a=1}^{N-1}{\hat \gamma}_{ai}\rho^r_a\approx\nonumber \\
&\approx& {\hat {\tilde x}}^{\mu}+{1\over {\epsilon}}
\epsilon^{\mu}_r(u(p)){\bar S}^{\bar or}-
{ {\eta^{\mu}_s{\bar S}^{sr}p^r}\over {\epsilon (p^o+\epsilon )} }
+{1\over {\sqrt{N}}}\epsilon^{\mu}_r(u(p))
\sum_{a=1}^{N-1}{\hat \gamma}_{ai}\rho^r_a\approx\nonumber \\
&\approx&{\tilde x}_s+{1\over {\epsilon} }[\eta^{\mu}_A({\bar S}_s
^{\bar oA}-{ {{\bar S}_s^{Ar}p_s^r}\over {p_s^o+\epsilon} })+
{ {p_s^{\mu}+2\epsilon\eta^{\mu o}}\over {\epsilon(p_s^o
+\epsilon)} }{\bar S}_s^{\bar or}p_s^r]
+{1\over {\sqrt{N}}}\epsilon^{\mu}_r(u(p))
\sum_{a=1}^{N-1}{\hat \gamma}_{ai}\rho^{{'}r}_a.\nonumber \\
\label{91}
\end{eqnarray}

By comparing the two expressions for $\mu =0$ and for $\mu =k$, we find

\begin{eqnarray}
{\tilde x}^{\mu}_s\approx {\hat {\tilde x}}^{\mu}\nonumber \\
{\vec \rho}_a^{'}={\vec \rho}_a
\label{92}
\end{eqnarray}

\noindent due to Eqs.(\ref{76}) and (\ref{77})
[for ${\vec \rho}^{'}_a={\vec \rho}_a$
we have ${\bar S}_s^{\bar or}={\bar S}^{\bar or}$ and ${\bar S}_s^{rs}={\bar
S}^{rs}$], consistently with the identification
of the two reduced phase spaces, whose final variables can be chosen as
${\tilde x}^{\mu}_s, p^{\mu}, {\vec \rho}_a, {\vec \pi}_a$.

\vfill\eject

\section{The mass spectrum and the introduction of interactions}

In the 1-time theory, by varying $\eta_i=sign\, p^o_i$ in Eq.(\ref{78}) we
obtain the $2^N$ branches of the mass spectrum of N free scalar particles
in the rest-frame instant form, when $p^2 > 0$ and the masses $m_i$'s are
generic, so that degeneracies are avoided. For each of these branches there is
a well defined Lagrangian, which gives rise to the constraints of
Eqs.(\ref{55})
on arbitrary spacelike hyperplanes: only the configurations with $p^2 > 0$
can be reduced to the rest-frame instant form. Being in classical relativistic
mechanics, we must take into account also negative masses and energies.

To see which other configurations may be allowed, we have to revert to a
study of the N-time theory with its first class constraints $\phi_i=p_i^2-
m_i^2\approx 0$.

Let us consider in detail the case N=2; when $p^2 > 0$ and $\eta_i=+$, i=1,2,
we are considering, in absence of interactions, the special configuration of
the kinematics of forward s-channel eleastic scattering, in which we have
$s={(p_1+p_2)}^2=p^2=\epsilon^2$, t=0, $u={(p_1-p_2)}^2=4Q^2$ as special values
of the Mandelstam variables
[see Ref.\cite{bk}], so that, given the following two combinations of
the $\phi_i$'s [see Appendix B]

\begin{eqnarray}
&&\chi_{+}=2(\phi_1+\phi_2)=p^2+4Q^2-2(m_1^2+m_2^2)\approx 0\nonumber \\
&&\chi_{-}={1\over 2}(\phi_1-\phi_2)=p\cdot Q-{1\over 2}(m_1^2-m_2^2)\approx 0,
\label{93}
\end{eqnarray}

\noindent we find that $\chi_{+}\approx 0$ is equivalent to the fundamental
relation $s+t+u=2(m_1^2+m_2^2)$.

Let us discuss the various possibilities of choice of the value of the masses.

1) $m_1\not= m_2$ ($m_1 > m_2 \geq 0$). In this case, the first class
constraints (\ref{93}) define a constraint submanifold $\bar \gamma$
(coisotropically embedded in phase space\cite{lusb}), which is the disjoint
union of three strata with $p^2 > 0$ (the main stratum dense in $\bar \gamma$),
$p^2 =0$ and $p^2 < 0$ respectively [the stratum with $p^{\mu}=0$ is
excluded by $\chi_{-}\approx 0$].

{}{}1a) Stratum with $p^2 > 0$. According to Appendix B, one has [${\hat Q}^2
\approx Q^2_{\perp}=-{\vec \pi}^2 < 0$; $Q^2\approx {\hat Q}^2+{{m_1^2-m_2^2}
\over {4p^2}}$]

\begin{eqnarray}
\chi_{-}&=&p\cdot {\hat Q}\approx 0 \nonumber \\
\chi_{+}&=&p^2+4Q^2_{\perp}+4{{(p\cdot Q)^2}\over {p^2}}-2(m_1^2+m_2^2)\approx
p^2+4{\hat Q}^2+{{m_1^2-m_2^2}\over {p^2}}-2(m_1^2+m_2^2)=\nonumber \\
&=&{1\over {p^2}}(p^2-M_{+}^2)(p^2-M_{-}^2)\approx 0.
\label{94}
\end{eqnarray}

\noindent with $M_{\pm}$ given in Eqs.(\ref{B4}).

There are four branches for $\epsilon =\eta \sqrt{p^2}$ in terms of ${\vec \pi}
^2=-Q^2_{\perp}\approx -{\hat Q}^2$

\begin{eqnarray}
&&\epsilon \approx \pm (\sqrt{m_1^2+{\vec \pi}^2}+\sqrt{m_2^2+{\vec \pi}^2}),
\quad\quad \eta_1=\eta_2\nonumber \\
&&\epsilon \approx \pm |\, (\sqrt{m_1^2+{\vec \pi}^2}-\sqrt{m_2^2+{\vec
\pi}^2},
|\, \quad\quad \eta_1=-\eta_2
\label{95}
\end{eqnarray}

\noindent For ${\vec \pi}^2\rightarrow 0$ [corresponding to ${\hat Q}^{\mu}
=0$ ($p_1^{\mu}\, //\, p_2^{\mu}$); ${\hat Q}^2=0$ is excluded by $\chi_{-}
\approx 0$] we have the (s-channel) thresholds $\epsilon \approx \pm (m_1+
m_2)$ and (u-channel) pseudo-thresholds $\epsilon \approx \pm |\, m_1-m_2\, |$.

{}{}1b) Stratum with $p^2=0$. It exists for $\eta_1=-\eta_2$: for $p^{\mu}_i$
timelike, i=1,2, we have $0=(p_1+p_2)^2=(\eta_1|\, p^o_1|+\eta_2|\, p^o_2|)^2
-{({\vec p}_1+{\vec p}_2)}^2\approx {(\eta_1\sqrt{m_1^2+{\vec p}_1^2}+\eta_2
\sqrt{m_2^2+{\vec p}_2^2})}^2-{({\vec p}_1+{\vec p}_2)}^2=m_1^2+m_2^2+2\eta_1
\eta_2\sqrt{m_1^2+{\vec p}_1^2}\sqrt{m_2^2+{\vec p}_2^2}-2{\vec p}_1\cdot
{\vec p}_2 \Rightarrow {}_{{\vec p}_2=0}\, -2\eta_1\eta_2m_2\sqrt{m_1^2+{\vec
p}_1^2}=m_1^2+m_2^2 \Rightarrow \eta_1\eta_2=-$.

Now the constraint $\chi_{+}\approx 0$ implies $Q^2\approx {1\over 2}(m_1^2+
m_2^2) > 0$ and two branches

\begin{equation}
Q^o\approx \pm \sqrt{{1\over 2}(m_1^2+m_2^2)+{\vec Q}^2}
\label{96}
\end{equation}

\noindent of excitations of relative energy with total energy given by

\begin{equation}
p^o\approx {1\over {Q^o}}[{1\over 2}(m_1^2-m_2^2)+\vec p\cdot \vec Q].
\label{97}
\end{equation}

{}{}1c) Stratum with $p^2 < 0$. It exists for $\eta_1=-\eta_2$;
the constraint $\chi_{+}
\approx 0$ implies $Q^2\approx {1\over 4}[2(m_1^2+m_2^2)-p^2] > 0$. In terms
of $Q^2_{\perp}=Q^2-{{(p\cdot Q)^2}\over {p^2}}\approx Q^2-{{(m_1^2-m_2^2)^2}
\over {4p^2}}$, we have two tachionic branches

\begin{equation}
p^2\approx -{(\sqrt{Q^2_{\perp}-m_1^2}\pm \sqrt{Q^2_{\perp}-m_2^2})}^2
\label{98}
\end{equation}

2) $m_1=m_2=m > 0$. We have

\begin{eqnarray}
&&\chi_{+}=p^2+4Q^2-4m^2\approx 0\nonumber \\
&&\chi_{-}=p\cdot Q\approx 0,
\label{99}
\end{eqnarray}

\noindent and there are three strata with $p^2 > 0$, $p^{\mu}=0$, $p^2 < 0$
respectively [$p^2=0$ is excluded by $Q^2\approx m^2 > 0$ and $p\cdot Q\approx
0$].

{}{}2a) Stratum with $p^2 > 0$. We have $Q^2\approx Q^2_{\perp}=-{\vec \pi}^2
< 0$ and there are two branches

\begin{equation}
\epsilon \approx \pm 2\sqrt{m^2+{\vec \pi}^2},\quad\quad [\epsilon \approx
\pm 2m\, for\, p^{\mu}_1=p^{\mu}_2,\, i.e.\, Q^{\mu}=0],
\label{100}
\end{equation}

\noindent which are the limit for $m_1-m_2\rightarrow 0$ of the two branches
with $\eta_1=\eta_2$ of the $p^2 > 0$ stratum with $m_1\not= m_2$.

{}{}2b) Stratum with $p^{\mu}=0$ [$p_1^{\mu}=-p_2^{\mu} \Rightarrow \eta_1=-
\eta_2$]. Since we have the constraint
$\chi_{+}=4(Q^2-m^2)\approx 0$, there are two branches

\begin{equation}
Q^o\approx \pm \sqrt{m^2+{\vec Q}^2}
\label{101}
\end{equation}

\noindent of excitations of relative energy: they correspond to two degenerate
$\epsilon =0$ levels obtained from the limit $m_1-m_2\rightarrow 0$ with
$\eta_1=-\eta_2$ of the $p^2 > 0$ stratum with $m_1\not= m_2$ [see Ref.\cite
{mosh} for the spin 1/2 case].

{}{}2c) Stratum with $p^2 < 0$. It exists for $\eta_1=-\eta_2$ [$p^2 < 0$ is
equivalent for ${\vec p}_2=0$ to $2m(m+\eta_1\eta_2\sqrt{m^2+{\vec p}_1^2})
< 0$], and we have $Q^2-{1\over 4}p^2 > 0$, so that there is a tachionic
branch

\begin{equation}
p^2\approx -4(Q^2-m^2)\approx -4(Q^2_{\perp}-m^2).
\label{102}
\end{equation}

3) $m_1=m_2=0$. The constraints $p_1^2\approx 0, p_2^2\approx 0$ give

\begin{eqnarray}
&&\chi_{+}=p^2+4Q^2\approx 0\nonumber \\
&&\chi_{-}=p\cdot Q\approx 0,
\label{103}
\end{eqnarray}

\noindent so that there are three strata
with $p^2 > 0$, $p^2=0$, $p^{\mu}=0$, respectively
[$p^2 < 0$ is excluded, because it would imply $p^2={(\eta_1|\, {\vec p}_1|+
\eta_2|\, {\vec p}_2|)}^2-{({\vec p}_1+{\vec p}_2)}^2=2|\, {\vec p}_1| |\,
{\vec p}_2| (\eta_1\eta_2-cos\, \theta_{12}) < 0$].

{}{}3a) Stratum with $p^2 > 0$. It has $Q^2\approx Q^2_{\perp}=-{\vec \pi}^2
< 0$ and there are two branches (with $\eta_1=\eta_2$)

\begin{equation}
\epsilon \approx \pm 2 \sqrt{{\vec \pi}^2}
\label{104}
\end{equation}

\noindent tangent at ${\vec \pi}^2=0$ ($Q^{\mu}=0$).

{}{}3b) Stratum with $p^2=0$. It implies $2|\, {\vec p}_1| |\, {\vec p}_2|
(\eta_1\eta_2-cos\, \theta_{12})=0$, i.e. ${\vec p}_1 // {\vec p}_2$ with
$\eta_1=\eta_2$ for $\theta_{12}=0$ and with $\eta_1=-\eta_2$ for $\theta_{12}
=\pi$. We have $Q^2\approx 0$, namely two branches

\begin{equation}
Q^o\approx \pm \sqrt{{\vec Q}^2}
\label{105}
\end{equation}

\noindent of excitations of relative energy with total energy

\begin{equation}
p^o\approx {{\vec p\cdot \vec Q}\over {Q^o}}\approx \pm {{\vec p\cdot \vec Q}
\over {\sqrt{{\vec Q}^2}}}.
\label{106}
\end{equation}

{}{}3c) Stratum with $p^{\mu}=0$ [$p^{\mu}_1=-p_2^{\mu}$, $\eta_1=-\eta_2$].
It has $Q^2 \approx 0$: there are two branches of excitations of relative
energy

\begin{equation}
Q^o\approx \pm \sqrt{{\vec Q}^2}.
\label{107}
\end{equation}

For $N > 2$ the discussion is more involved, since there are many degenerate
cases with complicated discussions of the allowed Poincar\'e orbits.

For N=3, $p^2 > 0$ and $m_1, m_2, m_3$ generic, there are 8 branches

\begin{eqnarray}
\epsilon &\approx& \pm [+\sqrt{m_1^2+3{({\hat \gamma}_{11}{\vec \pi}_1+{\hat
\gamma}_{21}{\vec \pi}_2)}^2}+\sqrt{m_2^2+3{({\hat \gamma}_{12}{\vec \pi}_1+
{\hat \gamma}_{22}{\vec \pi}_2)}^2}+\nonumber \\
&+&\sqrt{m_3^2+3{({\hat \gamma}_{13}{\vec \pi}
_1+{\hat \gamma}_{23}{\vec \pi}_2)}^2}]\nonumber \\
\epsilon &\approx& \pm |\, +\sqrt{m_1^2+3{({\hat \gamma}_{11}{\vec \pi}_1+{\hat
\gamma}_{21}{\vec \pi}_2)}^2}+\sqrt{m_2^2+3{({\hat \gamma}_{12}{\vec \pi}_1+
{\hat \gamma}_{22}{\vec \pi}_2)}^2}-\nonumber \\
&-&\sqrt{m_3^2+3{({\hat \gamma}_{13}{\vec \pi}
_1+{\hat \gamma}_{23}{\vec \pi}_2)}^2}\, | \nonumber \\
\epsilon &\approx& \pm |\, -\sqrt{m_1^2+3{({\hat \gamma}_{11}{\vec \pi}_1+{\hat
\gamma}_{21}{\vec \pi}_2)}^2}+\sqrt{m_2^2+3{({\hat \gamma}_{12}{\vec \pi}_1+
{\hat \gamma}_{22}{\vec \pi}_2)}^2}+\nonumber \\
&+&\sqrt{m_3^2+3{({\hat \gamma}_{13}{\vec \pi}
_1+{\hat \gamma}_{23}{\vec \pi}_2)}^2}\, | \nonumber \\
\epsilon &\approx& \pm |\, +\sqrt{m_1^2+3{({\hat \gamma}_{11}{\vec \pi}_1+{\hat
\gamma}_{21}{\vec \pi}_2)}^2}-\sqrt{m_2^2+3{({\hat \gamma}_{12}{\vec \pi}_1+
{\hat \gamma}_{22}{\vec \pi}_2)}^2}+\nonumber \\
&+&\sqrt{m_3^2+3{({\hat \gamma}_{13}{\vec \pi}
_1+{\hat \gamma}_{23}{\vec \pi}_2)}^2}\, | .
\label{108}
\end{eqnarray}

\noindent In the limit $m_1=m_2=m_3=m$, the first two branches start from
$\epsilon \approx \pm 3m$, while the other three positive (negative) branches
are tangent at $\epsilon \approx m$ ($\epsilon \approx -m$).

This pattern is valid for all N=2k+1.

For N=4, $p^2 > 0$ and $m_1, m_2, m_3, m_4$ generic, there are 16 branches

\begin{eqnarray}
\epsilon &\approx& \pm [ +\sqrt{m_1^2+4{({\hat \gamma}_{11}{\vec \pi}_1+{\hat
\gamma}_{21}{\vec \pi}_2+{\hat \gamma}_{31}{\vec \pi}_3)}^2}+
\sqrt{m_2^2+4{({\hat \gamma}_{12}{\vec \pi}_1+{\hat
\gamma}_{22}{\vec \pi}_2+{\hat \gamma}_{32}{\vec \pi}_3)}^2}+\nonumber \\
&+&\sqrt{m_3^2+4{({\hat \gamma}_{13}{\vec \pi}_1+{\hat
\gamma}_{23}{\vec \pi}_2+{\hat \gamma}_{33}{\vec \pi}_3)}^2}+
\sqrt{m_4^2+4{({\hat \gamma}_{14}{\vec \pi}_1+{\hat
\gamma}_{24}{\vec \pi}_2+{\hat \gamma}_{34}{\vec \pi}_3)}^2} ]\nonumber \\
\epsilon &\approx& \pm |\, +\sqrt{m_1^2+4{({\hat \gamma}_{11}{\vec \pi}_1+{\hat
\gamma}_{21}{\vec \pi}_2+{\hat \gamma}_{31}{\vec \pi}_3)}^2}+
\sqrt{m_2^2+4{({\hat \gamma}_{12}{\vec \pi}_1+{\hat
\gamma}_{22}{\vec \pi}_2+{\hat \gamma}_{32}{\vec \pi}_3)}^2}+\nonumber \\
&+&\sqrt{m_3^2+4{({\hat \gamma}_{13}{\vec \pi}_1+{\hat
\gamma}_{23}{\vec \pi}_2+{\hat \gamma}_{33}{\vec \pi}_3)}^2}-
\sqrt{m_4^2+4{({\hat \gamma}_{14}{\vec \pi}_1+{\hat
\gamma}_{24}{\vec \pi}_2+{\hat \gamma}_{34}{\vec \pi}_3)}^2}\, | \nonumber \\
\epsilon &\approx& \pm |\,-\sqrt{m_1^2+4{({\hat \gamma}_{11}{\vec \pi}_1+{\hat
\gamma}_{21}{\vec \pi}_2+{\hat \gamma}_{31}{\vec \pi}_3)}^2}+
\sqrt{m_2^2+4{({\hat \gamma}_{12}{\vec \pi}_1+{\hat
\gamma}_{22}{\vec \pi}_2+{\hat \gamma}_{32}{\vec \pi}_3)}^2}+\nonumber \\
&+&\sqrt{m_3^2+4{({\hat \gamma}_{13}{\vec \pi}_1+{\hat
\gamma}_{23}{\vec \pi}_2+{\hat \gamma}_{33}{\vec \pi}_3)}^2}+
\sqrt{m_4^2+4{({\hat \gamma}_{14}{\vec \pi}_1+{\hat
\gamma}_{24}{\vec \pi}_2+{\hat \gamma}_{34}{\vec \pi}_3)}^2}\, | \nonumber \\
\epsilon &\approx& \pm |\, +\sqrt{m_1^2+4{({\hat \gamma}_{11}{\vec \pi}_1+{\hat
\gamma}_{21}{\vec \pi}_2+{\hat \gamma}_{31}{\vec \pi}_3)}^2}-
\sqrt{m_2^2+4{({\hat \gamma}_{12}{\vec \pi}_1+{\hat
\gamma}_{22}{\vec \pi}_2+{\hat \gamma}_{32}{\vec \pi}_3)}^2}+\nonumber \\
&+&\sqrt{m_3^2+4{({\hat \gamma}_{13}{\vec \pi}_1+{\hat
\gamma}_{23}{\vec \pi}_2+{\hat \gamma}_{33}{\vec \pi}_3)}^2}+
\sqrt{m_4^2+4{({\hat \gamma}_{14}{\vec \pi}_1+{\hat
\gamma}_{24}{\vec \pi}_2+{\hat \gamma}_{34}{\vec \pi}_3)}^2}\, | \nonumber \\
\epsilon &\approx& \pm |\, +\sqrt{m_1^2+4{({\hat \gamma}_{11}{\vec \pi}_1+{\hat
\gamma}_{21}{\vec \pi}_2+{\hat \gamma}_{31}{\vec \pi}_3)}^2}+
\sqrt{m_2^2+4{({\hat \gamma}_{12}{\vec \pi}_1+{\hat
\gamma}_{22}{\vec \pi}_2+{\hat \gamma}_{32}{\vec \pi}_3)}^2}-\nonumber \\
&-&\sqrt{m_3^2+4{({\hat \gamma}_{13}{\vec \pi}_1+{\hat
\gamma}_{23}{\vec \pi}_2+{\hat \gamma}_{33}{\vec \pi}_3)}^2}+
\sqrt{m_4^2+4{({\hat \gamma}_{14}{\vec \pi}_1+{\hat
\gamma}_{24}{\vec \pi}_2+{\hat \gamma}_{34}{\vec \pi}_3)}^2}\, | \nonumber \\
\epsilon &\approx& \pm |\, +\sqrt{m_1^2+4{({\hat \gamma}_{11}{\vec \pi}_1+{\hat
\gamma}_{21}{\vec \pi}_2+{\hat \gamma}_{31}{\vec \pi}_3)}^2}+
\sqrt{m_2^2+4{({\hat \gamma}_{12}{\vec \pi}_1+{\hat
\gamma}_{22}{\vec \pi}_2+{\hat \gamma}_{32}{\vec \pi}_3)}^2}-\nonumber \\
&-&\sqrt{m_3^2+4{({\hat \gamma}_{13}{\vec \pi}_1+{\hat
\gamma}_{23}{\vec \pi}_2+{\hat \gamma}_{33}{\vec \pi}_3)}^2}-
\sqrt{m_4^2+4{({\hat \gamma}_{14}{\vec \pi}_1+{\hat
\gamma}_{24}{\vec \pi}_2+{\hat \gamma}_{34}{\vec \pi}_3)}^2}\, | \nonumber \\
\epsilon &\approx& \pm |\, -\sqrt{m_1^2+4{({\hat \gamma}_{11}{\vec \pi}_1+{\hat
\gamma}_{21}{\vec \pi}_2+{\hat \gamma}_{31}{\vec \pi}_3)}^2}+
\sqrt{m_2^2+4{({\hat \gamma}_{12}{\vec \pi}_1+{\hat
\gamma}_{22}{\vec \pi}_2+{\hat \gamma}_{32}{\vec \pi}_3)}^2}+\nonumber \\
&+&\sqrt{m_3^2+4{({\hat \gamma}_{13}{\vec \pi}_1+{\hat
\gamma}_{23}{\vec \pi}_2+{\hat \gamma}_{33}{\vec \pi}_3)}^2}-
\sqrt{m_4^2+4{({\hat \gamma}_{14}{\vec \pi}_1+{\hat
\gamma}_{24}{\vec \pi}_2+{\hat \gamma}_{34}{\vec \pi}_3)}^2}\, | \nonumber \\
\epsilon &\approx& \pm |\, +\sqrt{m_1^2+4{({\hat \gamma}_{11}{\vec \pi}_1+{\hat
\gamma}_{21}{\vec \pi}_2+{\hat \gamma}_{31}{\vec \pi}_3)}^2}-
\sqrt{m_2^2+4{({\hat \gamma}_{12}{\vec \pi}_1+{\hat
\gamma}_{22}{\vec \pi}_2+{\hat \gamma}_{32}{\vec \pi}_3)}^2}+\nonumber \\
&+&\sqrt{m_3^2+4{({\hat \gamma}_{13}{\vec \pi}_1+{\hat
\gamma}_{23}{\vec \pi}_2+{\hat \gamma}_{33}{\vec \pi}_3)}^2}-
\sqrt{m_4^2+4{({\hat \gamma}_{14}{\vec \pi}_1+{\hat
\gamma}_{24}{\vec \pi}_2+{\hat \gamma}_{34}{\vec \pi}_3)}^2}\, | .
\label{109}
\end{eqnarray}

\noindent In the equal mass limit, there are two branches starting at
$\epsilon \approx \pm 4m$, four positive (negative) branches starting at
$\epsilon \approx 2m$ ($\epsilon \approx -2m$) and six (three positive
and three negative) branches starting from $\epsilon \approx 0$.

This is the pattern for N=2k.

As we have seen, already in the free case there are generically tachionic
strata
corresponding to spacelike Poincar\'e orbits. At least at the classical level,
these strata have to be excluded to avoid problems with Einstein causality.
Instead, the rest-frame 1-time instant form is, by construction, free of
these tachionic strata.

For N=2, strata with $p^2 < 0$ appear always when $\eta_1=-\eta_2$; this fact
will extend to $N > 2$. Now there is the problem of the interpretation of
particles of negative mass and energy ($\eta_i=-$); they are present in
classical relativistic mechanics due to the two branches of timelike
Poincar\'e orbits and have a well defined nonrelativistic limit, connected
with the negative mass representations of the extended Galileo group
\cite{gal}.

If we make the first quantization of a scalar particle, the first class
constraint $p^2-m^2\approx 0$ goes into the one-particle Klein-Gordon
equation $(\Box +m^2)\phi (x)=0$ with $\phi (x)$ a complex wave function.
It is known\cite{nor} (see also Ref.\cite{longhi}) that this equation admits
two kinds of scalar products: i) the standard nondefinite positive  one
${(\phi_A,\phi_B)}_1=\int d^3\sigma_{\mu}\, \phi^{*}_A(x) {i\over 2}
\tensor{\partial}{}^{\mu}\phi_B(x)$ [$\tensor{\partial}{}^{\mu}=
\stackrel{\rightarrow}{\partial}{}^{\mu}-\stackrel{\leftarrow}{\partial}{}^{\mu}
$] associated with the conserved current $j_1^{\mu}(x)={i\over 2}\phi^{*}(x)
\tensor{\partial}{}^{\mu}\phi (x)$, which is interpreted as a charge (electric
charge, strangeness,..) current with the sign of the charge corresponding to
positive- and negative- norm (i.e. energy) states; ii) a positive definite
nonlocal one ${(\phi_A,\phi_B)}_2=\int d^3\sigma_{\mu}\, {i\over 4}\phi^{*}_A
(x)({\hat \eta}-{\hat \eta}^{\dagger})\tensor{\partial}{}^{\mu}\phi_B(x)$
[${\hat \eta}=i\partial^o/\sqrt{m^2-{\vec \partial}^2}$ is the nonlocal
Lorentz-scalar (under proper Lorentz transformations) ``sign of the energy"
operator satisfying $[\hat \eta ,x^{\mu}]=[\hat \eta ,i\partial^{\mu}]=0$],
associated with the conserved current $j^{\mu}_2(x)={i\over 4}\phi^{*}(x)
({\hat \eta}-{\hat \eta}^{\dagger})\tensor{\partial}{}^{\mu}\phi(x)$. The
negative mass (and energy) states are interpreted as describing antiparticles:
for charged scalar particles ($\pi^{+}\pi^{-}, K^{+}K^{-}, K^o{\bar K}^o$)
positive- (negative-) energy states describe particles with charge $+Q$
(antiparticles with charge $-Q$), which can be connected by a ``charge
conjugation" operation $\phi (x)\mapsto \eta_c\phi^{*}(x)$, $|\eta_c|=1$,
under which the action is invariant [for the electromagnetic potential one has
$A_{\mu}(x)\mapsto -A_{\mu}(x)$]; for neutral particles like $\pi^o$,
coinciding
with the antiparticle, see Ref.\cite{fv}.

In Ref.\cite{stuc}, it was shown at the classical level that, since the proper
time of a particle is defined by $ds=\pm \sqrt{dx^{\mu}\eta_{\mu\nu}dx^{\nu}}$
if $dx^o\, \stackrel{>}{<} 0$, we have $ds=\pm md\tau \sqrt{{\dot x}^{\mu}
\eta_{\mu\nu}{\dot x}^{\nu}}$, so that $d\tau > 0$ corresponds to
$dx^o\, \stackrel{>}{<} 0$ if ${\dot x}^o={{dx^o}\over {d\tau}}
\stackrel{>}{<} 0$; therefore, an evolution towards the future in $d\tau$ for
a particle with negative mass -m corresponds to an evolution towards the past
in $dx^o$ and this is in accord with the use of the complex Stueckelberg-
Feynman Green function $G_F(x)$ [see for instance Ref.\cite{iz}], which
propagates the positive- (negative-) frequencies forward (backward) in $x^o$,
in association with the first-quantized Klein-Gordon equation. Finally, the
equations of motion of a particle of mass m and electric charge e in presence
of gravitational and electromagnetic fields depend only on the ratio e/m, so
that an antiparticle of mass -m (propagating forward in $\tau$ but backwards
in $x^o$) and charge -e satisfies the same equations as the particle (m,e) if
parametrized in $\tau$, in accord with the charge conjugation invariance.

This is what happens with the parametrization in $\tau$ of the 1-time theory.
Let us remark that in the rest-frame instant form one has $\lbrace \epsilon
=\eta \sqrt{p^2}, T={{p\cdot x}\over {\eta \sqrt{p^2}}}\rbrace =1$, so that
$\eta m$ is associated with $\eta |\, T|$.

In contrast, in the Klein-Gordon quantum field theory, a hermitean quantum
field
$\hat \phi (x)$ [quantization of a classical real Klein-Gordon field $\phi
(x)=\phi^{*}(x)$ for which the complex Green function $G_F(x)$ cannot be
used, but only the real retarded and advanced ones] has the positive-
(negative-) frequency part associated with the creation (annihilation)
operators
of a scalar particle with positive energy. For a complex quantum field
$\hat \phi (x), {\hat \phi}^{\dagger}(x)$ [quantization of two real
Klein-Gordon fields, $\phi (x)={1\over {\sqrt{2}}}(\phi_1(x)+i\phi_2(x))$]
the two kinds of creation and annihilation operators are associated with a
particle of mass m and charge +1 and with an antiparticle of mass m and charge
-1 respectively.

The conclusion of this discussion is that, in the classical background of
relativistic particle physics, we have to consider only the branch of the
mass spectrum of N scalar particles with all the masses positive and with
particle and antiparticle distinguished by opposite charges [only neutral
scalar
particles cannot be described in this way, but they (by the way also the
charged scalar ones) are supposed to be bound states of spin 1/2 quarks];
the pseudothresholds of the lower positive branches are connected to the
thresholds of the other existing kinematical invariants (relevant for
scattering theory due to the crossing property).

Let us now consider the introduction of action-at-a-distance interactions.
This problem has been studied in the N-time theory (see the bibliography of
Refs.\cite{lusi,todor,taka}). In the framework of models with N first class
constraints, a closed form of the constraints is known for N=2: this is the
DrozVincent-Todorov-Komar model\cite{dv,todo,komar}, on which Ref.\cite{longhi}
is based. The constraints take the general form (only the case $V(R^2_{\perp})$
has been studied in detail; see Refs.\cite{longhi,pons} for the
nonrelativistic limit)

\begin{eqnarray}
\phi_i&=&p_i^2-m_i^2+V(R^2_{\perp}, p^2,Q^2,Q\cdot R_{\perp})\approx 0,
\quad i=1,2,\quad\quad
R^{\mu}_{\perp}=(\eta^{\mu}_{\nu}-{{p^{\mu}p_{\nu}}\over {p^2}})R^{\nu}
\nonumber \\
&&\lbrace \phi_1,\phi_2\rbrace =0.
\label{110}
\end{eqnarray}

\noindent By construction we have $p^2 > 0$. The potential V may be either
confining or separable (see Ref.\cite{long} for a finite-range potential, in
which the particles see each other only if their Cauchy data are restricted to
be compatible with the first class constraints in the interacting region).

For $N > 2$ there are models with a closed form of the constraints $\phi_i
\approx 0$, $\lbrace \phi_i,\phi_j\rbrace =0$, only for confining potentials
(see for instance Ref.\cite{lusf}). To find separable potentials such that
$\lbrace \phi_i,\phi_j\rbrace =0$ requires the solution of complicated
nonlinear partial differential equations connected with the necessity of
introducing 3-, 4-,.. N-body forces\cite{ror}, and no one succeeded in solving
them; it was only shown that solutions exist, in which the first class
constraints are expressed as series in the coupling constants\cite{sazt} and
that the solution is unique\cite{hk}. See also Refs.\cite{seplon,pons}.

In contrast, in the 1-time theory (the rest-frame instant form for $p^2 > 0$),
we can introduce the interactions like in Newtonian mechanics and use the
nonrelativistic definition of separability of the interactions. It is the
transition from the 1-time to the N-time theory that contains all the previous
difficulties.

The most general form of the final constraint (\ref{78}) of the 1-time theory
is

\begin{eqnarray}
{\hat {\cal H}}&=&\epsilon_s-\sum_{i=1}^N\eta_i\sqrt{ m_i^2+V_i+N{(\sum_{a=1}
^{N-1}{\hat \gamma}_{ai}{\vec \pi}_a)}^2 }-\sum_{i\not= j}^{1..N}\, U_{ij}
=\epsilon_s-H_R^{(T_s)}\approx 0\nonumber \\
&&V_i=V_i[\sum_{a=1}^{N-1}({\hat \gamma}_{ah}-{\hat \gamma}_{ak}){\vec \rho}_a]
\nonumber \\
&&U_{ij}=U_{ij}[\sum_{a=1}^{N-1}({\hat \gamma}_{ai}-{\hat \gamma}_{aj}){\vec
\rho}_a].
\label{111}
\end{eqnarray}

With $V_i=0$ and N=2, it is the form prescribed in Ref.\cite{baka} [see also
the bibliography of Ref.\cite{ppl}, but, as we shall see in the next Section,
the 2-body Coulomb potential coming from the longitudinal modes of the
electromagnetic field is of the type $U_{ij}$; therefore the models with
U-type potentials can be thought as deriving from couplings to gauge field
theories.

The action-at-a-distance interactions (instantaneous in the rest frame)
of models like the one of Refs.\cite{dv,todo,komar} (for instance the
relativistic harmonic oscillator) are
of the type $V_i$ (additive to $m_i^2$). If $V_i=\sum_{k\not= i}^{1..N}V_{ik}
[\sum_{a=1}^{N-1}({\hat \gamma}_{ai}-{\hat \gamma}_{ak}){\vec \rho}_a]+
\sum_{k,h\not= i}^{1..N}V_{ikh}[\sum_{a=1}^{N-1}({\hat \gamma}_{ai}-{\hat
\gamma}_{ak}){\vec \rho}_a, \sum_{a=1}^{N-1}({\hat \gamma}_{ai}-{\hat \gamma}
_{ah}){\vec \rho}_a]+...$, we have 2-body potentials $V_{ik}$, 3-body
potentials
$V_{ikh}$ and so on.
The potentials are separable if, whenever particle i decouples from the other
N-1 (the same must hold true for clusters of particles), we have $V_i
\rightarrow 0$ and $U_{ij}\rightarrow 0$, so that no one of the surviving
potentials depend on the index ``i".

Let us remark that we can have two different 1-time theories with the same
nonrelativistic limit $H_R$: i) $H^{(T_s)}_R=\sum_{i=1}^N\eta_i\sqrt{ m_i^2+
V_i+N{(\sum_{a=1}^{N-1}{\hat \gamma}_{ai}{\vec \pi}_a)}^2 }$; ii) $H_R^{(T_s)}=
\sum_{i=1}^N\eta_i\sqrt{ m_i^2+N{(\sum_{a=1}^{N-1}{\hat \gamma}_{ai}{\vec \pi}
_a)}^2 }\, +U$ if $U=\sum_{i=1}^N{{V_i}\over {2m_i}}$; instead, the N-time
theory with a unique potential\cite{lusf} [$p_i^2-m_i^2+{1\over {N^2}}{\tilde
V}\approx 0$] would give N nonrelativistic constraints [see Ref.\cite{pons}
for the case N=2] $\psi_i=E_i-{{{\vec p}^2_i}\over {2m_i}}-{{\tilde V}\over
{2m_iN}}\approx 0$ with $\sum_{i=1}^N\psi_i=E-H=E-{{{\vec p}^2}\over {2\sum
_{i=1}^Nm_i}}+H_R\approx 0$ with the same $H_R$ of the 1-time theories if $U=
\tilde V$.

Therefore, there are two 1-time models for the relativistic harmonic oscillator
when N=2: i) the 1-time version of the DrozVincent-Todorov-Komar model
$H^{(T_s)}_R=\eta_1\sqrt{m_1^2+V({\vec \rho}^2)+2{\vec \pi}^2}+\eta_2
\sqrt{m_2^2+V({\vec \rho}^2)+2{\vec \pi}^2}$; ii) a model with $H^{(T_s)}_R=
\eta_1\sqrt{m_1^2+2{\vec \pi}^2}+\eta_2\sqrt{m_2^2+2{\vec \pi}^2}+U({\vec
\rho}^2)$ with $U={1\over {\mu}}V$ [$\mu={{m_1m_2}\over {m_1+m_2}}$ is the
reduced mass]. The second model should be interpreted as coming from a gauge
field theory producing action-at-a-distance interparticle harmonic forces.

Let us also note that the number of branches of the mass spectrum in the free
case is a topological number: it is the dimension of the zeroth homotopy
group of the constraint hypersurface, counting how many disjoint components
are in it. While certain interactions  preserve this number, generic
interactions will change it; therefore, the interactions should be classified
according to the (lacking) theory of intersections of noncompact
hypersurfaces in phase space (when certain mass gaps disappear).

In the approach of this paper, based on the
theory of canonical realizations of the Poincar\'e group in phase space,
we cannot introduce external interactions, without destroying the
technology we are using, which presupposes the existence of ten finite
conserved Poincar\'e generators for every system under study.
External interactions should be thought as limits
from our isolated systems, when some its subsystem has the invariant mass
tending to infinity; alternatively one can try to introduce them in the final
reduced form of the constraints, like those in Eqs.(\ref{111}).

\vfill\eject

\section{The Electromagnetic Interaction}

In Ref.\cite{karp}, the case of a charged scalar particle interacting with the
electromagnetic field was considered. The 1-time theory was used because it
is not known to us how to develop a covariant N-time description, in which
the Poisson brackets of particle and field constraints can be evaluated
in a covariant way. Here, the case of N charged scalar
particles will be considered, with the electric charge of each particle
described in a pseudoclassical way\cite{casal} by means of a pair of
complex conjugate Grassmann variables\cite{casala} $\theta_i(\tau ), \theta
^{*}_i(\tau )$ satisfying [$I_i=I^{*}_i=\theta^{*}_i\theta_i$ is the generator
of the $U_{em}(1)$ group of particle i]

\begin{eqnarray}
&&\theta^2_i=\theta_i^{{*}2}=0,\quad\quad \theta_i\theta^{*}_i+\theta^{*}_i
\theta_i=0,\nonumber \\
&&\theta_i\theta_j=\theta_j\theta_i,\quad\quad \theta_i\theta^{*}_j=
\theta_j^{*}\theta_i,\quad\quad \theta^{*}_i\theta^{*}_j=\theta^{*}_j\theta
^{*}_i,\quad\quad i\not= j.
\label{112}
\end{eqnarray}

\noindent This amounts to assume that the electric charges $Q_i=e_i\theta^{*}_i
\theta_i$ are quatized with levels 0 and $e_i$\cite{casala}.

On the hypersurface $\Sigma (\tau )$, we describe the electromagnetic potential
and field strength with Lorentz-scalar variables $A_{\check A}(\tau ,\vec
\sigma
)$ and $F_{{\check A}{\check B}}(\tau ,\vec \sigma )$ respectively, defined by

\begin{eqnarray}
&&A_{\check A}(\tau ,\vec \sigma )=z^{\mu}_{\check A}(\tau ,\vec \sigma )
A_{\mu}(z(\tau ,\vec \sigma ))\nonumber \\
&&F_{{\check A}{\check B}}(\tau ,\vec \sigma )={\partial}_{\check A}A_{\check
B}(\tau ,\vec \sigma )-{\partial}_{\check B}A_{\check A}(\tau ,\vec \sigma )=
z^{\mu}_{\check A}(\tau ,\vec \sigma )z^{\nu}_{\check B}(\tau ,\vec \sigma )
F_{\mu\nu}(z(\tau ,\vec \sigma )).
\label{113}
\end{eqnarray}

The system is described by the action

\begin{eqnarray}
S&=& \int d\tau d^3\sigma \, {\cal L}(\tau ,\vec
\sigma )=\int d\tau L(\tau )\nonumber \\
L(\tau )&=&\int d^3\sigma {\cal L}(\tau ,\vec \sigma )\nonumber \\
{\cal L}(\tau ,\vec \sigma )&=&{i\over 2}\sum_{i=1}^N\delta^3(\vec \sigma
-{\vec \eta}_i(\tau ))[\theta^{*}_i(\tau ){\dot \theta}_i(\tau )-
{\dot \theta}^{*}_i(\tau )\theta_i(\tau )]-\nonumber \\
&&-\sum_{i=1}^N\delta^3(\vec \sigma -{\vec \eta}_i
(\tau ))[\eta_im_i\sqrt{ g_{\tau\tau}(\tau ,\vec \sigma )+2g_{\tau {\check r}}
(\tau ,\vec \sigma ){\dot \eta}^{\check r}_i(\tau )+g_{{\check r}{\check s}}
(\tau ,\vec \sigma ){\dot \eta}_i^{\check r}(\tau ){\dot \eta}_i^{\check s}
(\tau )  }+\nonumber \\
&&+e_i\theta^{*}_i(\tau )\theta_i(\tau )(A_{\tau }(\tau ,\vec \sigma )+
A_{\check r}(\tau ,\vec \sigma ){\dot \eta}^{\check r}_i(\tau ))]-\nonumber \\
&&-{1\over 4}\, \sqrt {g(\tau ,\vec \sigma )}
g^{{\check A}{\check C}}(\tau ,\vec \sigma )g^{{\check B}{\check D}}
(\tau ,\vec \sigma )F_{{\check A}{\check B}}(\tau ,\vec \sigma )
F_{{\check C}{\check D}}(\tau ,\vec \sigma ),
\label{114}
\end{eqnarray}

\noindent where the configuration variables are $z^{\mu}(\tau ,\vec \sigma )$
$A_{\check A}(\tau ,\vec \sigma )$, ${\vec \eta}_i(\tau )$, $\theta_i(\tau )$
and $\theta^{*}_i(\tau )$, i=1,..,N. We have

$-{1\over 4}\sqrt{g} g^{\check A\check C}g^{\check B\check D}F_{\check A
\check B}F_{\check C\check D}=-\sqrt{\gamma}[{1\over 2}\sqrt{ {{\gamma}\over
g} } F_{\tau \check r}\gamma^{\check r\check s}F_{\tau \check s}-{{\gamma}
\over g} g_{\tau \check v}\gamma^{\check v\check r}F_{\check r\check s}
\gamma^{\check s\check u}F_{\tau \check u}+{1\over 4}\sqrt{ {g\over {\gamma}}}
\gamma^{\check r\check s}F_{\check r\check u}F_{\check s\check
v}(\gamma^{\check
u\check v}+2{{\gamma}\over g}g_{\tau \check m}\gamma^{\check m\check u}
g_{\tau \check n}\gamma^{\check n\check v})].$

The action is invariant under separate
$\tau$- and $\vec \sigma$-reparametrizations, since $A_{\tau}(\tau ,\vec \sigma
)$ transforms as a $\tau$-derivative; moreover, it is invariant under the odd
phase transformations $\delta \theta_i\mapsto i\alpha \theta_i$, generated by
the $I_i$'s.

The canonical momenta are [$E_{\check r}=F_{{\check r}\tau}$ and $B_{\check r}
=\epsilon_{{\check r}{\check s}{\check t}}F_{{\check s}{\check t}}$ are the
electric and magnetic fields respectively]

\begin{eqnarray}
\rho_{\mu}(\tau ,\vec \sigma )&=&-{ {\partial {\cal L}(\tau ,\vec \sigma )}
\over {\partial z^{\mu}_{\tau}(\tau ,\vec \sigma )} }=\sum_{i=1}^N\delta^3
(\vec \sigma -{\vec \eta}_i(\tau ))\eta_im_i\nonumber \\
&&{ {z_{\tau\mu}(\tau ,\vec \sigma )+z_{{\check r}\mu}(\tau ,\vec \sigma )
{\dot \eta}_i^{\check r}(\tau )}\over {\sqrt{g_{\tau\tau}(\tau ,\vec \sigma )+
2g_{\tau {\check r}}(\tau ,\vec \sigma ){\dot \eta}_i^{\check r}(\tau )+
g_{{\check r}{\check s}}(\tau ,\vec \sigma ){\dot \eta}_i^{\check r}(\tau
){\dot
\eta}_i^{\check s}(\tau ) }} }+\nonumber \\
&&+{ {\sqrt {g(\tau ,\vec \sigma )}}\over 4}[(g^{\tau \tau}z_{\tau \mu}+
g^{\tau {\check r}}z_{\check r\mu})(\tau ,\vec \sigma )g^{{\check A}{\check C}}
(\tau ,\vec \sigma )g^{{\check B}{\check D}}(\tau ,\vec \sigma )F_{{\check A}
{\check B}}(\tau ,\vec \sigma )F_{{\check C}{\check D}}(\tau ,\vec \sigma )
-\nonumber \\
&-&2[z_{\tau \mu}(\tau ,\vec \sigma )(g^{\check A\tau}g^{\tau \check C}
g^{{\check B}{\check D}}+g^{{\check A}{\check C}}g^{\check B\tau}g^{\tau
\check D})(\tau ,\vec \sigma )+\nonumber \\
&+&z_{\check r\mu}(\tau ,\vec \sigma )
(g^{{\check A}{\check r}}g^{\tau {\check C}}+g^{{\check A}\tau}g^{{\check r}
{\check C}})(\tau ,\vec \sigma )g^{{\check B}{\check D}}
(\tau ,\vec \sigma )]F_{{\check A}{\check B}}(\tau ,\vec \sigma )
F_{{\check C}{\check D}}(\tau ,\vec \sigma )]=
\nonumber \\
&&=[(\rho_{\nu}l^{\nu})l_{\mu}+(\rho_{\nu}z^{\nu}_{\check r})\gamma^{{\check r}
{\check s}}z_{{\check s}\mu}](\tau ,\vec \sigma )\nonumber \\
&&{}\nonumber \\
\pi^{\tau}(\tau ,\vec \sigma )&=&{ {\partial L}\over {\partial \partial_{\tau}
A_{\tau}(\tau ,\vec \sigma )} }=0,\nonumber \\
\pi^{\check r}(\tau ,\vec \sigma )&=&{ {\partial L}\over {\partial \partial
_{\tau}A_{\check r}(\tau ,\vec \sigma )} }=-{ {\gamma (\tau ,\vec \sigma )}
\over {\sqrt {g(\tau ,\vec \sigma )}} }\gamma^{{\check r}{\check s}}(\tau ,
\vec \sigma )(F_{\tau {\check s}}-g_{\tau {\check v}}\gamma^{{\check v}
{\check u}}F_{{\check u}{\check s}})(\tau ,\vec \sigma )=\nonumber \\
&&={ {\gamma (\tau ,\vec \sigma )}\over {\sqrt {g(\tau ,\vec \sigma )}} }
\gamma^{{\check r}{\check s}}(\tau ,\vec \sigma )(E_{\check s}(\tau ,\vec
\sigma )-g_{\tau {\check v}}(\tau ,\vec \sigma )\gamma^{{\check v}{\check u}}
(\tau ,\vec \sigma )\epsilon_{{\check u}{\check s}{\check t}} B_{\check t}
(\tau ,\vec \sigma )),\nonumber \\
&&{}\nonumber \\
\kappa_{i{\check r}}(\tau )&=&-{ {\partial L(\tau )}\over {\partial {\dot
\eta}_i^{\check r}(\tau )} }=\nonumber \\
&=&\eta_im_i{ {g_{\tau {\check r}}(\tau ,{\vec \eta}_i(\tau ))+g_{{\check r}
{\check s}}(\tau ,{\vec \eta}_i(\tau )){\dot \eta}_i^{\check s}(\tau )}\over
{ \sqrt{g_{\tau\tau}(\tau ,{\vec \eta}_i(\tau ))+
2g_{\tau {\check r}}(\tau ,{\vec \eta}_i(\tau )){\dot \eta}_i^{\check r}(\tau
)+
g_{{\check r}{\check s}}(\tau ,{\vec \eta}_i(\tau )){\dot \eta}_i^{\check r}
(\tau ){\dot \eta}_i^{\check s}(\tau ) }} }+\nonumber \\
&+&e_i\theta^{*}_i(\tau )\theta_i(\tau )A_{\check r}(\tau ,{\vec \eta}_i
(\tau ),\nonumber \\
&&{}\nonumber \\
\pi_{\theta \,i}(\tau )&=&{{\partial L(\tau )}\over {\partial {\dot \theta}_i
(\tau )}}=-{i\over 2}\theta^{*}_i(\tau )\nonumber \\
\pi_{\theta^{*} \, i}(\tau )&=&{{\partial L(\tau )}\over {\partial {\dot
\theta}^{*}_i(\tau )}}=-{i\over 2}\theta_i(\tau ),
\label{115}
\end{eqnarray}

\noindent and the following Poisson brackets are assumed

\begin{eqnarray}
&&\lbrace z^{\mu}(\tau ,\vec \sigma ),\rho_{\nu}(\tau ,{\vec \sigma}^{'}\rbrace
=-\eta^{\mu}_{\nu}\delta^3(\vec \sigma -{\vec \sigma}^{'})\nonumber \\
&&\lbrace A_{\check A}(\tau ,\vec \sigma ),\pi^{\check B}(\tau ,\vec
\sigma^{'} )\rbrace =\eta^{\check B}_{\check A}
\delta^3(\vec \sigma -\vec \sigma^{'})\nonumber \\
&&\lbrace \eta^{\check r}_i(\tau ),\kappa_j^{\check s}(\tau )\rbrace =
\delta_{ij}\delta^{{\check r}{\check s}}\nonumber \\
&&\lbrace \theta_i(\tau ),\pi_{\theta \, j}(\tau )\rbrace =-\delta_{ij}
\nonumber \\
&&\lbrace \theta^{*}_i(\tau ),\pi_{\theta^{*} \, j}(\tau )\rbrace
=-\delta_{ij}.
\label{116}
\end{eqnarray}

The Grassmann momenta give origin to the second class constraints $\pi_{\theta
\, i}+{i\over 2}\theta^{*}_i\approx 0$, $\pi_{\theta^{*}\, i}+{i\over 2}
\theta_i\approx 0$ [$\lbrace \pi_{\theta \, i}+{i\over 2}\theta^{*}_i,
\pi_{\theta^{*}\, j}+{i\over 2}\theta_j\rbrace =-i\delta_{ij}$]; $\pi
_{\theta \, i}$ and $\pi_{\theta^{*}\, i}$ are then eliminated with the
Dirac brackets

\begin{equation}
\lbrace A,B\rbrace {}^{*}=\lbrace A,B\rbrace -i[\lbrace A,\pi_{\theta \, i}+
{i\over 2}\theta^{*}_i\rbrace \lbrace \pi_{\theta^{*}\, i}+{i\over 2}
\theta_i,B\rbrace  +\lbrace A,\pi_{\theta^{*}\, i}+{i\over 2}
\theta_i \rbrace \lbrace \pi_{\theta \, i}+{i\over 2}\theta^{*}_i,B\rbrace ]
\label{117}
\end{equation}

\noindent so that the remaining Grassmann variables have the fundamental
Dirac brackets [which we will still denote $\lbrace .,.\rbrace$ for the sake of
simplicity]

\begin{eqnarray}
&&\lbrace \theta_i(\tau ),\theta_j(\tau )\rbrace = \lbrace \theta_i^{*}(\tau ),
\theta_j^{*}(\tau )\rbrace =0\nonumber \\
&&\lbrace \theta_i(\tau ),\theta_j^{*}(\tau )\rbrace =-i\delta_{ij}.
\label{118}
\end{eqnarray}

Again, we obtain four primary constraints

\begin{eqnarray}
&{\cal H}_{\mu}&(\tau ,\vec \sigma )= \rho_{\mu}(\tau ,\vec \sigma )-
l_{\mu}(\tau ,\vec \sigma )[T_{\tau\tau}(\tau ,\vec \sigma )+
\nonumber \\
&+&\sum_{i=1}^N\delta^3(\vec \sigma -{\vec \eta}_i(\tau ))\times
\nonumber \\
&\eta_i&\sqrt{ m^2_i-\gamma^{{\check r}{\check s}}(\tau ,\vec \sigma )
[\kappa_{i{\check r}}(\tau )-e_i\theta^{*}_i(\tau )\theta_i(\tau )A_{\check r}
(\tau ,\vec \sigma )][\kappa_{i{\check s}}(\tau ) -e_i\theta^{*}_i(\tau )
\theta_i(\tau )A_{\check s}(\tau ,\vec \sigma )]  }]-\nonumber \\
&-&z_{{\check r}\mu}(\tau ,\vec \sigma )\gamma^{{\check r}{\check s}}
(\tau ,\vec \sigma )\lbrace
T_{\tau \check s}(\tau ,\vec \sigma )+\sum_{i=1}^N\delta^3(\vec \sigma -
{\vec \eta}_i(\tau ))[\kappa_{i{\check s}}-e_i\theta_i^{*}(\tau )\theta_i
(\tau )A_{\check s}(\tau ,\vec \sigma )] \rbrace \approx 0,
\label{119}
\end{eqnarray}

\noindent where

\begin{eqnarray}
T_{\tau \tau}(\tau ,\vec \sigma )
&=&-{1\over 2}({1\over {\sqrt {\gamma}} }\pi^{\check r}g_{{\check r}{\check s}}
\pi^{\check s}-{ {\sqrt {\gamma}}\over 2}\gamma^{{\check r}{\check s}}
\gamma^{{\check u}{\check v}}F_{{\check r}{\check u}}F_{{\check s}{\check v}})
(\tau ,\vec \sigma ),\nonumber \\
T_{\tau {\check s}}(\tau ,\vec \sigma )&=&F_{{\check s}{\check t}}(\tau ,\vec
\sigma )\pi^{\check t}(\tau ,\vec \sigma )={[\vec B(\tau ,\vec \sigma )
\times \vec \pi (\tau ,\vec \sigma )]}_{\check s},
\label{120}
\end{eqnarray}

\noindent are the energy density and the Poynting vector respectively.

Since the canonical Hamiltonian is (we assume boundary conditions for the
electromagnetic potential such that all the surface terms can be neglected;
see Ref.\cite{lusc})

\begin{eqnarray}
H_c&=&-\sum_{i=1}^N\kappa_{i{\check r}}(\tau ){\dot \eta}_i^{\check r}(\tau )+
\int d^3\sigma [\pi^{\check A}(\tau ,\vec \sigma )\partial_{\tau}A_{\check A}
(\tau ,\vec \sigma )-\rho_{\mu}(\tau ,\vec \sigma )z^{\mu}_{\tau}(\tau ,\vec
\sigma )-{\cal L}(\tau ,\vec \sigma )]=\nonumber \\
&=&\int d^3\sigma [\partial_{\check r}(\pi^{\check r}(\tau ,\vec \sigma )
A_{\tau}(\tau ,\vec \sigma )-A_{\tau}(\tau ,\vec \sigma )\Gamma
(\tau ,\vec \sigma )]=-\int d^3\sigma A_{\tau}(\tau ,\vec \sigma )
\Gamma (\tau ,\vec \sigma ),
\label{121}
\end{eqnarray}

\noindent with

\begin{equation}
\Gamma(\tau ,\vec \sigma )=-\partial^r\pi^r(\tau ,\vec \sigma )+\sum_{i=1}
^Ne_i\theta^{*}_i(\tau )\theta_i(\tau )\delta^3(\vec \sigma -{\vec \eta}_i
(\tau )),
\label{122}
\end{equation}

\noindent we have the Dirac Hamiltonian ($\lambda^{\mu}(\tau ,\vec \sigma )$
and $\lambda_{\tau}(\tau ,\vec \sigma )$  are Dirac's multipliers)

\begin{equation}
H_D=\int d^3\sigma \lambda^{\mu}(\tau ,\vec \sigma ){\cal H}_{\mu}(\tau ,\vec
\sigma )+
\lambda_{\tau}(\tau ,\vec \sigma )\pi^{\tau}(\tau ,\vec \sigma )-
A_{\tau}(\tau ,\vec \sigma )\Gamma (\tau ,\vec \sigma )].
\label{123}
\end{equation}

The Lorentz scalar constraint $\pi^{\tau}(\tau ,\vec \sigma )\approx 0$ is
generated by the gauge invariance of S; its time constancy will produce the
only secondary constraint (Gauss law)

\begin{equation}
\Gamma (\tau ,\vec \sigma )\approx 0.
\label{124}
\end{equation}

The six constraints ${\cal H}_{\mu}(\tau ,\vec \sigma )\approx 0$, $\pi^{\tau}
(\tau ,\vec \sigma )\approx 0$, $\Gamma (\tau ,\vec \sigma )\approx 0$ are
first class with the only non vanishing Poisson brackets

\begin{eqnarray}
\lbrace {\cal H}_{\mu}(\tau ,\vec \sigma )&,&{\cal H}_{\nu}(\tau ,{\vec
\sigma}^{'} )\rbrace =\nonumber \\
&=&\lbrace [l_{\mu}(\tau ,\vec \sigma )z_{{\check r}\nu}(\tau ,\vec \sigma )
-l_{\nu}(\tau ,\vec \sigma )z_{{\check r}\mu}(\tau ,\vec \sigma )]
{ {g_{{\check r}{\check s}}(\tau ,\vec \sigma )\pi^{\check s}(\tau ,\vec
\sigma )}\over {\sqrt{\gamma (\tau ,\vec \sigma )}} }+\nonumber \\
&+&z^{\check r}_{\mu}(\tau ,\vec \sigma )F_{{\check r}{\check s}}(\tau ,\vec
\sigma )z^{\check s}_{\nu}(\tau ,\vec \sigma )\rbrace
\Gamma (\tau ,\vec \sigma )\delta^3(\vec \sigma -{\vec \sigma}^{'})\approx 0.
\label{125}
\end{eqnarray}

\noindent Let us remark that the simplicity of Eqs.(\ref{125}) is due to the
use of Cartesian coordinates: if we had used the constraints ${\cal H}_l(\tau ,
\vec \sigma )=l^{\mu}(\tau ,\vec \sigma ){\cal H}_{\mu}(\tau ,\vec \sigma )$,
${\cal H}_{\check r}(\tau ,\vec \sigma )=z^{\mu}_{\check r}(\tau ,\vec \sigma )
{\cal H}_{\mu}(\tau ,\vec \sigma )$ (i.e. nonholonomic coordinates), so that
their associated Dirac multipliers $\lambda_l(\tau ,\vec \sigma )$, $\lambda
_{\check r}(\tau ,\vec \sigma )$ would have been
the lapse and shift functions of general relativity, one would have obtained
the universal algebra of Ref.\cite{diraca}.

The ten conserved Poincar\'e generators are

\begin{eqnarray}
P_s^{\mu}&=&\int d^3\sigma \, \rho^{\mu}(\tau ,\vec \sigma ),\nonumber \\
J_s^{\mu\nu}&=&\int d^3\sigma \, (z^{\mu}(\tau ,\vec \sigma )\rho^{\nu}
(\tau ,\vec \sigma )-z^{\nu}(\tau ,\vec \sigma )\rho^{\mu}(\tau ,\vec \sigma
)),
\label{126}
\end{eqnarray}

\noindent so that the total momentum is built starting from the existing
energy momentum densities on the hypersurface

\begin{eqnarray}
&\int& d^3\sigma {\cal H}^{\mu}(\tau ,\vec \sigma )=p^{\mu}_s-
\int d^3\sigma l_{\mu}(\tau ,\vec \sigma )[T_{\tau\tau}(\tau ,\vec \sigma )+
\nonumber \\
&+&\sum_{i=1}^N\delta^3(\vec \sigma -{\vec \eta}_i(\tau ))\times
\nonumber \\
&\eta_i&\sqrt{ m^2_i-\gamma^{{\check r}{\check s}}(\tau ,\vec \sigma )
[\kappa_{i{\check r}}(\tau )-e_i\theta^{*}_i(\tau )\theta_i(\tau )A_{\check r}
(\tau ,\vec \sigma )][\kappa_{i{\check s}}(\tau )-e_i\theta^{*}_i(\tau )
\theta_i(\tau )A_{\check s}(\tau ,\vec \sigma )]  }]-\nonumber \\
&-&\int d^3\sigma
z_{{\check r}\mu}(\tau ,\vec \sigma )\gamma^{{\check r}{\check s}}
(\tau ,\vec \sigma )\lbrace
T_{\tau \check s}(\tau ,\vec \sigma )+\nonumber \\
&+&\sum_{i=1}^N\delta^3(\vec \sigma -
{\vec \eta}_i(\tau ))[\kappa_{i{\check s}}(\tau )-e_i\theta_i^{*}(\tau
)\theta_i
(\tau )A_{\check s}(\tau ,\vec \sigma )] \rbrace \approx 0.
\label{127}
\end{eqnarray}

If we add the gauge-fixings (\ref{47}) and the Dirac brackets (\ref{51}),
(\ref{57}), we remain with the variables $x^{\mu}_s, p^{\mu}_s,b^{\mu}_{\check
A}, S_s^{\mu\nu}, A_{\check A}, \pi^{\check A}, {\vec \eta}_i, {\vec \kappa}_i,
\theta_i, \theta^{*}_i$ and the twelve constraints

\begin{eqnarray}
{\tilde {\cal H}}^{\mu}(\tau )
&=&\int d^3\sigma {\cal H}^{\mu}(\tau ,\vec \sigma )=
p^{\mu}_s-l_{\mu}\lbrace {1\over 2}\int d^3\sigma [{\vec \pi}^2(\tau ,\vec
\sigma )+{\vec B}^2(\tau ,\vec \sigma )]+\nonumber \\
&+&\sum_{i=1}^N\eta_i\sqrt{ m^2_i+
{[{\vec \kappa}^2_i(\tau )-e_i\theta^{*}_i(\tau )\theta_i(\tau ){\vec A}
(\tau ,{\vec \eta}_i(\tau ))]}^2 }\rbrace -\nonumber \\
&-&b_{{\check r}\mu}(\tau )\lbrace \int d^3\sigma {[\vec B(\tau ,\vec \sigma )
\times \vec \pi (\tau ,\vec \sigma )]}_{\check r}+\sum_{i=1}^N
[\kappa_{i{\check r}}(\tau )-e_i\theta_i^{*}(\tau )\theta_i
(\tau )A_{\check r}(\tau ,{\vec \eta}_i(\tau ))] \rbrace
\approx 0\nonumber \\
{\tilde {\cal H}}^{\mu\nu}(\tau )&=& b^{\mu}_{\check r}(\tau )\int d^3\sigma
\sigma^{\check r}\, {\cal H}^{\nu}(\tau ,\vec \sigma )-b^{\nu}_{\check r}(\tau
)
\int d^3\sigma \sigma^{\check r}\, {\cal H}^{\mu}(\tau ,\vec \sigma )=
\nonumber \\
&=&S_s^{\mu\nu}(\tau )-[b^{\mu}_{\check r}(\tau )b^{\nu}_{\tau}-b^{\nu}_{\check
r}(\tau )b^{\mu}_{\tau}]\, [{1\over 2}\int d^3\sigma \sigma^{\check r}\,
[{\vec \pi}^2(\tau ,\vec \sigma )+{\vec B}^2(\tau ,\vec \sigma )]+\nonumber \\
&+&\sum_{i=1}^N\eta_i^{\check r}(\tau )\eta_i
\sqrt{m^2_i+[{\vec \kappa}_i(\tau )-e_i\theta^{*}_i(\tau )\theta_i(\tau )
\vec A(\tau ,{\vec \eta}_i(\tau ))]{}^2 }]-\nonumber \\
&-&[b^{\mu}_{\check r}(\tau )b^{\nu}_{\check s}(\tau )-b^{\nu}_{\check r}(\tau
)
b^{\mu}_{\check s}(\tau )]\, [\int d^3\sigma \sigma^{\check r}\,
{[\vec B(\tau ,\vec \sigma )\times \vec \pi (\tau ,\vec \sigma )]}_{\check s}+
\nonumber \\
&+&\sum_{i=1}^N\eta_i^{\check r}(\tau )[\kappa_i^{\check s}(\tau )-e_i\theta
^{*}_i(\tau )\theta_i(\tau )A^{\check s}(\tau ,{\vec \eta}_i(\tau ))]\,\, ]
\approx 0\nonumber \\
\pi^{\tau}(\tau ,\vec \sigma )&\approx& 0\nonumber \\
\Gamma (\tau ,\vec \sigma )&\approx& 0,
\label{128}
\end{eqnarray}

\noindent with Poisson algebra

\begin{eqnarray}
&&\lbrace {\tilde {\cal H}}^{\mu}(\tau ),{\tilde {\cal H}}^{\nu}
(\tau )\rbrace {}^{*}=\int d^3\sigma \lbrace [b^{\mu}_{\tau}b^{\nu}_{\check r}
(\tau )-b^{\nu}_{\tau}b^{\mu}_{\check r}(\tau )]  \pi_{\check r}(\tau ,\vec
\sigma )+\nonumber \\
&&+b^{\mu}_{\check r}(\tau )F_{{\check r}{\check s}}(\tau ,\vec \sigma )
b^{\nu}_{\check s}(\tau )\rbrace \Gamma (\tau ,\vec \sigma )\nonumber \\
&&\lbrace {\tilde {\cal H}}^{\mu}(\tau ),{\tilde {\cal H}}^{\alpha
\beta}(\tau )\rbrace {}^{*}=\int d^3\sigma \, \sigma^{\check t}\, \lbrace
[b^{\alpha}_{\tau}b^{\beta}_{\check t}(\tau
)-b^{\beta}_{\tau}b^{\alpha}_{\check
t}(\tau )]b^{\mu}_{\check r}(\tau )\pi_{\check r}(\tau ,\vec \sigma )+
\nonumber \\
&&+b^{\mu}_{\check r}(\tau )F_{{\check r}{\check s}}(\tau ,\vec \sigma )
[b^{\alpha}_{\check t}(\tau )b^{\beta}_{\check s}(\tau )-b^{\beta}_{\check t}
(\tau )b^{\alpha}_{\check s}(\tau )]\rbrace
\Gamma (\tau ,\vec \sigma )\nonumber \\
&&\lbrace {\tilde {\cal H}}^{\mu\nu}(\tau ),{\tilde {\cal H}}
^{\alpha\beta}(\tau )\rbrace {}^{*}=C^{\mu\nu\alpha\beta}
_{\gamma\delta}{\tilde {\cal H}}^{\gamma\delta}(\tau )+\int d^3\sigma \,
\sigma^{\check u}\, \sigma^{\check v}\nonumber \\
&&\lbrace b^{\mu}_{\check u}(\tau )b^{\alpha}_{\check v}(\tau )([b^{\nu}
_{\tau}b^{\beta}_{\check r}(\tau )-b^{\beta}_{\tau}b^{\nu}_{\check r}(\tau )]
\pi_{\check r}(\tau ,\vec \sigma )+b^{\nu}_{\check r}(\tau )F_{{\check
r}{\check
s}}(\tau ,\vec \sigma )b^{\beta}_{\check s}(\tau ))-\nonumber \\
&&-b^{\mu}_{\check u}(\tau )b^{\beta}_{\check v}(\tau )([b^{\nu}
_{\tau}b^{\alpha}_{\check r}(\tau )-b^{\alpha}_{\tau}b^{\nu}_{\check r}(\tau )]
\pi_{\check r}(\tau ,\vec \sigma )+b^{\nu}_{\check r}(\tau )F_{{\check
r}{\check
s}}(\tau ,\vec \sigma )b^{\alpha}_{\check s}(\tau ))-\nonumber \\
&&-b^{\nu}_{\check u}(\tau )b^{\alpha}_{\check v}(\tau )([b^{\mu}
_{\tau}b^{\beta}_{\check r}(\tau )-b^{\beta}_{\tau}b^{\mu}_{\check r}(\tau )]
\pi_{\check r}(\tau ,\vec \sigma )+b^{\mu}_{\check r}(\tau )F_{{\check
r}{\check
s}}(\tau ,\vec \sigma )b^{\beta}_{\check s}(\tau ))+\nonumber \\
&&+b^{\nu}_{\check u}(\tau )b^{\beta}_{\check v}(\tau )([b^{\mu}
_{\tau}b^{\alpha}_{\check r}(\tau )-b^{\alpha}_{\tau}b^{\mu}_{\check r}(\tau )]
\pi_{\check r}(\tau ,\vec \sigma )+b^{\mu}_{\check r}(\tau )F_{{\check
r}{\check
s}}(\tau ,\vec \sigma )b^{\alpha}_{\check s}(\tau ))\rbrace
\Gamma (\tau ,\vec \sigma ),
\label{129}
\end{eqnarray}

\noindent and the form of Eq.(\ref{56}) for the Poincar\'e generators.

Then we make the canonical transformation of Eqs.(\ref{59}), so that the
new variables are ${\tilde x}^{\mu}_s, p_s^{\mu}, b^A_{\check A}, {\tilde S}_s
^{\mu\nu}, A_{\check A}, \pi^{\check A}, {\vec \eta}_i, {\vec \kappa}_i,
\theta_i, \theta_i^{*}$ and one has

\begin{eqnarray}
{\bar S}_s^{AB}&=&\epsilon^A_{\mu}(u(p_s))\epsilon^B_{\nu}(u(p_s))S_s^{\mu\nu}
\approx [b^A_{\check r}(\tau )b^B_{\tau}-b^B_{\check r}(\tau )b^A_{\tau}]\,
[{1\over 2}\int d^3\sigma \sigma^{\check r}\,
[{\vec \pi}^2(\tau ,\vec \sigma )+{\vec B}^2(\tau ,\vec \sigma )]+\nonumber \\
&+&\sum_{i=1}^N\eta_i^{\check r}(\tau )\eta_i
\sqrt{m^2_i+[{\vec \kappa}_i(\tau )-e_i\theta^{*}_i(\tau )\theta_i(\tau )
\vec A(\tau ,{\vec \eta}_i(\tau ))]{}^2 }]+\nonumber \\
&+&[b^A_{\check r}(\tau )b^B_{\check s}(\tau )-b^B_{\check r}(\tau )b^A
_{\check s}(\tau )]\, [\int d^3\sigma \sigma^{\check r}\,
{[\vec B(\tau ,\vec \sigma )\times \vec \pi (\tau ,\vec \sigma )]}_{\check s}+
\nonumber \\
&+&\sum_{i=1}^N\eta_i^{\check r}(\tau )[\kappa_i^{\check s}(\tau )-e_i\theta
^{*}_i(\tau )\theta_i(\tau )A^{\check s}(\tau ,{\vec \eta}_i(\tau ))]\,\, ].
\label{130}
\end{eqnarray}

The final gauge-fixings (\ref{62}) reduce the variables to ${\tilde x}^{\mu}_s,
p^{\mu}_s, A_{\tau}, \pi^{\tau}, A_r, \pi^r, \eta^r_i, \kappa_{ir}, \theta_i,
\theta^{*}_i,$ with ``r" being a Wigner spin-1 index and $\bar o=\tau$ a
Lorentz-scalar one.

We have

\begin{eqnarray}
{\bar S}_s^{AB}&\approx& (\eta^A_{\check r}\eta^B_{\tau}-\eta^B_{\check
r}\eta^A
_{\tau})\, [{1\over 2}\int d^3\sigma \sigma^{\check r}\,
[{\vec \pi}^2 (\tau ,\vec \sigma )+{\vec B}^2(\tau ,\vec \sigma )]+\nonumber \\
&+&\sum_{i=1}^N\eta_i^{\check r}(\tau )\eta_i
\sqrt{m^2_i+[{\vec \kappa}_i(\tau )-e_i\theta^{*}_i(\tau )\theta_i(\tau )
\vec A(\tau ,{\vec \eta}_i(\tau ))]{}^2 }]+\nonumber \\
&+&(\eta^A_{\check r}\eta^B_{\check s}-\eta^B_{\check r}\eta^A_{\check s})\,
[\int d^3\sigma \sigma^{\check r}\,
{[\vec B(\tau ,\vec \sigma )\times \vec \pi (\tau ,\vec \sigma )]}_{\check s}+
\nonumber \\
&+&\sum_{i=1}^N\eta_i^{\check r}(\tau )[\kappa_i^{\check s}(\tau )-e_i\theta
^{*}_i(\tau )\theta_i(\tau )A^{\check s}(\tau ,{\vec \eta}_i(\tau ))]\,\, ]
\nonumber \\
{\bar S}_s^{rs}&\approx&
\sum_{i=1}^N(\eta_i^{\check r}(\tau )[\kappa_i^{\check s}(\tau )-e_i\theta
^{*}_i(\tau )\theta_i(\tau )A^{\check s}(\tau ,{\vec \eta}_i(\tau ))]-
\nonumber \\
&-&\sum_{i=1}^N\eta_i^{\check s}(\tau )[\kappa_i^{\check r}(\tau )-e_i\theta
^{*}_i(\tau )\theta_i(\tau )A^{\check r}(\tau ,{\vec \eta}_i(\tau ))]\, )-
\nonumber \\
&-&\int d^3\sigma \, (\sigma^r\,
{[\vec B(\tau ,\vec \sigma )\times \vec \pi (\tau ,\vec \sigma )]}^s-
\sigma^s\,
{[\vec B(\tau ,\vec \sigma )\times \vec \pi (\tau ,\vec \sigma )]}^r)
\nonumber \\
{\bar S}_s^{\bar or}&\approx& -{\bar S}_s^{r\bar o}=-
\sum_{i=1}^N\eta_i^r(\tau )\eta_i
\sqrt{m^2_i+[{\vec \kappa}_i(\tau )-e_i\theta^{*}_i(\tau )\theta_i(\tau )
\vec A(\tau ,{\vec \eta}_i(\tau ))]{}^2 }-\nonumber \\
&-&{1\over 2}\int d^3\sigma \sigma^r\,
[{\vec \pi}^2(\tau ,\vec \sigma )+{\vec B}^2(\tau ,\vec \sigma )]
\nonumber \\
&&{}\nonumber \\
J_s^{ij}&\approx& {\tilde x}^i_sp_s^j-{\tilde x}_s^jp^i_s+\delta^{ir}\delta
^{js}{\bar S}_s^{rs}\nonumber \\
J_s^{oi}&\approx&{\tilde x}_s^op^i_s-{\tilde x}_s^ip^o_s-{ {\delta^{ir}{\bar
S}_s^{rs}p^s_s}\over {p^o_s+\eta_s\sqrt{p^2_s}} }.
\label{131}
\end{eqnarray}

The Poincar\'e generators now have the form of Eqs.(\ref{71}) and only six
first class constraints are left

\begin{eqnarray}
{\tilde {\cal H}}^{\mu}(\tau )&=&p_s^{\mu}-u^{\mu}(u(p_s))\,
[{1\over 2}\int d^3\sigma
[{\vec \pi}^2(\tau ,\vec \sigma )+{\vec B}^2(\tau ,\vec \sigma )]+\nonumber \\
&+&\sum_{i=1}^N\eta_i
\sqrt{m^2_i+[{\vec \kappa}_i(\tau )-e_i\theta^{*}_i(\tau )\theta_i(\tau )
\vec A(\tau ,{\vec \eta}_i(\tau ))]{}^2 }]-\nonumber \\
&-&\epsilon^{\mu}_r(u(p_s))\, [-\int d^3\sigma
{[\vec B(\tau ,\vec \sigma )\times \vec \pi (\tau ,\vec \sigma )]}^r+
\nonumber \\
&+&\sum_{i=1}^N[\kappa_i^r(\tau )-e_i\theta
^{*}_i(\tau )\theta_i(\tau )A^r(\tau ,{\vec \eta}_i(\tau ))]\,\, ]
\approx 0\nonumber \\
\pi^{\tau}(\tau ,\vec \sigma )&\approx& 0\nonumber \\
\Gamma (\tau ,\vec \sigma )&\approx& 0,\nonumber \\
&&{}\nonumber \\
&&\lbrace {\tilde {\cal H}}^{\mu},{\tilde {\cal H}}^{\nu}\rbrace {}^{**}=
\int d^3\sigma \lbrace [\epsilon^{\mu}_{\tau}(u(p_s))\epsilon^{\nu}_r
(u(p_s))-\epsilon^{\nu}_{\tau}(u(p_s))\epsilon^{\mu}_r(u(p_s))] \pi_r(\tau
,\vec
\sigma )+\nonumber \\
&&+\epsilon^{\mu}_r(u(p_s))F_{rs}(\tau ,\vec \sigma )
\epsilon^{\nu}_s(u(p_s))\rbrace \Gamma (\tau ,\vec \sigma )
\label{132}
\end{eqnarray}

\noindent or

\begin{eqnarray}
{\cal H}(\tau )&=&\eta_s\sqrt{p^2_s}-
[\sum_{i=1}^N\eta_i
\sqrt{m^2_i+[{\vec \kappa}_i(\tau )-e_i\theta^{*}_i(\tau )\theta_i(\tau )
\vec A(\tau ,{\vec \eta}_i(\tau ))]{}^2 }+\nonumber \\
&+&{1\over 2}\int d^3\sigma
[{\vec \pi}^2(\tau ,\vec \sigma )+{\vec B}^2(\tau ,\vec \sigma ) ]
\approx 0\nonumber \\
{\vec {\cal H}}_p(\tau )&=&
\sum_{i=1}^N[{\vec \kappa}_i(\tau )-e_i\theta
^{*}_i(\tau )\theta_i(\tau )\vec A(\tau ,{\vec \eta}_i(\tau ))]-
\nonumber \\
&-&\int d^3\sigma
\vec B(\tau ,\vec \sigma )\times \vec \pi (\tau ,\vec \sigma )
\approx 0\nonumber \\
\pi^{\tau}(\tau ,\vec \sigma )&\approx& 0\nonumber \\
\Gamma (\tau ,\vec \sigma )&\approx& 0,
\label{133}
\end{eqnarray}

Then, for $N \geq 2$, we do the canonical transformation (\ref{73}),
(\ref{74}): the new form of the constraints is

\begin{eqnarray}
&&{\cal H}(\tau )=\epsilon_s-\lbrace \sum_{i=1}^N\eta_i\times \nonumber \\
&& \sqrt{m^2_i+[{1\over N}{\vec \kappa}_{+}(\tau )+\sqrt{N}
\sum_{a=1}^{N-1}{\hat \gamma}_{ai}{\vec \pi}_a(\tau )-e_i\theta^{*}_i(\tau )
\theta_i(\tau )\vec A(\tau ,{\vec \eta}_{+}(\tau )+{1\over {\sqrt{N}}}
\sum_{a=1}^{N-1}{\hat \gamma}_{ai}{\vec \rho}_a(\tau ))]{}^2 }+\nonumber \\
&+&{1\over 2}\int d^3\sigma
[{\vec \pi}^2(\tau ,\vec \sigma )+{\vec B}^2(\tau ,\vec \sigma ) ]\rbrace
=\epsilon_s-E_{(P+I)s}-E_{(F)s}\approx 0\nonumber \\
\nonumber \\
&&{\vec {\cal H}}_p(\tau )={\vec \kappa}_{+}(\tau )-\sum_{i=1}^N
e_i\theta^{*}_i(\tau )
\theta_i(\tau )\vec A(\tau ,{\vec \eta}_{+}(\tau )+{1\over {\sqrt{N}}}
\sum_{a=1}^{N-1}{\hat \gamma}_{ai}{\vec \rho}_a(\tau ))-
\nonumber \\
&-& \int d^3\sigma \vec B(\tau ,\vec \sigma )\times \vec \pi (\tau ,\vec
\sigma )={\vec P}_{(P+I)s}+{\vec P}_{(F)s}\approx 0\nonumber \\
&&\pi^{\tau}(\tau ,\vec \sigma )\approx 0\nonumber \\
&&\Gamma (\tau ,\vec \sigma )\approx 0,
\label{134}
\end{eqnarray}

\noindent where $E_{(F)s}={1\over 2}\int d^3\sigma [{\vec \pi}^2(\tau ,\vec
\sigma )+{\vec B}^2(\tau ,\vec \sigma )]$ and ${\vec P}_{(F)s}=-
\int d^3\sigma \vec \pi (\tau ,\vec \sigma )\times \vec B(\tau ,\vec
\sigma )$ are the rest-frame field energy and three-momentum respectively
[now we have $\vec \pi (\tau ,\vec \sigma )=\vec E(\tau ,\vec \sigma )$],
while $E_{(P+I)s}$ and ${\vec P}_{(P+I)s}$ denote the particle+interaction
total rest-frame energy and three-momentum, before the decoupling from the
electromagnetic gauge degrees of freedom.

The final form of the rest-frame spin tensor is

\begin{eqnarray}
{\bar S}_s^{rs}&=&-\sum_{i=1}^Ne_i\theta^{*}_i(\tau )\theta_i(\tau )
[\eta^r_{+}(\tau )A^s(\tau ,{\vec \eta}_{+}(\tau )+{1\over {\sqrt{N}}}
\sum_{b=1}^{N-1}{\hat \gamma}_{bi}{\vec \rho}_b(\tau ))-\nonumber \\
&-&\eta^s_{+}(\tau )A^r(\tau ,{\vec \eta}_{+}(\tau )+{1\over {\sqrt{N}}}
\sum_{b=1}^{N-1}{\hat \gamma}_{bi}{\vec \rho}_b(\tau ))]+\nonumber \\
&+&\sum_{a=1}^{N-1}[\rho^r_a(\tau )(\pi^s_a(\tau )-{1\over {\sqrt{N}}}
\sum_{i=1}^N{\hat \gamma}_{ai}e_i\theta^{*}_i(\tau )\theta_i(\tau )
A^s(\tau ,{\vec \eta}_{+}(\tau )+{1\over {\sqrt{N}}}
\sum_{b=1}^{N-1}{\hat \gamma}_{bi}{\vec \rho}_b(\tau ))-\nonumber \\
&-&\rho^s_a(\tau )(\pi^r_a(\tau )-{1\over {\sqrt{N}}}
\sum_{i=1}^N{\hat \gamma}_{ai}e_i\theta^{*}_i(\tau )\theta_i(\tau )
A^r(\tau ,{\vec \eta}_{+}(\tau )+{1\over {\sqrt{N}}}
\sum_{b=1}^{N-1}{\hat \gamma}_{bi}{\vec \rho}_b(\tau ))]-\nonumber \\
&-&\int d^3\sigma \, (\sigma^r\, {[\vec B(\tau ,\vec \sigma )
\times \vec \pi (\tau ,\vec \sigma )]}^s-\sigma^s\,
{[\vec B(\tau ,\vec \sigma )\times \vec \pi (\tau ,\vec \sigma )]}^r)=
{\bar S}_{(P+I)s}^{rs}+{\bar S}_{(F)s}^{rs}\nonumber \\
&&{}\nonumber \\
{\bar S}_s^{\bar or}&=&-{\bar S}_s^{r\bar o}=-\sum_{i=1}^N(\eta^r_{+}(\tau )
+{1\over {\sqrt{N}}}\sum_{a=1}^{N-1}{\hat \gamma}_{ai}\rho^r_a(\tau ))
\eta_i \times \nonumber \\
&&\sqrt{m^2_i+[{1\over N}{\vec \kappa}_{+}(\tau )+\sqrt{N}
\sum_{a=1}^{N-1}{\hat \gamma}_{ai}{\vec \pi}_a(\tau )-e_i\theta^{*}_i(\tau )
\theta_i(\tau )\vec A(\tau ,{\vec \eta}_{+}(\tau )+{1\over {\sqrt{N}}}
\sum_{a=1}^{N-1}{\hat \gamma}_{ai}{\vec \rho}_a(\tau ))]{}^2 }\nonumber \\
&-&{1\over 2}\int d^3\sigma \sigma^r\,
[{\vec \pi}^2(\tau ,\vec \sigma )+{\vec B}^2(\tau ,\vec \sigma )],
\label{135}
\end{eqnarray}

\noindent while the Dirac Hamiltonian is

\begin{equation}
H_D=\lambda (\tau ){\cal H}-\vec \lambda (\tau ){\vec {\cal H}}_p+\int d^3
\sigma [\lambda_{\tau}(\tau ,\vec \sigma )\pi^{\tau}(\tau ,\vec \sigma )-
A_{\tau}(\tau ,\vec \sigma )\Gamma (\tau ,\vec \sigma )].
\label{136}
\end{equation}

\noindent We can check that $\theta^{*}_i(\tau )\theta_i(\tau )$ is a
constant of motion for each i=1,..,N, so that we will write it as
$\theta^{*}_i\theta_i$.

Let us now look at the electromagnetic Dirac observables, namely at the
functions on phase space invariant under electromagnetic gauge transformations.
Referring to Ref.\cite{lusc} for the detailed calculations, for the
electromagnetic field we have the following decompositions

\begin{eqnarray}
A^r(\tau ,\vec \sigma )&=&{ {\partial}\over {\partial \sigma^r} }\eta_{em}
(\tau ,\vec \sigma )+A^r_{\perp}(\tau ,\vec \sigma )\nonumber \\
\pi^r(\tau ,\vec \sigma )&=&\pi^r_{\perp}(\tau ,\vec
\sigma )+{1\over {\triangle_{\sigma}} }{ {\partial}\over {\partial \sigma^r}}
[\Gamma (\tau ,\vec \sigma )-\sum_{i=1}^Ne_i\theta^{*}_i\theta_i\delta^3
(\vec \sigma -{\vec \eta}_i(\tau ))]\nonumber \\
&&{}\nonumber \\
\eta_{em}(\tau ,\vec \sigma )&=&-{1\over {\triangle_{\sigma}} }{ {\partial}
\over {\partial \vec \sigma} }\cdot \vec A(\tau ,\vec \sigma )\nonumber \\
&&{}\nonumber \\
&&\lbrace \eta_{em}(\tau ,\vec \sigma ),\Gamma (\tau ,{\vec \sigma}^{'} )
\rbrace {}^{**}=-\delta^3(\vec \sigma -{\vec \sigma}^{'})\nonumber \\
&&\lbrace A^r_{\perp}(\tau ,\vec \sigma ),\pi^s_{\perp}(\tau ,{\vec \sigma}
^{'})\rbrace {}^{**}=-(\delta^{rs}+{{\partial^r_{\sigma}\partial^s_{\sigma}}
\over {\triangle_{\sigma} }})\delta^3(\vec \sigma -{\vec \sigma}^{'}),
\label{137}
\end{eqnarray}

\noindent with the pairs of conjugate variables $A_{\tau}(\tau ,\vec \sigma ),
\pi^{\tau}(\tau ,\vec \sigma )\approx 0, \eta_{em}(\tau ,\vec \sigma ),
\Gamma (\tau ,\vec \sigma )\approx 0$ spanning the Lorentz-scalar gauge
subspace of phase space, and  with ${\vec A}_{\perp}(\tau ,\vec \sigma ),
{\vec \pi}_{\perp}(\tau ,\vec \sigma )$ [${\vec \partial}_{\sigma}\cdot
{\vec A}_{\perp}(\tau ,\vec \sigma )={\vec \partial}_{\sigma}\cdot {\vec
\pi}_{\perp}(\tau ,\vec \sigma )=0$] being a canonical basis of
electromagnetic Dirac observables, transforming as Wigner spin-1 3-vectors.

Since we have

\begin{eqnarray}
&&\lbrace \eta^r_i(\tau ), \Gamma (\tau ,\vec \sigma )\rbrace {}^{**}=0
\nonumber \\
&&\lbrace \kappa^r_i(\tau ), \Gamma (\tau ,\vec \sigma )\rbrace {}^{**}=
-e_i\theta^{*}_i\theta_i{{\partial}\over {\partial \eta^r_i}}\delta^3
(\vec \sigma -{\vec \eta}_i(\tau )),
\label{138}
\end{eqnarray}

\noindent the particle momenta ${\vec \kappa}_i(\tau )$ are not gauge
invariant.
Also the Grassmann variables $\theta_i(\tau ), \theta^{*}_i(\tau )$ are not
gauge invariant

\begin{eqnarray}
&&\lbrace \theta_i(\tau ),\Gamma (\tau ,\vec \sigma )\rbrace {}^{**}=-i
e_i\theta_i(\tau )\delta^3(\vec \sigma -{\vec \eta}_i(\tau ))\nonumber \\
&&\lbrace \theta^{*}_i(\tau ),\Gamma (\tau ,\vec \sigma )\rbrace {}^{**}=i
e_i\theta^{*}_i(\tau )\delta^3(\vec \sigma -{\vec \eta}_i(\tau )).
\label{139}
\end{eqnarray}

The Grassmann Dirac observables are

\begin{eqnarray}
&&{\hat \theta}_i(\tau )=e^{-i\eta_{em}(\tau ,{\vec \eta}_i(\tau ))}\theta_i
(\tau ),\quad \quad \lbrace {\hat \theta}_i(\tau ),\Gamma (\tau ,\vec \sigma )
\rbrace {}^{**}=0,\nonumber \\
&&{\hat \theta}_i^{*}(\tau )=e^{i\eta_{em}(\tau ,{\vec \eta}_i(\tau ))}\theta_i
^{*}(\tau ),\quad \quad \lbrace {\hat \theta}^{*}_i(\tau ),\Gamma (\tau ,\vec
\sigma )\rbrace {}^{**}=0,\nonumber \\
&&\Rightarrow {\hat \theta}^{*}_i{\hat
\theta}_i=\theta^{*}_i\theta_i,\quad\quad
\lbrace \theta^{*}_i\theta_i,\Gamma (\tau ,\vec \sigma )\rbrace {}^{**}=0.
\label{140}
\end{eqnarray}

Since we have

\begin{equation}
{\vec \kappa}_i(\tau )-e_i\theta^{*}_i\theta_i\vec A(\tau ,{\vec \eta}_i(\tau
))={\hat {\vec \kappa}}_i(\tau )-e_i{\hat \theta}^{*}_i{\hat \theta}_i{\vec
A}_{\perp}(\tau ,{\vec \eta}_i(\tau )),
\label{141}
\end{equation}

\noindent with

\begin{eqnarray}
&&{\hat {\vec \kappa}}_i(\tau )={\vec \kappa}_i(\tau )-e_i\theta^{*}_i\theta_i
\vec \partial \eta_{em}(\tau ,{\vec \eta}_i(\tau )),\nonumber \\
&&{}\nonumber \\
&&\lbrace {\hat \kappa}^r_i(\tau ),\Gamma (\tau ,\vec \sigma )\rbrace {}^{**}
=0\nonumber \\
&&\lbrace \eta^r_i(\tau ),{\hat \kappa}^s_j(\tau )\rbrace {}^{**}=\delta
_{ij}\delta^{rs},\nonumber \\
&&\lbrace {\hat \kappa}^r_i(\tau ),{\hat \theta}_j(\tau )\rbrace {}^{**}=0,
\quad\quad \lbrace {\hat \kappa}^r_i(\tau ),{\hat \theta}^{*}_j(\tau )\rbrace
{}^{**}=0,
\label{142}
\end{eqnarray}

\noindent the particle Dirac observables are ${\vec \eta}_i(\tau ), {\hat
{\vec \kappa}}_i(\tau )$: the scalar particles have been dressed with the
Coulomb cloud (like the fermion fields in Ref.\cite{lusc}).

{}From Equations (\ref{137}), we have

\begin{eqnarray}
{\vec \pi}^2(\tau ,\vec \sigma )&=&\lbrace {\vec \pi}_{\perp}(\tau ,\vec
\sigma )+{1\over {\triangle_{\sigma}} }{{\partial}\over {\partial \vec \sigma}}
[\Gamma (\tau ,\vec \sigma )-\sum_{i=1}^Ne_i{\hat \theta}^{*}_i{\hat \theta}_i
\delta^3(\vec \sigma -{\vec \eta}_i(\tau ))]\rbrace {}^2\approx \nonumber \\
&\approx& {\vec \pi}^2_{\perp}(\tau ,\vec \sigma )-2{\vec \pi}_{\perp}
(\tau ,\vec \sigma )\cdot {1\over {\triangle_{\sigma}}}{{\partial}\over
{\partial \vec \sigma}}\sum_{i=1}^Ne_i{\hat \theta}^{*}_i{\hat \theta}_i
\delta^3(\vec \sigma -{\vec \eta}_i(\tau ))+\nonumber \\
&+&\sum_{i,j=1}^N[e_i{\hat \theta}^{*}_i{\hat \theta}_i][e_j{\hat \theta}^{*}_j
{\hat \theta}_j]{1\over {\triangle_{\sigma}}}{{\partial}\over {\partial \vec
\sigma}}\delta^3(\vec \sigma -{\vec \eta}_i(\tau ))\, \cdot \, {1\over
{\triangle_{\sigma}}}{{\partial}\over {\partial \vec \sigma}}\delta^3(\vec
\sigma -{\vec \eta}_j(\tau )),
\label{143}
\end{eqnarray}

\noindent so that by integrating by parts and using the property $Q^2_i=0$,
$Q_i=e_i{\hat \theta}^{*}_i{\hat \theta}_i$, of Grassmann variables, we
obtain

\begin{eqnarray}
&&\int d^3\sigma {\vec \pi}^2(\tau ,\vec \sigma )=\nonumber \\
&&=\int d^3\sigma \lbrace {\vec \pi}^2_{\perp}(\tau ,\vec \sigma )-\sum
_{i\not= j}^{1..N}Q_iQ_j\delta^3(\vec \sigma -{\vec \eta}_i(\tau )){1\over
{\triangle_{\sigma}}}\delta^3(\vec \sigma -{\vec \eta}_j(\tau ))\rbrace =
\nonumber \\
&&=\int d^3\sigma {\vec \pi}^2_{\perp}(\tau ,\vec \sigma )-\sum_{i\not= j}
^{1..N}Q_iQ_j\, c({\vec \eta}_i(\tau )-{\vec \eta}_j(\tau )),
\label{144}
\end{eqnarray}

\noindent where we have introduced the Green function $c(\vec x)$ defined by

\begin{equation}
c(\vec x-\vec y)={{-1}\over {4\pi |\, \vec x-\vec y\, |}}={1\over {\triangle
_x}}\delta^3(\vec x-\vec y).
\label{145}
\end{equation}

Having decoupled the electromagnetic gauge variables, the canonical basis of
Dirac observables ${\tilde x}^{\mu}_s(\tau ),
p^{\mu}_s, {\vec A}_{\perp}(\tau ,\vec
\sigma ), {\vec \pi}_{\perp}(\tau ,\vec \sigma ), {\vec \eta}_i(\tau ),
{\hat {\vec \kappa}}_i(\tau ), {\hat \theta}_i(\tau ), {\hat \theta}_i^{*}
(\tau )$ spans the phase space, where the remaining gauge invariant four
first class constraints in Eqs.(\ref{134}) have the form

\begin{eqnarray}
{\hat {\cal H}}(\tau )&=&\epsilon_s-\lbrace \sum_{i=1}^N\eta_i\nonumber \\
&&\sqrt{m^2_i+[{1\over N}{\hat {\vec \kappa}}_{+}
(\tau )+\sqrt{N}\sum_{a=1}^{N-1}{\hat \gamma}_{ai}{\hat {\vec \pi}}_a(\tau )
-e_i{\hat \theta}^{*}_i{\hat \theta}_i{\vec A}_{\perp}(\tau ,{\vec \eta}_{+}
(\tau )+{1\over {\sqrt{N}}}\sum_{a=1}^{N-1}{\hat \gamma}_{ai}{\vec \rho}_a
(\tau ))]{}^2 }-\nonumber \\
&-&{1\over 2}\sum_{i\not= j}^{1..N}[e_i{\hat \theta}^{*}_i{\hat \theta}_i]
[e_j{\hat \theta}^{*}_j{\hat \theta}_j]\, c[{1\over {\sqrt{N}}}\sum_{a=1}^{N-1}
({\hat \gamma}_{ai}-{\hat \gamma}_{aj}){\vec \rho}_a(\tau )]
+\nonumber \\
&+&{1\over 2}\int d^3\sigma [{\vec \pi}_{\perp}^2(\tau ,\vec \sigma )+{\vec B}
^2[{\vec A}_{\perp}(\tau ,\vec \sigma )]\, ]\rbrace \approx 0\nonumber \\
&&{}\nonumber \\
{\hat {\vec {\cal H}}}_p(\tau )&=&
{\hat {\vec \kappa}}_{+}(\tau )-\sum_{i=1}^N
e_i{\hat \theta}^{*}_i{\hat \theta}_i{\vec A}_{\perp}(\tau ,{\vec \eta}_{+}
(\tau )+{1\over {\sqrt{N}}}\sum_{a=1}^{N-1}{\hat \gamma}_{ai}{\vec \rho}_a
(\tau ))-\nonumber \\
&-&\int d^3\sigma \, \vec B[{\vec A}_{\perp}(\tau ,\vec \sigma )]\times
{\vec \pi}_{\perp}(\tau ,\vec \sigma )\approx 0\nonumber \\
&&{}\nonumber \\
&&\lbrace {\hat {\cal H}},{\hat {\cal H}}^r_p\rbrace {}^{**}=\lbrace {\hat
{\cal H}}^r_p,{\hat {\cal H}}^s_p\rbrace {}^{**}=0;
\label{146}
\end{eqnarray}

\noindent the Bianchi identity $\vec \partial \cdot \vec B(\tau ,\vec \sigma )
=0$ has been used to obtain ${\hat {\cal H}}^r_p$.
The variables ${\hat {\vec \kappa}}_{+},
{\hat {\vec \pi}}_a$ are the Dirac observables induced by ${\hat {\vec \kappa}}
_i$. Now the Dirac Hamiltonian is $H_D=\lambda (\tau ){\hat {\cal H}}(\tau )-
{\vec \lambda}(\tau )\cdot {\hat {\vec {\cal H}}}_p(\tau )$.

The rest-frame spin tensor becomes

\begin{eqnarray}
{\bar S}_s^{rs}&=&-\sum_{i=1}^Ne_i{\hat \theta}^{*}_i{\hat \theta}_i
[\eta^r_{+}(\tau )A_{\perp}^s(\tau ,{\vec \eta}_{+}(\tau )+{1\over {\sqrt{N}}}
\sum_{b=1}^{N-1}{\hat \gamma}_{bi}{\vec \rho}_b(\tau ))-\nonumber \\
&-&\eta^s_{+}(\tau )A_{\perp}^r(\tau ,{\vec \eta}_{+}(\tau )+{1\over
{\sqrt{N}}}
\sum_{b=1}^{N-1}{\hat \gamma}_{bi}{\vec \rho}_b(\tau ))]+\nonumber \\
&+&\sum_{a=1}^{N-1}[\rho^r_a(\tau )({\hat \pi}^s_a(\tau )-{1\over {\sqrt{N}}}
\sum_{i=1}^N{\hat \gamma}_{ai}e_i{\hat \theta}^{*}_i{\hat \theta}_i
A_{\perp}^s(\tau ,{\vec \eta}_{+}(\tau )+{1\over {\sqrt{N}}}
\sum_{b=1}^{N-1}{\hat \gamma}_{bi}{\vec \rho}_b(\tau ))-\nonumber \\
&-&\rho^s_a(\tau )({\hat \pi}^r_a(\tau )-{1\over {\sqrt{N}}}
\sum_{i=1}^N{\hat \gamma}_{ai}e_i{\hat \theta}^{*}_i{\hat \theta}_i
A_{\perp}^r(\tau ,{\vec \eta}_{+}(\tau )+{1\over {\sqrt{N}}}
\sum_{b=1}^{N-1}{\hat \gamma}_{bi}{\vec \rho}_b(\tau ))]-\nonumber \\
&-&\sum_{i=1}^Ne_i{\hat \theta}^{*}_i{\hat \theta}_i\int d^3\sigma
(\sigma^r\epsilon^{suv}-\sigma^s\epsilon^{ruv})[{1\over {\triangle_{\sigma}}}
{{\partial}\over {\partial \sigma^u}}\delta^3(\vec \sigma -{\vec \eta}_i(\tau
))]\, B^v[{\vec A}_{\perp}(\tau ,\vec \sigma )]-\nonumber \\
&-&\int d^3\sigma \, (\sigma^r\, {[\vec B[{\vec A}_{\perp}(\tau ,\vec \sigma )]
\times {\vec \pi}_{\perp}(\tau ,\vec \sigma )\, ]}^s-\sigma^s\,
{[\vec B[{\vec A}_{\perp}(\tau ,\vec \sigma )]\times
{\vec \pi}_{\perp}(\tau ,\vec \sigma )\, ]}^r)=\nonumber \\
&=&\sum_{a=1}^{N-1}[\rho^r_a(\tau ){\hat \pi}_a^s(\tau )-\rho^s_a(\tau ){\hat
\pi}_a^r(\tau )]-\nonumber \\
&-&\int d^3\sigma \, (\sigma^r\, {[\vec B[{\vec A}_{\perp}(\tau ,\vec \sigma )]
\times {\vec \pi}_{\perp}(\tau ,\vec \sigma )\, ]}^s-\sigma^s\,
{[\vec B[{\vec A}_{\perp}(\tau ,\vec \sigma )]\times
{\vec \pi}_{\perp}(\tau ,\vec \sigma )\, ]}^r)-\nonumber \\
&-&\sum_{i=1}^Ne_i{\hat \theta}^{*}_i{\hat \theta}_i
[\eta^r_{+}(\tau )A_{\perp}^s(\tau ,{\vec \eta}_{+}(\tau )+{1\over {\sqrt{N}}}
\sum_{b=1}^{N-1}{\hat \gamma}_{bi}{\vec \rho}_b(\tau ))-\nonumber \\
&-&\eta^s_{+}(\tau )A_{\perp}^r(\tau ,{\vec \eta}_{+}(\tau )+{1\over
{\sqrt{N}}}
\sum_{b=1}^{N-1}{\hat \gamma}_{bi}{\vec \rho}_b(\tau ))]+\nonumber \\
&-&{1\over {\sqrt{N}}}\sum_{a=1}^{N-1}[\rho^r_a(\tau )
\sum_{i=1}^N{\hat \gamma}_{ai}e_i{\hat \theta}^{*}_i{\hat \theta}_i
A_{\perp}^s(\tau ,{\vec \eta}_{+}(\tau )+{1\over {\sqrt{N}}}
\sum_{b=1}^{N-1}{\hat \gamma}_{bi}{\vec \rho}_b(\tau ))-\nonumber \\
&-&\rho^s_a(\tau )
\sum_{i=1}^N{\hat \gamma}_{ai}e_i{\hat \theta}^{*}_i{\hat \theta}_i
A_{\perp}^r(\tau ,{\vec \eta}_{+}(\tau )+{1\over {\sqrt{N}}}
\sum_{b=1}^{N-1}{\hat \gamma}_{bi}{\vec \rho}_b(\tau ))]-\nonumber \\
&-&\epsilon^{rsk} \sum_{i=1}^N e_i{\hat \theta}^{*}_i{\hat \theta}_i\int
d^3\sigma c(\vec \sigma -{\vec \eta}_i(\tau )) \lbrace 2B^k[{\vec A}_{\perp}
(\tau ,\vec \sigma )]-\nonumber \\
&-&\vec \sigma \cdot {{\partial}\over {\partial \vec \sigma}}
B^k[{\vec A}_{\perp}(\tau ,\vec \sigma )]+\vec \sigma \cdot {{\partial}\over
{\partial \sigma^k}}\vec B[{\vec A}_{\perp}(\tau ,\vec \sigma )]\rbrace =
\nonumber \\
&=&{\bar S}^{rs}_{s,FREE}+{\bar S}^{rs}_{s,INT}\nonumber \\
&&{}\nonumber \\
{\bar S}_s^{\bar or}&=&-{\bar S}_s^{r\bar o}=-\sum_{i=1}^N(\eta^r_{+}(\tau )
+{1\over {\sqrt{N}}}\sum_{a=1}^{N-1}{\hat \gamma}_{ai}\rho^r_a(\tau ))
\eta_i \times \nonumber \\
&&\sqrt{m^2_i+[{1\over N}{\hat {\vec \kappa}}_{+}(\tau )+\sqrt{N}
\sum_{a=1}^{N-1}{\hat \gamma}_{ai}{\hat {\vec \pi}}_a(\tau )-e_i{\hat \theta}
^{*}_i{\hat \theta}_i{\vec A}_{\perp}(\tau ,{\vec \eta}_{+}(\tau )+{1\over
{\sqrt{N}}}\sum_{a=1}^{N-1}{\hat \gamma}_{ai}{\vec \rho}_a(\tau ))]{}^2 }-
\nonumber \\
&-&{1\over 2}\int d^3\sigma \sigma^r\, [-2{\vec \pi}_{\perp}
(\tau ,\vec \sigma )\cdot {1\over {\triangle_{\sigma}}}{{\partial}\over
{\partial \vec \sigma}}\sum_{i=1}^Ne_i{\hat \theta}^{*}_i{\hat \theta}_i
\delta^3(\vec \sigma -{\vec \eta}_i(\tau ))+\nonumber \\
&+&\sum_{i,j=1}^N[e_i{\hat \theta}^{*}_i{\hat \theta}_i][e_j{\hat \theta}^{*}_j
{\hat \theta}_j]{1\over {\triangle_{\sigma}}}{{\partial}\over {\partial \vec
\sigma}}\delta^3(\vec \sigma -{\vec \eta}_i(\tau ))\, \cdot \, {1\over
{\triangle_{\sigma}}}{{\partial}\over {\partial \vec \sigma}}\delta^3(\vec
\sigma -{\vec \eta}_j(\tau ))]-\nonumber \\
&-&{1\over 2}\int d^3\sigma \sigma^r\,
[{\vec \pi}_{\perp}^2(\tau ,\vec \sigma )+{\vec B}^2[{\vec A}_{\perp}(\tau ,
\vec \sigma )]=\nonumber \\
&=&-\sum_{i=1}^N(\eta^r_{+}(\tau )
+{1\over {\sqrt{N}}}\sum_{a=1}^{N-1}{\hat \gamma}_{ai}\rho^r_a(\tau ))
\eta_i \times \nonumber \\
&&\sqrt{m^2_i+[{1\over N}{\hat {\vec \kappa}}_{+}(\tau )+\sqrt{N}
\sum_{a=1}^{N-1}{\hat \gamma}_{ai}{\hat {\vec \pi}}_a(\tau )-e_i{\hat \theta}
^{*}_i{\hat \theta}_i{\vec A}_{\perp}(\tau ,{\vec \eta}_{+}(\tau )+{1\over
{\sqrt{N}}}\sum_{a=1}^{N-1}{\hat \gamma}_{ai}{\vec \rho}_a(\tau ))]{}^2 }+
\nonumber \\
&+&\sum_{i=1}^Ne_i{\hat \theta}^{*}_i{\hat \theta}_i\lbrace \sum^{1..N}
_{j\not= i} e_j{\hat \theta}^{*}_j{\hat \theta}_j[{1\over {\triangle
_{{\vec \eta}_j}} }{{\partial}\over {\partial \eta^r_j}} c({\vec \eta}_i(\tau
)-
{\vec \eta}_j(\tau ))-\eta^r_j(\tau ) c({\vec \eta}_i(\tau )-{\vec \eta}_j
(\tau ))]+\nonumber \\
&+&\int d^3\sigma \pi^r_{\perp}(\tau ,\vec \sigma ) c(\vec \sigma -{\vec \eta}
_i(\tau ))\rbrace-{1\over 2}\int d^3\sigma \sigma^r\,
({\vec \pi}_{\perp}^2(\tau ,\vec \sigma )+{\vec B}^2[{\vec A}_{\perp}(\tau ,
\vec \sigma )]).
\label{147}
\end{eqnarray}

Therefore the final result is the extraction of the static
action-at-a-distance
Coulomb potential from the electromagnetic field theory, which is
reduced to the transverse radiation field.This is obtained at the
pseudoclassical level, where the hypothesis of charge quantization is
reflected in the property $Q^2_i=0$, which regularizes the classical
electromagnetic self-energy and produces the rule $\sum_{i\not= j}$
without the need to introduce Feuynman-Wheeler\cite{feyw} theory of the
absorbers. Clearly all the effects of order $e_i^2$ (like those producing the
runaway solutions in the Abraham-Lorentz-Dirac equation;
see Refs.\cite{teitel,grandy}) are killed: only the
interference effects of order $e_ie_j$, $i\not= j$ are preserved.
These problems will be investigated in more detail elsewhere.
In any case, it is clear that in this formulation the initial data problem
consists in specifying Cauchy data for the radiation field and
Newton-like action-at-a-distance initial data for the charged particles with
mutual Coulomb interaction.

In particular, if we put equal to zero the radiation field, ${\vec A}_{\perp}
(\tau ,\vec \sigma )={\vec \pi}_{\perp}(\tau ,\vec \sigma )=0$, we obtain a
well defined relativistic 1-time reduced N-body problem with the Coulomb
potential extracted from field theory and with a decoupling of the motion of
the center of mass, whose constraints are

\begin{eqnarray}
{\hat {\cal H}}(\tau )&=&\epsilon_s-\lbrace \sum_{i=1}^N\eta_i
\sqrt{m^2_i+[{1\over N}{\hat {\vec \kappa}}_{+}
(\tau )+\sqrt{N}\sum_{a=1}^{N-1}{\hat \gamma}_{ai}{\hat {\vec \pi}}_a(\tau )
]{}^2 }-\nonumber \\
&-&{1\over 2}\sum_{i\not= j}^{1..N}[e_i{\hat \theta}^{*}_i{\hat \theta}_i]
[e_j{\hat \theta}^{*}_j{\hat \theta}_j]\, c[{1\over {\sqrt{N}}}\sum_{a=1}^{N-1}
({\hat \gamma}_{ai}-{\hat \gamma}_{aj}){\vec \rho}_a(\tau )]\rbrace
\approx \nonumber \\
&\approx& \epsilon_s-\lbrace \sum_{i=1}^N\eta_i
\sqrt{m^2_i+N[\sum_{a=1}^{N-1}{\hat \gamma}_{ai}{\hat {\vec \pi}}_a(\tau )
]{}^2 }-\nonumber \\
&-&{1\over 2}\sum_{i\not= j}^{1..N}[e_i{\hat \theta}^{*}_i{\hat \theta}_i]
[e_j{\hat \theta}^{*}_j{\hat \theta}_j]\, c[{1\over {\sqrt{N}}}\sum_{a=1}^{N-1}
({\hat \gamma}_{ai}-{\hat \gamma}_{aj}){\vec \rho}_a(\tau )]\rbrace \approx 0
\nonumber \\
&&{}\nonumber \\
{\hat {\vec {\cal H}}}_p(\tau )&=&{\hat {\vec \kappa}}_{+}(\tau )\approx 0
\nonumber \\
&&{}\nonumber \\
&&\lbrace {\hat {\cal H}},{\hat {\cal H}}^r_p\rbrace {}^{**}=\lbrace {\hat
{\cal H}}^r_p,{\hat {\cal H}}^s_p\rbrace {}^{**}=0;
\label{148}
\end{eqnarray}

\noindent whose natural gauge-fixing for decoupling the last three constraints
is ${\vec \eta}_{+}(\tau )\approx 0$ as in the free case [the nonrelativistic
limit, using the methods of Refs.\cite{pons}, would give the first class
constraint ${\hat {\cal H}}=E_R-H_R\approx 0$ (where $E_R$ and $H_R$ are the
reduced energy and Hamiltonian respectively) for the parametrized
Newtonian N-body problem with Coulomb interaction]. See also Ref.\cite{ppl}
and its bibliography for models with additive potentials like the Coulomb
one in Eq.(\ref{148}). For the N=2 case, Eqs.(\ref{148}) were already found
in Refs.\cite{gst}, where there is a study of the classical equations of
motion and their explicit solution in the equal mass case.

It is difficult to find the gauge-fixings replacing ${\vec \eta}_{+}\approx 0$,
i.e. which is the point of the system to be identified with ${\tilde x}_s^{\mu}
$, for the full theory, and then the final form of the rest-frame spin tensor
(\ref{147}). Given these gauge-fixings, we have to find a
canonical basis of Dirac's observables with vanishing Poisson brackets with
the constraints ${\hat {\vec {\cal H}}}_p\approx 0$ of Eqs.(\ref{146}) and with
the gauge-fixings, and to reexpress ${\hat {\cal H}}\approx 0$ of
Eqs.(\ref{146}) in terms of them: this will give the final 1-time reduced
Hamiltonian for N charged scalar particles interacting through the Coulomb
potential and coupled to the radiation field.
However, an implicit form of the three gauge-fixings may be found by
remembering
that in any instant form of the dynamics there are only four generators of the
Poincar\'e group which depend on the interaction. In the rest-frame instant
form they are $\epsilon_s=\eta_s\sqrt{p_s^2}$, given by the first of Eqs.(
\ref{146}), and the three boosts $J^{oi}_s$,
given in the last of Eqs.(\ref{131})
[they depend on the interaction through the term $\eta_s\sqrt{p_s^2}$]. The
other six, i.e. ${\vec p}_s$ and $J_s^{ij}$ [namely ${\bar S}^{rs}_s$], must be
identical to the analogous generators in absence of interactions, namely to
the ganerators for N neutral scalar particles plus the electromagnetic field.
Therefore, from Eqs.(\ref{147}) we must require the vanishing of the
interacting part ${\bar S}^{rs}_{s,INT}$ of the rest-frame three-spin

\begin{eqnarray}
\chi^k(\tau )&=&{1\over 2}\epsilon^{krs}{\bar S}^{rs}_{s,INT}=
-\epsilon^{krs}\eta^r_{+}(\tau )\sum_{i=1}^Ne_i{\hat \theta}^{*}_i{\hat \theta}
_iA_{\perp}^s(\tau ,{\vec \eta}_{+}(\tau )+{1\over {\sqrt{N}}}
\sum_{b=1}^{N-1}{\hat \gamma}_{bi}{\vec \rho}_b(\tau ))-\nonumber \\
&-&{{\epsilon^{krs}}\over {\sqrt{N}}}\sum_{a=1}^{N-1}\rho^r_a(\tau )
\sum_{i=1}^N{\hat \gamma}_{ai}e_i{\hat \theta}^{*}_i{\hat \theta}_i
A_{\perp}^s(\tau ,{\vec \eta}_{+}(\tau )+{1\over {\sqrt{N}}}
\sum_{b=1}^{N-1}{\hat \gamma}_{bi}{\vec \rho}_b(\tau ))-\nonumber \\
&-&\sum_{i=1}^N e_i{\hat \theta}^{*}_i{\hat \theta}_i\int
d^3\sigma c(\vec \sigma -{\vec \eta}_i(\tau )) \lbrace 2B^k[{\vec A}_{\perp}
(\tau ,\vec \sigma )]-\nonumber \\
&-&\vec \sigma \cdot {{\partial}\over {\partial \vec \sigma}}
B^k[{\vec A}_{\perp}(\tau ,\vec \sigma )]+\vec \sigma \cdot {{\partial}\over
{\partial \sigma^k}}\vec B[{\vec A}_{\perp}(\tau ,\vec \sigma )]\rbrace
\approx 0.
\label{new148}
\end{eqnarray}

The time constancy of these constrains determines the Dirac multipliers
$\vec \lambda (\tau)$, so that the final Dirac Hamiltonian becomes
$H_D=\lambda (\tau ){\hat {\cal H}}(\tau )$.
The problem of evaluating the Dirac brackets of the three pairs of second class
constraints ${\hat {\vec {\cal H}}}_p\approx 0$ and $\chi^r
\approx 0$ as well as the attempt of diagonalizing these brackets will be
studied elsewhere.

Let us remark
that, if in Eqs.(\ref{146}) we disregard the constraints ${\hat {\vec {\cal
H}}}_p\approx 0$ and eliminate the field energy term in ${\hat {\cal H}}
\approx 0$, we get a constraint ${\hat {\cal H}}^{'}\approx 0$ in which ${\vec
A}_{\perp}$ is an external radiation field coupled to the system.

If we eliminate the particles and we remain with only the field, the four
constraints are

\begin{eqnarray}
{\hat {\cal H}}(\tau )&=&\epsilon_s-
{1\over 2}\int d^3\sigma [{\vec \pi}_{\perp}^2(\tau ,\vec \sigma )+{\vec B}
^2[{\vec A}_{\perp}(\tau ,\vec \sigma )]\, ]\rbrace \approx 0\nonumber \\
&&{}\nonumber \\
{\hat {\vec {\cal H}}}_p(\tau )&=&
\int d^3\sigma \, \vec B[{\vec A}_{\perp}(\tau ,\vec \sigma )]\times
{\vec \pi}_{\perp}(\tau ,\vec \sigma )={\vec P}_{(F)s}\approx 0\nonumber \\
&&{}\nonumber \\
&&\lbrace {\hat {\cal H}},{\hat {\cal H}}^r_p\rbrace {}^{**}=\lbrace {\hat
{\cal H}}^r_p,{\hat {\cal H}}^s_p\rbrace {}^{**}=0.
\label{149}
\end{eqnarray}

\noindent Now the gauge-fixing should be the vanishing of a center-of-mass
field three-coordinate conjugate to the total three-momentum ${\vec P}_{(F)s}
={\hat {\cal H}}_p\approx 0$. Therefore, we have indirectly
shown that there must exist a center-of-mass decomposition also for classical
gauge field theory and that this is the lacking ingredient to get
a reformulation of classical field theory as a ``1-time rest-frame field
theory" with Wigner covariance of the variables relative to the center of mass
when restricted to configurations belonging to timelike representations of the
Poincar\'e group. Again, only three center-of-mass coordinates would not be
covariant.

\vfill\eject

\section{The 1-time Theory versus the Feshbach-Villars Hamiltonian Theory}

In Ref.\cite{fv} (see also Ref.\cite{cors}), the first quantized one-particle
second order Klein-Gordon equation coupled to an external electromagnetic
field, $(D^2+m^2)\phi (x)=0$, $D_{\mu}=\partial_{\mu}-ieA_{\mu}(x)$, was put
in Hamiltonian form after having rewritten it in the first order formalism:
$\phi_o(x)-{i\over m}D_o\phi (x)=0$, $({\vec D}^2-m^2)\phi (x)+imD_o\phi_o
(x)=0$. If we put $\phi (x)={1\over {\sqrt{2}}}[\varphi (x)+\chi (x)]$,
$\phi_o(x)={1\over {\sqrt{2}}}[\varphi (x)-\chi (x)]$, we obtain

\begin{eqnarray}
&&i\partial_o\varphi (x)={1\over {2m}}{(-i\vec \partial -e\vec A(x))}^2
(\varphi (x)+\chi (x))+(eA_o(x)+m)\varphi (x)\nonumber \\
&&i\partial_o\chi (x)=-{1\over {2m}}{(-i\vec \partial -e\vec A(x))}^2
(\varphi (x)+\chi (x))+(eA_o(x)-m)\chi (x).
\label{150}
\end{eqnarray}

\noindent In a $2\times 2$ matrix formalism we have ($\tau_i$ are the Pauli
matrices)

\begin{eqnarray}
i\partial_o \Psi (x)&=& H\, \Psi (x)\nonumber \\
&&\Psi (x)=\left( \begin{array}{c} \varphi (x)\\ \chi (x) \end{array} \right)
\nonumber \\
&&H={1\over {2m}}{(-i\vec \partial -e\vec A(x))}^2\, (\tau_3+i\tau_2)+m\tau_3
+eA_o(x) 1\nonumber \\
&&{\rightarrow}_{A_{\mu}\rightarrow 0}\,\, H_o={ {{(-i\vec \partial )}^2}
\over {2m}}(\tau_3+i\tau_2)+m\tau_3.
\label{151}
\end{eqnarray}

In the momentum representation we have $H_o={ {{\vec p}^2}\over {2m}}(\tau_3+
i\tau_2)+m\tau_3$ and this Hamiltonian can be diagonalized ($p^o=+\sqrt{m^2+
{\vec p}^2}$)

\begin{eqnarray}
H_{o,U}&=&U^{-1}(\vec p)H_oU(\vec p)=p^o\tau_3=\left( \begin{array}{cc}
\sqrt{m^2+{\vec p}^2}&0\\ 0&-\sqrt{m^2+{\vec p}^2} \end{array} \right)
\nonumber \\
&&\Psi_U(p)=U^{-1}(\vec p)\Psi (p)\nonumber \\
&&i\partial_o\Psi_U=H_{o,U}\Psi_u\nonumber \\
&&{}\nonumber \\
&&U(\vec p)={1\over {2\sqrt{mp^o}}}[(m+p^o) 1-(m-p^o)\tau_1]\nonumber \\
&&U^{-1}(\vec p)={1\over {2\sqrt{mp^o}}}[(m+p^o) 1+(m-p^o)\tau_1].
\label{152}
\end{eqnarray}

Like in the case of the Foldy-Wouthuysen transformation for particles of spin
1/2, also in the spin 0 case the exact diagonalization of the Hamiltonian
cannot be achieved in presence of an arbitrary external electromagnetic field
\cite{fv}.

At the classical level, the first class constraint ${(p-eA(x))}^2-m^2\approx 0$
may be resolved as $(p^o-eA^o(x)-\sqrt{m^2+{(\vec p-e\vec A(x))}^2})
(p^o-eA^o(x)+\sqrt{m^2+{(\vec p-e\vec A(x))}^2})\approx 0$, which can be
replaced by the $2\times 2$ matrix

$\left( \begin{array}{cc} (p^o-eA^o(x)-
\sqrt{m^2+{(\vec p-e\vec A(x))}^2}) & 0\\ 0& (p^o-eA^o(x)+
\sqrt{m^2+{(\vec p-e\vec A(x))}^2}) \end{array} \right) =(p^o-eA^o(x)) 1+
\sqrt{m^2+{(\vec p-e\vec A(x))}^2} \tau_3 \approx 0$.

\noindent This shows that the
$2\times 2$ matrix formalism has a topological origin in the two disjoint
branches of the timelike orbits. However, in presence of the external
electromagnetic field $p^{\mu}$ is no more a constant of the motion and it is
not clear how to apply the theory of the canonical realizations of the
Poincar\'e group.

In contrast,
in the 1-time theory for N=1 in presence of a dynamical (not external)
electromagnetic field, from Eqs.(\ref{133}), (\ref{134}) [in the canonical
transformation (\ref{73}), (\ref{74}), $\vec \eta (\tau )$ and $\vec \kappa
(\tau )$ remain fixed], we have the following rest-frame first class
constraints

\begin{eqnarray}
&&{\cal H}(\tau )
=\epsilon_s-\eta \sqrt{m^2+{[\vec \kappa (\tau )-e\theta^{*}\theta
\vec A(\tau ,\vec \eta (\tau ))]}^2}-{1\over 2}\int d^3\sigma [{\vec \pi}^2
(\tau ,\vec \sigma )+{\vec B}^2(\tau ,\vec \sigma )]\approx 0\nonumber \\
&&{\vec {\cal H}}_p(\tau )
=\vec \kappa (\tau )-e\theta^{*}\theta \vec A(\tau ,\vec
\eta (\tau ))+\int d^3\sigma \vec \pi (\tau ,\vec \sigma )\times \vec B(\tau ,
\vec \sigma )\approx 0\nonumber \\
&&\pi^{\tau}(\tau ,\vec \sigma )\approx 0\nonumber \\
&&\Gamma(\tau ,\vec \sigma )=-\vec \partial \cdot \vec \pi (\tau ,\vec \sigma )
+e\theta^{*}\theta \delta^3(\vec \sigma -\vec \eta (\tau ))\approx 0,
\label{153}
\end{eqnarray}

\noindent or in terms of Dirac observables [see Eq.(\ref{146}); again we have
${(e\theta^{*}\theta )}^2=0$]

\begin{eqnarray}
&&{\hat {\cal H}}(\tau )
=\epsilon_s-\eta \sqrt{m^2+{[{\hat {\vec \kappa}}(\tau )-
e{\hat \theta}^{*}\hat \theta {\vec A}_{\perp}(\tau ,\vec \eta (\tau ))]}^2}-
{1\over 2}\int d^3\sigma \, ({\vec \pi}^2_{\perp}(\tau ,\vec \sigma )+
{\vec B}^2[{\vec A}_{\perp}(\tau ,\vec \sigma )])\approx 0\nonumber \\
&&{\hat {\vec {\cal H}}}_p(\tau )={\hat {\vec \kappa}}(\tau )-e{\hat
\theta}^{*}
\hat \theta {\vec A}_{\perp}(\tau ,\vec \eta (\tau ))+\int d^3\sigma {\vec \pi}
_{\perp}(\tau ,\vec \sigma )\times \vec B[{\vec A}_{\perp}(\tau ,\vec \sigma )]
\approx 0.
\label{154}
\end{eqnarray}

\noindent Even if it is not yet clear which are the natural gauge-fixings to
decouple the constraints ${\hat {\vec {\cal H}}}_p\approx 0$ [i.e. which is
the point of the plane to be identified with ${\tilde x}_s^{\mu}$; in absence
of the field, the gauge-fixing is $\vec \eta (\tau )\approx 0$, namely the
particle position is described by ${\tilde x}_s^{\mu}$] so to
remain only with one final first class constraint ${\check {\cal H}}\approx
0$ for the final Dirac observables, some remarks may be done.

There is a well defined coupling of the electromagnetic field to each branch
$\eta =\pm$ of the particle mass spectrum, independently from the other: in a
sense for $p_s^2 > 0$ we have obtained a diagonalization of the overall
particle +field mass spectrum. However this has been achieved by treating the
electromagnetic field as a dynamical field and not as an external one.
Moreover,
we have used a definition of rest frame based on the conserved Poincar\'e
generator $p_s^{\mu}$, which is associated with the isolated system
particle+field and which does not exist when the field is external. Finally,
even if we can form a $2\times 2$ matrix constraint $\left( \begin{array}{cc}
{\hat {\cal H}}{|}_{\eta =+} & 0 \\ 0 & {\hat {\cal H}}{|}_{\eta =-}
\end{array} \right) \approx 0$, both ${\hat {\cal H}}{|}_{\eta}$ contain the
term in the field energy: only after having found the rest-frame field
theory based on a decomposition of the fields in center-of-mass and relative
variables we will have the tools to try to understand better this matrix
formulation and its connection with the Klein-Gordon equation (the coupling
of the 1-time theory could be a nonminimal nonlocal coupling of the
original Klein-Gordon equation to the electromagnetic field).

In Ref.\cite{fv}, there is also a discussion of the first-quantized effect of
zitterbewegung of the canonical noncovariant Newton-Wigner 3-position
operator ${\vec x}_{op}$ for a scalar particle and of its nonlocalizability.
Since the positive energy states (with the standard scalar product) do not
satisfy a completeness relation, the narrowest wave packet containing only
positive frequencies has a width of the order of the Compton wavelength 1/m.
Equivalently, the eigenstates of ${\vec x}_{op}=i\vec \partial /\partial \vec
p$
contain both positive and negative energies: in the diagonal $2\times 2$
representation we define a ``mean position operator" [which has $\delta^3
(\vec x-{\vec x}^{'})$ as eigenfunctions] as ${\vec x}_{U,op}=U^{-1}(\vec p)
i{{\vec \partial}\over {\partial \vec p}}\, U(\vec p)=i
{{\vec \partial}\over {\partial \vec p}}\, 1-{{i\vec p}\over {2(m^2+{\vec
p}^2)}
}\tau_1 ={\vec x}^{(+)}_{U,op}+{\vec x}^{(-)}_{U,op}$, where the odd matrix
$\tau_1$ couples positive and negative energies.
The operator ${\vec x}^{(+)}_{U,op}$,
with velocity $\partial_o{\vec x}^{(+)}_{U,op}={{\vec p}\over {\sqrt{m^2+
{\vec p}^2}}} \tau_3$, describes the steady motion of the particle, is the
Newton-Wigner position operator in this representation but has eigenfunctions
delocalized over a region of radius of the order 1/m (they have the previous
velocity as group velocity); instead a wavepacket with a localization more
accurate than 1/m (built with eigenfunctions of ${\vec x}_{U,op}$) will
contain both kinds of energies and will have a trembling motion
(zitterbewegung)
superimposed to the steady motion in a region of the order 1/m.

In Klein-Gordon quantum field theory\cite{lurie}, the byproducts of these
effects are the nonexistence of perfectly localized states (eigenstates of a
local occupation number operator $N_{V,op}(x^o)=\int_Vd^3x\, N_{op}(\vec x,
x^o)$ for an infinitesimally small volume V; $[N_{V,op}(\vec x,x^o),
N_{V,op}({\vec x}^{'},x^o)]$ is different from zero for distances $|\, \vec x-
{\vec x}^{'}|$ of the order 1/m\cite{ht}): we cannot fix the particle number
in a volume smaller than 1/m and the number of particles in two volumes V and
$V^{'}$ cannot be measured simultaneously unless V and $V^{'}$ are separated
by distances greater than 1/m. In contrast,
we can define a local charge operator
${\hat Q}_V$ and build states with perfect localization of charge in
infinitesimally small volumes; however $[{\hat Q}_V,N_{op}]\not= 0$, i.e.
the eigenstates of ${\hat Q}_V$ do not have a sharp particle number.

In contrast,
in the classical rest-frame 1-time theory of a scalar particle, positive
and negative energy states are completely decoupled (the upper and lower
branches), and, instead of the zitterbewegung effect, we have the
impossibility to localize the center-of-mass canonical 3-position ${\vec
{\tilde x}}_s$ [to which collapses the particle position after the
gauge-fixings $\vec \eta (\tau )\approx 0$; the first class constraint becomes
${\cal H}=\epsilon_s-\eta m\approx 0$ and we have a Hamilton-Jacobi
description of the free particle, with ${\vec z}_s/\eta m$ as Jacobi data]
inside the noncovariance worldtube
of radius $|\, {\vec {\bar S}}|\, /m$ in a frame independent way.

\vfill\eject

\section{1-time Relativistic statistical mechanics}

In Ref.\cite{hakim}, there is a a review with rich bibliography till 1966 of
the
two types of approaches to classical relativistic statistical mechanics: i) the
nonmanifestly covariant and ii) the covariant ones. While the former approaches
started in 1911 with Refs.\cite{jutt}, where the equilibrium of a relativistic
(classical Boltzmann and also quantum either Bose or Fermi) gas was studied,
only in 1940 in Ref.\cite{lichn} a first relativistic Boltzmann equation was
given in the framework of relativistic kinetic theory; then the studies in
plasma physics focused on equal-time Hamiltonians

\begin{equation}
H=\sum_{i=1}^N\lbrace \sqrt{m_i^2+{({\vec p}_i-e_i\vec A({\vec x}_i,x^o))}^2}+
V({\vec x}_i,x^o)\rbrace +{1\over 4}\int d^3z [{\vec \pi}^2(\vec z,x^o)+
{\vec B}^2(\vec z,x^o)],
\label{155}
\end{equation}

\noindent derived a relativistic Liouville equation for a density involving
both
particles and fields and were extensively studied in the sixties [see the
bibliography of Ref.\cite{hakim}].

In contrast,
the covariant approach started with Ref.\cite{berg}, where, taking into
account constraints, there are the first steps in defining covariant
relativistic kinetic theory, relativistic statistical mechanics (canonical
ensembles) and relativistic thermodynamics following the first developments of
relativistic hydrodynamics [see the bibliography of Ref.\cite{hakim}];
the Vlasov-, Boltzmann-, Landau-, Fokker-Planck- relativistic kinetic equations
came out as ad hoc semiphenomenological equations [again see the bibliography
of Ref.\cite{hakim}].

In Ref.\cite{hakim}, there is a critical analysis of the problems in the
foundations of relativistic kinetic theory and relativistic statistical
mechanics stemming from the covariant relativistic description of particles
with either action-at-a-distance or field-mediated interactions. Due to the
absence of an absolute time, relativistic kinetic theory ($\mu$-space) is a
statistical theory of curves (worldlines) and not of points (Cauchy data)
as at the nonrelativistic level. Analogously, due to the presence of N times
in the description of N particle systems, relativistic statistical mechanics
($\Gamma$-space, Gibbs ensembles) was a statistical theory of N-dimensional
manifolds (the gauge orbits spanned by the N times in the 7N constraint
manifold defined in the 8N-dimensional phase space by the N mass-shell
first class constraints $p_i^2-m^2\approx 0$) instead of 1-dimensional
manifolds as in the nonrelativistic case (so that one has N and not 1
relativistic Liouville equations in $\Gamma$-space and associated problems
with the definition of measures). These problems were already posed in Ref.
\cite{berg}. However, there was a not clear understanding that the transition
from the configuration space [worldlines with coordinates $q^{\mu}_i(\tau )$ or
${\vec q}_i(q^o_i)$; see predictive mechanics\cite{bel} and the predictive
conditions on the relativistic forces for the implementation of a
realization of the Poincar\'e group\cite{ch}] to phase space [with coordinates
$x_i^{\mu}(\tau ), p_i^{\mu}(\tau )$] had to face the No-Interaction-Theorem
\cite{nit,nita}, so that $q^{\mu}_i(\tau )=x_i^{\mu}(\tau )$ is allowed only
in the free case. Moreover, in Ref.\cite{hakim} there is the recognition that
to
build relativistic kinetic theory and statistical mechanics the initial data
must be chosen as the mathematical Cauchy data on a spacelike hypersurface
[it is not clear whether they can be measured] and not for instance as data
on the backward light-cone of an observer [in the case of the electromagnetic
field, the light cone is a characteristic surface of Maxwell equations and
does not define a well posed Cauchy problem: the future field is not
completely and uniquely determined]: in this case, in $\mu$-space, starting
from the predictive equations of motion in the first-order formalism [${\dot q}
^{\mu}(\tau )=u^{\mu}(\tau )$, $m{\dot u}^{\mu}(\tau )=F^{\mu}$], from the
density $R(q^{\mu},u^{\mu},q^{\mu}_o,u^{\mu}_o,\tau )=\delta^4[q^{\mu}-q^{\mu}
(\tau ,q_o,u_o)]\times \delta^4[u^{\mu}-u^{\mu}(\tau ,q_o,u_o)]$ and from
an ensemble of such systems with a random distribution $D_o(q_o,u_o)$ of
Cauchy data, it is possible to define the density in $\mu$-space
$D(q^{\mu},u^{\mu},\tau )=\int_{\mu}R(q^{\mu},u^{\mu},q_o^{\mu},u_o^{\mu},\tau
)
D_o(q_o,u_o) d^4q_od^4u_o$, to find the one-particle relativistic equation
for D [and then, by adding ad hoc collision terms, various kinetic equations],
to show that ${\cal N}(q^{\mu},u^{\mu})=\int^{+\infty}_{-\infty}d\tau
D(q^{\mu},u^{\mu},\tau )$ is a distribution function such that $j^{\mu}(q)=
\int d^4p\, {\cal N}(q^{\mu},p^{\mu}=mu^{\mu})\, u^{\mu}2m\theta(p^o)
\delta^4(p^2-m^2)$ [if one chooses $u^2=1$] is the particle four-current, that
${\cal N}(q^{\mu},u^{\mu})$ satisfies the one-particle Liouville equation,
which in turn, by adding ad hoc collision terms ${\cal C}({\cal N})$, becomes
a kinetic transport equation. When the external force $F^{\mu}$ vanishes, the
equation ${\cal C}({\cal N})=0$ is satisfied by the relativistic version of the
Maxwell-Boltzmann distribution function, i.e. the classical
J\"uttner\cite{jutt}
-Synge\cite{synge} distribution [$K_2(x)$ is the modified Bessel function of
order two]

\begin{equation}
{\cal N}(q^{\mu},p^{\mu})={ {N\rho (q)}\over {4\pi m^2K_2(m\beta )} }
e^{-\beta^{\mu}p_{\mu}}.
\label{156}
\end{equation}

\noindent In this equation $\beta^{\mu}$ is the reciprocal temperature timelike
fourvector\cite{synge}
[$\beta =\sqrt{\beta^2}=1/kT$ with k being the Boltzmann constant].

In Refs.\cite{groot}, there was a further development of relativistic kinetic
theory and a discussion of how to formulate covariant relativistic
thermodynamics [see Ref.\cite{haar} for the bibliography on the subject and
for a review of the associated problems], whose final formulation [containing
the solution of the problem of having causal propagation of heat flow
(parabolic transport equations replaced by hyperbolic ones) with some
molecular speed less than the velocity of light] was given in Refs.\cite{isra}
for fluids both in equilibrium and out of it. The conclusion was that at
equilibrium all relevant fourvectors [the reciprocal temperature $\beta^{\mu}$,
the entropy four-current $s^{\mu}$, the volume fourvector
$V^{\mu}$, the number density of particles $n^{\mu}$, the four-momentum
$p^{\mu}$] should be all proportional, through their rest-frame values, to the
hydrodynamical four-velocity $u^{\mu}$ [$u^2=1$]: $\beta^{\mu}=\beta u^{\mu}=
u^{\mu}/kT$, $V^{\mu}=Vu^{\mu}$, $s^{\mu}=su^{\mu}$, $n^{\mu}=nu^{\mu}$,
$p^{\mu}=pu^{\mu}$; $u^{\mu}$ is a timelike eigenvector of the energy-momentum
tensor $T^{\mu\nu}$ if it satisfies certain positivity conditions\cite{synge}.

At equilibrium, starting from the J\"uttner-Synge distribution, we can derive
the covariant canonical partition function $Q_N(V^{\mu},\beta^{\mu})$ for an
ideal classical relativistic Boltzmann gas of N scalar particles of mass m
(and going to the quantum level of Bose and Fermi ideal quantum gases) and then
the covariant grand-canonical partition function $\Xi(V^{\mu},\beta^{\mu},
i^{\mu})=\sum_Ne^{i_{\mu}n^{\mu}}Q_N(V^{\mu},\beta^{\mu})$ [$i^{\mu}=\mu \beta
^{\mu}$ with $\mu$ the relativistic chemical potential, related to the
nonrelativistic one by $\mu=\mu_{NR}+m$] and the covariant thermodynamical
potential $\Omega(V^{\mu},\beta^{\mu},i^{\mu})=-kT\, ln\, \Xi(V^{\mu},\beta
^{\mu},i^{\mu})$. This has been done in Ref.\cite{chai} following previous
results in Refs.\cite{lepore,maga}, after the identification of the
invariant microcanonical density of states

\begin{equation}
\sigma_N(p^2,V^2,p_{\mu}V^{\mu})=\int \delta^4(p-\sum_{i=1}^Np_i) \prod_{i=1}^N
d\sigma_i(p_i,m),
\label{157}
\end{equation}

\noindent based on the invariant momentum space measure\cite{tousc,jabs}

\begin{equation}
d\sigma (p,m)=2V_{\mu}p^{\mu} \theta(p^o)\delta (p^2-m^2) d^4p.
\label{158}
\end{equation}

Then the canonical partition function for the ideal Boltzmann gas of N free
scalar particles is [$1/N!$ is the Boltzmann counting factor for identical
particles]

\begin{eqnarray}
Q_N(V^{\mu},\beta^{\mu})&=&{1\over {N!}}\, \int d^4p e^{-\beta^{\mu}p_{\mu}}
\sigma_N(p^2,V^2,p_{\mu}V^{\mu})=\nonumber \\
&=&{1\over {N!}}[ {2\over {(2\pi )^3}} \int d^4p \theta (p^o)\delta (p^2-m^2)
p_{\mu}V^{\mu} e^{-\beta^{\mu}p_{\mu}} ]{}^N=\nonumber \\
&=&{1\over {N!}}\prod_{i=1}^N [{{2V}\over {(2\pi )^3}} \int d^3p_i e^{-\beta
\sqrt{m^2+{\vec p}^2_i}}]{}^N\nonumber \\
&=&{1\over {N!}} [ { {Vm^2}\over {2\pi^2\beta}} K_2(m\beta ) ]{}^N,\quad\quad
\beta =1/kT,
\label{159}
\end{eqnarray}

\noindent where the last evaluation has been made in the rest frame of the
volume V.

This same result may be obtained\cite{karsch}
from the phase space description of N scalar
particles with N first class constraints, by using a Faddeev-Popov measure in
which N gauge-fixings [of the kind $p_i\cdot x_i/\sqrt{p_i^2}-\tau_i\approx 0$]
have been added to the N first class constraints $\phi_i\approx 0$ to
eliminate particles' times and energies (the many-time problem). The resulting
form of $Q_N$ for a system of N interacting particles is [$p^{\mu}=\sum_{i=1}^N
p_i^{\mu}$]

\begin{equation}
Q_N(V^{\mu},\beta^{\mu})={1\over {N!}}\int_{V^{\mu}} e^{-\beta^{\mu}p_{\mu}}
\prod_{i=1}^N \theta(p_i^o)\delta (\phi_i)\delta (\chi_i) det\, |\, \lbrace
\phi_i,\chi_j\rbrace |\, {{d^4x_id^4p_i}\over {(2\pi )^3}},
\label{160}
\end{equation}

\noindent where $\phi_i=p^2_i-m^2_i-...\approx 0$ [the dots are for the
(generically unknown) interaction terms in the N-time theory] are the
original first class constraints and $\chi_i\approx 0$ are the N gauge-fixings.
In Ref.\cite{karsch}, there is also a study of a gas of N 2-body bound states
based on the model of Refs.\cite{todo,komar}.

Moreover, Feynman nonrelativistic result\cite{feynm} that the partition
function of quantum statistical mechanics can be obtained by evaluating the
density matrix for an N particle system from the continuation to imaginary
times ($x^o\rightarrow -i\hbar/kT$) of the path integral used for the
evaluation of the kernel of the evolution operator, may be extended to free
relativistic scalar particles ($p^2-m^2\approx 0$) in the proper time gauge
and the result is again $Q_N$\cite{lusm}.

See for instance Refs.\cite{hakimb} for applications of relativistic
statistical
mechanics to relativistic plasmas.

Coming back, after this sketchy review, to the 1-time description of N
relativistic scalar particles, we see that now we have a natural kinematical
framework (when the system is isolated and the total momentum is timelike) for
developing an 1-time covariant formulation of relativistic statistical
mechanics of systems of interacting scalar particles along the lines of the
nonrelativistic one; it could be termed ``rest-frame instant form relativistic
statistical mechanics".

To arrive at this formulation, let us start from the ideal gas in the
formulation of Eq.(\ref{160}) and let us try to reproduce the
canonical partition function of Eq.(\ref{159}) with the new formulation.
Let us make the canonical transformation from
the basis $x_i^{\mu}, p_i^{\mu}$ of the N-time theory to the basis ${\hat
{\tilde x}}^{\mu}, p^{\mu}, {\hat T}_{Ra}, {\hat \epsilon}_{Ra}, {\hat {\vec
\rho}}_a, {\vec \pi}_a$ of Eqs.(\ref{31}) and then to the basis ${\hat T},
\epsilon ,{\hat {\vec z}}, \vec k, {\hat T}_{Ra}, {\hat \epsilon}_{Ra}, {\hat
{\vec \rho}}_a, {\vec \pi}_a$ of Eqs.(\ref{35}). We have

\begin{eqnarray}
\prod_{i=1}^Nd^4x_id^4p_i&=&d^4{\hat {\tilde x}}d^4p\prod_{a=1}^{N-1}
d{\hat T}_{Ra}d{\hat \epsilon}_{Ra},d^3{\hat \rho}_ad^3\pi_a=\nonumber \\
&=&d\epsilon d^3k d\hat T d^3{\hat z} \prod_{a=1}^{N-1}
d{\hat T}_{Ra}d{\hat \epsilon}_{Ra}d^3{\hat \rho}_ad^3\pi_a,
\label{161}
\end{eqnarray}

\noindent since $d^4{\hat {\tilde x}}=\sqrt{1+{\vec k}^2}d\hat Td^3\hat z
/{|\, \epsilon |}^3$ and $d^4p={|\, \epsilon |}^3d\epsilon d^3k/\sqrt{1+
{\vec k}^2}$\cite{longhi}.

The gauge-fixings needed to go to the rest-frame instant form are $\hat T -\tau
\approx 0$ and ${\hat T}_{Ra}\approx 0$. Since we do not know explicitely the
inverse canonical transformations, we could expect to have a Jacobian coming
out in the following transformation [to avoid the problem of degenaracies we
start by making calculations with different masses $m_i$]

\begin{eqnarray}
&&\prod_{i=1}^N \theta(p^o_i)\delta(p_i^2-m_i^2)\delta(\hat T(x,p)-\tau )
\prod_{a=1}^{N-1}\delta [{\hat T}_{Ra}(x,p)]=\nonumber \\
&&=J\, \prod_{i=1}^N \theta (\eta_i)\delta (p^2-...)\delta (\hat T-\tau )
\prod_{a=1}^{N-1}\delta ({\hat \epsilon}_{Ra})\delta ({\hat T}_{Ra})=
\nonumber \\
&&=J\, \prod_{i=1}^N \theta (\eta_i)\delta (\epsilon^2-...)\delta (\hat T-
\tau ) \prod_{a=1}^{N-1}\delta ({\hat \epsilon}_{Ra})\delta ({\hat T}_{Ra}).
\label{162}
\end{eqnarray}

\noindent We have written $p^2-...=\epsilon^2-..\approx 0$ for the (explicitely
unknown) mass spectrum constraint; but with the help of Eq.(\ref{82}) and due
to
the $\theta (\eta_i)$, Eq.(\ref{162}) may be rewritten as

\begin{equation}
\tilde J\, \delta (\epsilon -\sum_{i=1}^N\sqrt{m^2+N{(\sum_{a=1}^{N-1}{\hat
\gamma}_{ai}{\vec \pi}_a)}^2}\delta (\hat T-\tau ) \prod_{a=1}^{N-1}
\delta ({\hat T}_{Ra})\delta ({\hat \epsilon}_{Ra}),
\label{163}
\end{equation}

\noindent where we have put $m_i=m$, because now there is no ambiguity with the
uppest positive branch of the mass spectrum.

By using Equations
(\ref{161}), (\ref{163}), and our gauge-fixings, the analogue of
Eq.(\ref{160}) is [the Jacobian $\tilde J$ is to be fixed by comparison with
Eq,(\ref{159})]

\begin{eqnarray}
&&Q_N(V^{\mu},\beta^{\mu})={1\over {N!}}\int_{V^{\mu}}\prod_{i=1}^N e^{-\beta
_{\mu}p_i^{\mu}} \theta (p_i^o)\delta (p_i^2-m^2)\delta (\hat T(x_k,p_k)-\tau )
\prod_{a=1}^{N-1}\delta ({\hat T}_{Ra}(x_k,p_k)) \nonumber \\
&&\prod_{i=1}^N{{d^4x_id^4p_i}
\over {(2\pi )^3}}=\nonumber \\
&&={1\over {N!}}{{\tilde J}\over {(2\pi )^{3N}}} \int_{V^{\mu}} \delta
(\epsilon -\sum_{i=1}^N\sqrt{m^2+N{(\sum_{a=1}^{N-1}{\hat \gamma}_{ai}{\vec
\pi}_a)}^2}) \delta (\hat T-\tau ) \prod_{a=1}^{N-1}\delta ({\hat T}_{Ra})
\delta ({\hat \epsilon}_{Ra}) \prod_{i=1}^N e^{-\beta_{\mu}p_i^{\mu}}
\nonumber \\
&&d\epsilon d^3k d\hat T d^3\hat z \prod_{a=1}^{N-1} d{\hat T}_{Ra}d{\hat
\epsilon}_{Ra} d^3\rho_a d^3\pi_a=\nonumber \\
&&={1\over {N!}}{{\tilde J}\over {(2\pi )^{3N}}} \int_{V^{\mu}} d^3\hat z
d^3k \prod_{a=1}^{N-1}d^3\rho_a d^3\pi_a \prod_{i=1}^N e^{-\beta_{\mu}p_i
^{\mu}}.
\label{164}
\end{eqnarray}

\noindent What is lacking in the N-time theory is the evaluation of
$\beta_{\mu}
p_i^{\mu}$, because we do not know explicitely the inverse canonical
transformation.

Let us shift to the 1-time description on the rest-frame hyperplane: in the
rest
frame one had the constraints ${\vec \kappa}_{+}\approx 0$ and the
gauge-fixings
${\vec \eta}_{+}\approx 0$ [identifying the rest frame center of mass with
${\tilde x}^{\mu}_s$], so that $\sqrt{m^2+{\vec \kappa}_i^2}\approx \sqrt{m^2+
N{(\sum_{a=1}^{N-1}{\hat \gamma}_{ai}{\vec \pi}_a)}^2}$ and $\beta_{\mu}
p_i^{\mu}\approx \beta \sqrt{m^2+N{(\sum_{a=1}^{N-1}{\hat \gamma}_{ai}{\vec
\pi}_a)}^2}$ due to Eq.(\ref{80}). But if we look at the rest-frame hyperplane
from an arbitrary frame in Minkowski spacetime, the Minkowski canonical
3-position and 3-momentum of the center of mass are ${\vec z}_s/\epsilon_s$
and $\epsilon_s{\vec k}_s$ respectively with $d^3z_sd^3k_s=(d^3z_s/{|\,
\epsilon_s|}^3)({|\, \epsilon_s|}^3d^3k_s$) [it is a Lorentz scalar
\cite{longhi}: $d^3z_sd^3k_s=(\sqrt{1+{\vec k}_s^2}d^3z_s)(d^3k_s/\sqrt{1+
{\vec k}_s^2})$] and we would like to identify $d^3{\hat z}d^3k=d^3z_sd^3k_s$
with the Lorentz scalar one $d^3\eta_{+}d^3\kappa_{+}$. In this way ${\vec
\eta}
_{+}$ and ${\vec \kappa}_{+}$ would be Wigner spin-1 quantities simulating on
the rest-frame hyperplane [relaxing the constraints ${\vec \kappa}_{+}
\approx 0$] the motion of the three-dimensional center of mass in an arbitrary
Minkowski frame: the 3-coordinates and the 3-momenta of the particles in this
descritpion would be the Wigner spin-1 3-vectors ${\vec \eta}_i$ and ${\vec
\kappa}_i$ in Eq.(\ref{74}).

This suggests the replacement of the unknown $\beta_{\mu}p_i^{\mu}$ with
$\beta \sqrt{m^2+{\vec \kappa}_i^2}$, which is a Lorentz scalar [$\beta$ is
the rest-frame reciprocal temperature]. With this prescription, by using
$d^3\eta_{+}d^3\kappa_{+}\prod_{a=1}^{N-1}d^3\rho_ad^3\pi_a=\prod_{i=1}^N
d^3\eta_id^3\kappa_i$ and by replacing $V^{\mu}$ with its rest-frame value
$V=\int_Vd^3\eta_i$, Eq.(\ref{164}) becomes

\begin{eqnarray}
Q_N(V,\beta )&=&{1\over {N!}}{{\tilde J}\over {(2\pi )^{3N}}}\prod_{i=1}^N
\int_V d^3\eta_id^3\kappa_i e^{-\beta \sqrt{m^2+{\vec \kappa}_i^2}}=
\nonumber \\
&=&{{\tilde J}\over {N!}} \prod_{i=1}^N [ {V\over {(2\pi )^3}} \int d^3\kappa_i
e^{-\beta \sqrt{m^2+{\vec \kappa}_i^2}} ]=\nonumber \\
&=&{1\over {N!}}{{\tilde J}\over {2^N}} [ {{2V}\over {(2\pi )^3}} \int
d^3\kappa
e^{-\beta \sqrt{m^2+{\vec \kappa}^2_i}} ]{}^N=\nonumber \\
&=&{1\over {N!}}{{\tilde J}\over {2^N}} [ {{Vm^2}\over {2\pi^2\beta}}
K_2(m\beta )]{}^N.
\label{165}
\end{eqnarray}

Therefore, if we choose $\tilde J=2^N$, we get the following definition of the
Lorentz scalar canonical partition function of an ideal Boltzmann gas in the
rest-frame instant form

\begin{equation}
Q_N(V,\beta )={1\over {N!}}({2\over {(2\pi )^3}}){}^N \int_V \prod_{i=1}^N
d^3\eta_id^3\kappa_i e^{-\beta \sqrt{m^2+{\vec \kappa}_i^2}}.
\label{166}
\end{equation}

If, following Ref.\cite{karsch}, we consider a relativistic gas formed by N
2-body bound states defined by the following 2-body reduced Hamiltonian
[obtained from Eq.(\ref{111}) with the previous prescriptions and where
${\vec R}_{12}={\vec \eta}_1-{\vec \eta}_2$]

\begin{equation}
H_R=\sqrt{m_1^2+V_1({\vec R}_{12}^2)+{\vec \kappa}_1^2}+
\sqrt{m_2^2+V_2({\vec R}_{12}^2)+{\vec \kappa}_2^2}+U({\vec R}_{12}^2)'
\label{167}
\end{equation}

\noindent then the definition of the Lorentz-scalar canonical partition
function is [${\vec \eta}_{12}={1\over 2}({\vec \eta}_1+{\vec \eta}_2)$]

\begin{eqnarray}
Q_N(V,\beta )&=&{1\over {(2\pi )^{6N} N!}} [\int_Ve^{-\beta H_R} d^3\eta_1
d^3\eta_2d^3\kappa_1d^3\kappa_2 ]{}^N=\nonumber \\
&=&{1\over {(2\pi )^{6N} N!}} [\int_Ve^{-\beta H_R} d^3\eta_{12}d^3R_{12}
d^3\kappa_1d^3\kappa_2 ]{}^N=\nonumber \\
&=&{{V^N}\over {(2\pi )^{6N} N!}} [\int_Ve^{-\beta H_R} d^3R_{12}d^3\kappa_1
d^3\kappa_2 ]{}^N.
\label{168}
\end{eqnarray}

Clearly this can be extended to more complicated situations. When a better
understanding of the gauge-fixings for the case of charged particles
interacting with the electromagnetic field will be obtained, Eqs.(\ref{146})
will serve to define the covariant analogue of the canonical partition function
associated with the noncovariant Hamiltonian of Eq.(\ref{155}).
Finally, Feynman path integral approach should now be based on the continuation
of the scalar rest-frame time $T_s$ to imaginary values $-i\hbar /kT$.

\vfill\eject

\section{Conclusions}

In this paper a covariant 1-time formulation, the rest-frame instant form,
of classical relativistic dynamics has been developed for those isolated
systems
of both N scalar particles with action-at-a-distance interactions and of
charged scalar particles plus the electromagnetic field, which belong to
timelike irreducible Poincar\'e representations. While with only particles
the kinematics is completely understood, in the case of particles plus fields
more work will be needed to identify which point of the system has to be
identified with the canonical 3-position of the center of mass on the
spacelike hyperplane orthogonal to the total momentum, due to the presence
of the interaction term in the constraints ${\hat {\vec {\cal H}}}_p\approx 0$
of Eqs.(\ref{146}). This problem is non trivial and is connected with the
action-reaction problem and with the
classical basis of the electromagnetic interaction vertex of a scalar particle.
In turn this point is associated with the pseudoclassical regularization of
the classical electromagnetic self-energies of charged particles. Again more
work is needed to clarify the meaning of the pseudoclassical theory with
Dirac observables where the effects of order $e_i^2$ are absent and only the
$e_ie_j$, $i\not= j$, effects are present: one has to compare the theory
with Dirac observables to the standard methods of separation of the Coulomb
and radiation effects in the Lienard-Wiechert potentials\cite{teitel,grandy},
to revisit the Abraham-Lorentz-Dirac equation and the problem of classical
bound states (Eqs.(\ref{148}) define a regularized bound system with Coulomb
interactions without emission of radiation and without fall on the centre).

For free particles a quasi-Shanmugadhasan canonical basis has been found, which
is suited to the description of the relativistic Cauchy problem for bound
states, because the relative energies and times can be explicitely eliminated
in a covariant way. It is under investigation\cite{luc} how to introduce the
second Poincar\'e Casimir $W^2=-p^2{\vec {\bar S}}^2$ of timelike Poincar\'e
representations in the basis and how to present in a clear form the gauge
character of the classical zitterbewegung. The next step will be the
introduction of the spin 1/2 degrees of freedom by means of Grassmann
variables\cite{casal,crater} both in the N- and 1-time theories and the
comparison with the results coming from the Dirac equation. A further
development will be to replace the electromagnetic field with a
non-Abelian Yang-Mills one and the search of the associated interparticle
potential.

When the electromagnetic field is considered, the 1-time theory shows clearly
the existence of a new formulation of the classical theory of free linear
fields, i.e. the rest-frame field theory based on a center-of-mass and relative
variables decomposition of field configurations. The formulation of
rest-frame field theory, which is under active investigation, will be the
necessary tool for attempting a quantization of the nonlocal field theories
with Dirac's observables, for utilizing the ultraviolet cutoff induced by the
noncovariance of the canonical center-of-mass 3-position and for a fresh
start to the problem of putting relativistic bound states among its
Tomonaga-Schwinger-like asymptotic states.

Furthermore, we have to understand the quantization of the 1-time covariant
relativistic mechanics in the rest-frame instant form and to compare it with
relativistic quantum mechanics [see the rich bibliography Refs.\cite{lev,kei}
for the status of this theory,which also originated from Ref.\cite{diracc}],
which is used in low energy (below the pion mass threshold) nuclear physics,
where at the order $1/c^2$ one can define a theory with a fixed number of
nucleons.

On the other hand, 1-time covariant relativistic statistical mechanics should
be developed to a stage to be compared with the existing applications of
relativistic statistical mechanics to relativistic (either astrophysical or
nuclear) plasmas\cite{hakimb}.

In particular one should revisit the case of charged particles plus the
electromagnetic field in the pseudoclassical approach with Grassmann charges.
One should obtain simplifications of the transport equations evaluated by
taking into account the Abraham-Lorentz-Dirac equation with its dependence
on higher accelerations\cite{hakim}. More in general, it is still completely
open the problem of how to get a Hamiltonian formulation of the
integro-differential equations of motion coming from the
Tetrode-Fokker-Feynman-Wheeler actions, in which the radiation electromagnetic
degrees of freedom have been eliminated: these actions admit reparametrization
invariance of each particle worldline notwithstanding the interaction, are
supposed to replace the nonexisting (due to the No-Interaction-Theorem)
Lagrangians for predictive mechanics\cite{pons} and are formally equivalent to
an infinite system of differential eqations for accelerations of every order,
whose study with constraint theory is just at the beginning\cite{gaida}.

\acknowledgments

I wish to thank Dr.R.De Pietri for let me know Ref.\cite{ku}.

\vfill\eject

\appendix
\section{}

The rest frame form of the timelike fourvector $p^{\mu}$ is $\stackrel
{\circ}{p}{}^{\mu}=\eta \sqrt{p^2} (1;\vec 0)= \eta^{\mu o}\eta \sqrt{p^2}$,
$\stackrel{\circ}{p}{}^2=p^2$, where $\eta =sign\, p^o$.

The standard Wigner boost transforming $\stackrel{\circ}{p}{}^{\mu}$ into
$p^{\mu}$ is

\begin{eqnarray}
L^{\mu}{}_{\nu}(p,\stackrel{\circ}{p})&=&\epsilon^{\mu}_{\nu}(u(p))=
\nonumber \\
&=&\eta^{\mu}_{\nu}+2{ {p^{\mu}{\stackrel{\circ}{p}}_{\nu}}\over {p^2}}-
{ {(p^{\mu}+{\stackrel{circ}{p}}^{\mu})(p_{\nu}+{\stackrel{\circ}{p}}_{\nu})}
\over {p\cdot \stackrel{\circ}{p} +p^2} }=\nonumber \\
&=&\eta^{\mu}_{\nu}+2u^{\mu}(p)u_{\nu}(\stackrel{\circ}{p})-{ {(u^{\mu}(p)+
u^{\mu}(\stackrel{\circ}{p}))(u_{\nu}(p)+u_{\nu}(\stackrel{\circ}{p}))}
\over {1+u^o(p)} }\nonumber \\
&&{} \nonumber \\
\nu =0 &&\epsilon^{\mu}_o(u(p))=u^{\mu}(p)=p^{\mu}/\eta \sqrt{p^2} \nonumber \\
\nu =r &&\epsilon^{\mu}_r(u(p))=(-u_r(p); \delta^i_r-{ {u^i(p)u_r(p)}\over
{1+u^o(p)} }).
\label{A1}
\end{eqnarray}

The inverse of $L^{\mu}{}_{\nu}(p,\stackrel{\circ}{p})$ is $L^{\mu}{}_{\nu}
(\stackrel{\circ}{p},p)$, the standard boost to the rest frame, defined by

\begin{equation}
L^{\mu}{}_{\nu}(\stackrel{\circ}{p},p)=L_{\nu}{}^{\mu}(p,\stackrel{\circ}{p})=
L^{\mu}{}_{\nu}(p,\stackrel{\circ}{p}){|}_{\vec p\rightarrow -\vec p}.
\label{A2}
\end{equation}

Therefore, we can define the following vierbeins [the
$\epsilon^{\mu}_r(u(p))$'s
are also called polarization vectors; the indices r, s will be used for A=1,2,3
and $\bar o$ for A=0]

\begin{eqnarray}
&&\epsilon^{\mu}_A(u(p))=L^{\mu}{}_A(p,\stackrel{\circ}{p})\nonumber \\
&&\epsilon^A_{\mu}(u(p))=L^A{}_{\mu}(\stackrel{\circ}{p},p)=\eta^{AB}\eta
_{\mu\nu}\epsilon^{\nu}_B(u(p))\nonumber \\
&&{} \nonumber \\
&&\epsilon^{\bar o}_{\mu}(u(p))=\eta_{\mu\nu}\epsilon^{\nu}_o(u(p))=u_{\mu}(p)
\nonumber \\
&&\epsilon^r_{\mu}(u(p))=-\delta^{rs}\eta_{\mu\nu}\epsilon^{\nu}_r(u(p))=
(\delta^{rs}u_s(p);\delta^r_j-\delta^{rs}\delta_{jh}{{u^h(p)u_s(p)}\over
{1+u^o(p)} })\nonumber \\
&&\epsilon^A_o(u(p))=u_A(p),
\label{A3}
\end{eqnarray}

\noindent which satisfy

\begin{eqnarray}
&&\epsilon^A_{\mu}(u(p))\epsilon^{\nu}_A(u(p))=\eta^{\mu}_{\nu}\nonumber \\
&&\epsilon^A_{\mu}(u(p))\epsilon^{\mu}_B(u(p))=\eta^A_B\nonumber \\
&&\eta^{\mu\nu}=\epsilon^{\mu}_A(u(p))\eta^{AB}\epsilon^{\nu}_B(u(p))=u^{\mu}
(p)u^{\nu}(p)-\sum_{r=1}^3\epsilon^{\mu}_r(u(p))\epsilon^{\nu}_r(u(p))
\nonumber \\
&&\eta_{AB}=\epsilon^{\mu}_A(u(p))\eta_{\mu\nu}\epsilon^{\nu}_B(u(p))\nonumber
\\
&&p_{\alpha}{{\partial}\over {\partial p_{\alpha}} }\epsilon^{\mu}_A(u(p))=
p_{\alpha}{{\partial}\over {\partial p_{\alpha}} }\epsilon^A_{\mu}(u(p))
=0.
\label{A4}
\end{eqnarray}

The Wigner rotation corresponding to the Lorentz transformation $\Lambda$ is

\begin{eqnarray}
R^{\mu}{}_{\nu}(\Lambda ,p)&=&{[L(\stackrel{\circ}{p},p)\Lambda^{-1}L(\Lambda
p,\stackrel{\circ}{p})]}^{\mu}{}_{\nu}=\left(
\begin{array}{cc}
1 & 0 \\
0 & R^i{}_j(\Lambda ,p)
\end{array}
\right) \nonumber \\
{} && {}\nonumber \\
R^i{}_j(\Lambda ,p)&=&{(\Lambda^{-1})}^i{}_j-{ {(\Lambda^{-1})^i{}_op_{\beta}
(\Lambda^{-1})^{\beta}{}_j}\over {p_{\rho}(\Lambda^{-1})^{\rho}{}_o+\eta
\sqrt{p^2}} }-\nonumber \\
&-&{{p^i}\over {p^o+\eta \sqrt{p^2}} }[(\Lambda^{-1})^o{}_j- {
{((\Lambda^{-1})^o
{}_o-1)p_{\beta}(\Lambda^{-1})^{\beta}{}_j}\over {p_{\rho}(\Lambda^{-1})^{\rho}
{}_o+\eta \sqrt{p^2}} }].
\label{A5}
\end{eqnarray}

The polarization vectors and the variables of Eq.(\ref{13}) transform under the
Poincar\'e transformations $(a,\Lambda )$ in the following way

\begin{eqnarray}
\epsilon^{\mu}_r(u(\Lambda p))&=&(R^{-1})_r{}^s\, \Lambda^{\mu}{}_{\nu}\,
\epsilon^{\nu}_s(u(p))\nonumber \\
&& {}\nonumber \\
p^{'\mu}&=&\Lambda^{\mu}{}_{\nu} p^{\nu}\nonumber \\
{\tilde x}^{'\mu}&=&\Lambda^{\mu}{}_{\nu}[{\tilde x}^{\nu}+{1\over 2}{\bar S}
_{rs}R^r{}_k(\Lambda ,p){{\partial}\over {\partial p_{\nu}} }R^s{}_k(\Lambda ,
p)]+a^{\mu}=\nonumber \\
&=&\Lambda^{\mu}{}_{\nu}\{ {\tilde x}^{\nu}+{ {{\bar S}_{rs}}\over {\Lambda^o
{}_{\alpha}p^{\alpha}+\eta \sqrt{p^2}} }[\eta^{\nu}_r(\Lambda^o{}_s-
{ {(\Lambda^o{}_o-1)p_s}\over {p^o+\eta \sqrt{p^2}} })-\nonumber \\
&-&{ {(p^{\nu}+\eta^{\nu}_o\eta \sqrt{p^2})p_r\Lambda^o{}_s}\over
{\eta \sqrt{p^2}(p^o+\eta \sqrt{p^2})} }] \} +a^{\mu}\nonumber \\
T^{'}_{Ra}&=&T_{Ra}\nonumber \\
\epsilon^{'}_{Ra}&=&\epsilon_{Ra}\nonumber \\
\rho^{'}_{ar}&=&\rho_{as}R^s{}_r(\Lambda ,p)\nonumber \\
\pi^{'}_{ar}&=&\pi_{as}R^s{}_r(\Lambda ,p).
\label{A6}
\end{eqnarray}

Therefore, ${\tilde x}^{\mu}$ is not a fourvector and ${\vec \rho}_a, {\vec
\pi}_a$ are Wigner spin-1 3-vectors; their infinitesimal transformation
properties under Lorentz transformations generated by $J^{\mu\nu}={\tilde L}
^{\mu\nu}+{\tilde S}^{\mu\nu}$ of Eq.(\ref{14}), are

\begin{eqnarray}
\lbrace {\tilde x}^{\mu},J^{\alpha\beta}\rbrace &=&\eta^{\mu\alpha}{\tilde x}
^{\beta}-\eta^{\mu\beta}{\tilde x}^{\alpha}+\lbrace {\tilde x}^{\mu},
{\tilde S}^{\alpha\beta}\rbrace \nonumber \\
&&\lbrace {\tilde x}^{\mu},{\tilde S}^{oi}\rbrace =-{1\over {p^o+\eta
\sqrt{p^2}} }[\eta^{\mu j}{\tilde S}^{ji}+{ {(p^{\mu}+\eta^{\mu o}\eta
\sqrt{p^2}){\tilde S}^{ik}p^k}\over {\eta \sqrt{p^2}(p^o+\eta \sqrt{p^2})} }]
\nonumber \\
&&\lbrace {\tilde x}^{\mu},{\tilde S}^{ij}\rbrace =0\nonumber \\
\lbrace \rho^r_a,J^{oi}\rbrace &=&-{ {\delta^{is}(p^r\rho^s_a-p^s\rho^r_a)}
\over {p^o+\eta \sqrt{p^2}} }\nonumber \\
\lbrace \rho^r_a,J^{ij}\rbrace &=&\delta^{is}\delta^{jt}\lbrace \rho^r_a,{\bar
S}^{st}\rbrace =(\delta^{is}\delta^{jr}-\delta^{ir}\delta^{js})\rho^s_a
\nonumber \\
\lbrace \pi^r_a,J^{oi}\rbrace &=&-{ {\delta^{is}(p^r\pi^s_a-p^s\pi^r_a)}
\over {p^o+\eta \sqrt{p^2}} }\nonumber \\
\lbrace \pi^r_a,J^{ij}\rbrace
&=&(\delta^{is}\delta^{jr}-\delta^{ir}\delta^{js})
\pi^s_a.
\label{A7}
\end{eqnarray}

In contrast, the variables of Eqs.(\ref{22}) transform under the Poincar\'e
transformations $(a,\Lambda )$ in the following way

\begin{eqnarray}
&&T^{'}=T+k_{\mu}{(\Lambda^{-1}a)}^{\mu}\nonumber \\
&&\epsilon^{'}=\epsilon \nonumber \\
&&z^{{'}i}=(\Lambda^i{}_j-{ {\Lambda^i{}_{\mu}k^{\mu}}\over {\Lambda^o{}_{\nu}
k^{\nu}} }\Lambda^o{}_j)+\epsilon (\Lambda^i{}_{\mu}-{ {\Lambda^i{}_{\nu}
k^{\nu}}\over {\Lambda^o{}_{\rho}k^{\rho}} }\Lambda^o{}_{\mu}){(\Lambda^{-1}a
)}^{\mu}\nonumber \\
&&k^{{'}\mu}=u^{\mu}(p^{'})=\Lambda^{\mu}{}_{\nu}k^{\nu}\nonumber \\
&&{}\nonumber \\
&&\lbrace T,p^o\rbrace =-\sqrt{1+{\vec k}^2},\quad\quad \lbrace T,p^i\rbrace =
-k^i\nonumber \\
&&\lbrace k^h,J^{oi}\rbrace =-\delta^{hi}\sqrt{1+{\vec k}^2},\quad\quad
\lbrace k^h,J^{ij}\rbrace =\delta^{hj}k^i-\delta^{hi}k^j\nonumber \\
&&\lbrace z^h,p^o\rbrace ={{\epsilon k^h}\over {\sqrt{1+{\vec k}^2}} },\quad
\quad \lbrace z^h,p^i\rbrace =\delta^{hi}\epsilon \nonumber \\
&&\lbrace z^h,J^{oi}\rbrace ={{z^ik^h}\over {\sqrt{1+{\vec k}^2}} },\quad\quad
\lbrace z^h,J^{ij}\rbrace =\delta^{hj}z^i-\delta^{hi}z^j.
\label{A8}
\end{eqnarray}

Some further useful formulas are ($\epsilon =\eta \sqrt{p^2}$)

\begin{eqnarray}
{{\partial}\over {\partial p_{\mu}} }&\epsilon^B_{\rho}&(u(p))=
{{\partial}\over {\partial p_{\mu}} }L^B{}_{\rho}(\stackrel{\circ}{p},p)=
\nonumber \\
&=&{2\over {\epsilon}}\eta^B_o(\eta^{\mu}_{\rho}-{{p^{\mu}p_{\rho}}\over
{\epsilon^2}})-{1\over {\epsilon^2(p^o+\epsilon)}}[(p^B+\epsilon \eta^B_o)
(p^{\mu}\eta^o_{\rho}+\epsilon \eta^{\mu}_{\rho})+(p^{\mu}\eta^B_o+
\epsilon \eta^{\mu B})(p_{\rho}+\epsilon \eta^o_{\rho})]+\nonumber \\
&+&{ {(p^B+\epsilon \eta^B_o)(p_{\rho}+\epsilon \eta^o_{\rho})[(p^o+2\epsilon )
p^{\mu}+\epsilon^2\eta^{\mu}_o]}\over {\epsilon^3{(p^o+\epsilon )}^2} },
\label{A9}
\end{eqnarray}

\begin{eqnarray}
{1\over 2}\epsilon^{\nu}_A(u(p))\eta^{AB}&&{{\partial \epsilon^{\rho}_B(u(p))}
\over {\partial p_{\mu}} }S_{\rho\nu}=-{1\over 2}\eta_{AC}{{\partial \epsilon
^C_{\rho}(u(p))}\over {\partial p_{\mu}} }\epsilon^{\rho}_B(u(p)){\bar S}
^{AB}=\nonumber \\
&=&-{1\over {\epsilon}}[\eta^{\mu}_A({\bar S}^{\bar oA}-{{{\bar S}^{Ar}p^r}
\over {p^o+\epsilon}})+{\bar S}^{\bar or}p^r{{p^{\mu}+2\epsilon \eta^{\mu}_o}
\over {\epsilon (p^o+\epsilon )}}]=\nonumber \\
&=&-{1\over {\epsilon (p^o+\epsilon )}}[p_{\rho}S^{\rho\mu}+\epsilon (S^{o\mu}
-S^{o\rho}{{p_{\rho}p^{\mu}}\over {\epsilon^2}})],
\label{A10}
\end{eqnarray}

\begin{eqnarray}
{{\partial}\over {\partial p_{\nu}} }&R^i{}_k&(\Lambda ,p)={{(\Lambda^{-1})^i
{}_o}\over {(p_{\rho}(\Lambda^{-1})^{\rho}{}_o+\epsilon )^2}}\times \nonumber
\\
&&\times [{1\over
{\epsilon}}(p^{\nu}+\epsilon (\Lambda^{-1})^{\nu}{}_o)p_{\beta}(\Lambda^{-1})
^{\beta}{}_k-(p_{\rho}(\Lambda^{-1})^{\rho}{}_o+\epsilon )(\Lambda^{-1})
^{\nu}{}_k]-\nonumber \\
&-&{{\eta^{i\nu}}\over {p^o+\epsilon}}[(\Lambda^{-1})^o{}_k-{ {((\Lambda^{-1})
^o{}_o-1)p_{\beta}(\Lambda^{-1})^{\beta}{}_k}\over {p_{\rho}(\Lambda^{-1})
^{\rho}{}_o+\epsilon} }]+{{p^i}\over {(p^o+\epsilon )^2}}\{ {1\over
{\epsilon}}(p^{\nu}+\epsilon \eta^{\nu o})(\Lambda^{-1})^o{}_k+
\nonumber \\
&+& {{p^o+\epsilon}\over {p_{\rho}(\Lambda^{-1})^{\rho}{}_o+\epsilon}}
((\Lambda^{-1})^o{}_o-1)\times \nonumber \\
&&\times [(\Lambda^{-1})^{\nu}{}_k-{1\over {\epsilon}}
({{p^{\nu}+\epsilon \eta^{\nu o}}\over {p^o+\epsilon}}+{{p^{\nu}+
\epsilon (\Lambda^{-1})^{\nu}{}_o}\over {p_{\rho}(\Lambda^{-1})^{\rho}{}_o+
\epsilon}})p_{\beta}(\Lambda^{-1})^{\beta}{}_k]\} .
\label{A11}
\end{eqnarray}

\vfill\eject

\section{}

In the N=2 case\cite{longhi} Eqs.(\ref{6}), (\ref{9}), (\ref{13}), (\ref{26}),
(\ref{28}), (\ref{31}) become
respectively

\begin{eqnarray}
&&x^{\mu}={1\over 2}(x^{\mu}_1+x^{\mu}_2)\nonumber \\
&&p^{\mu}=p^{\mu}_1+p^{\mu}_2\nonumber \\
&&R^{\mu}=x^{\mu}_1-x^{\mu}_2\nonumber \\
&&Q^{\mu}={1\over 2}(p^{\mu}_1-p^{\mu}_2);
\label{B1}
\end{eqnarray}

\begin{eqnarray}
&&x^{\mu}_1=x^{\mu}+{1\over 2}R^{\mu}\nonumber \\
&&x^{\mu}_2=x^{\mu}-{1\over 2}R^{\mu}\nonumber \\
&&p^{\mu}_1={1\over 2}p^{\mu}+Q^{\mu}\nonumber \\
&&p^{\mu}_2={1\over 2}p^{\mu}-Q^{\mu};
\label{B2}
\end{eqnarray}

\begin{eqnarray}
&&{\tilde x}^{\mu}=x^{\mu}+{1\over 2}\, \epsilon^A_{\nu}(u(p))\eta_{AB}
{ {\partial \epsilon^B_{\rho}(u(p))}\over {\partial p_{\mu}} }\, S^{\nu\rho}
=\nonumber \\
&&=x^{\mu}-{ 1\over {\eta \sqrt{p^2}(p^o+\eta \sqrt{p^2})} }\, [p_{\nu}
S^{\nu\mu}+\eta \sqrt{p^2} (S^{o\mu}-S^{o\nu}{ {p_{\nu}p^{\mu}}\over {p^2} })]
\nonumber \\
&&{}\nonumber \\
&&p^{\mu}\nonumber \\
&&{}\nonumber \\
&&T_{R}={ {p\cdot R}\over {\eta \sqrt{p^2}} }
\nonumber \\
&&\epsilon_{R}=
{ {p\cdot Q}\over {\eta \sqrt{p^2}} }\nonumber \\
&&\rho^r=
R^r-{ {p^r}\over {\eta \sqrt{p^2}} }
(R^o-{ {{\vec p}\cdot {{\vec R}} }\over {p^o+\eta \sqrt{p^2}} })
\nonumber \\
&&\pi^r=Q^r-{ {p^r}\over {\eta \sqrt{p^2}} }
(Q^o-{ {{\vec p}\cdot {{\vec Q}} }\over {p^o+\eta \sqrt{p^2}} }),
\label{B3}
\end{eqnarray}

\begin{eqnarray}
\chi_{-}&=&{1\over 2}(\phi_1-\phi_2)=p\cdot Q-{1\over 2}(m_1^2-m_2^2)=\epsilon
\epsilon_R-{1\over 2}(m^2_1-m_2^2)\approx 0\nonumber \\
\chi_{+}&=&2(\phi_1+\phi_2)={1\over {p^2}}(p^2-M^2_{+})(p^2-M^2_{-})+4\chi_{-}
(\chi_{-}+m_1^2-m_2^2)=\nonumber \\
&=&{1\over {\epsilon^2}}(\epsilon^2-M^2_{+})(\epsilon^2-M^2_{-})+4\chi_{-}
(\chi_{-}+m_1^2-m_2^2)\approx 0\nonumber \\
&&M_{\pm}=\sqrt{m_1^2-Q^2_{\perp}}\pm\sqrt{m_2^2-Q^2_{\perp}}=\sqrt{m_1^2+
{\vec \pi}^2}\pm \sqrt{m_2^2+{\vec \pi}^2},\nonumber \\
&&Q^{\mu}_{\perp}=
(\eta^{\mu\nu}-{{p^{\mu}p^{\nu}}\over {p^2}})Q_{\nu};
\label{B4}
\end{eqnarray}

\begin{eqnarray}
&&{\hat x}^{\mu}=x^{\mu}+{ {m^2_1-m^2_2}\over {2p^2}}(\eta^{\mu\nu}-
{{2p^{\mu}p^{\nu}}\over {p^2}})R_{\nu}\nonumber \\
&&p^{\mu}\nonumber \\
&&{\hat R}^{\mu}=R^{\mu}\nonumber \\
&&{\hat Q}^{\mu}=Q^{\mu}-{{m_1^2-m_2^2}\over {2p^2}}p^{\mu},\quad\quad
{\hat Q}^{\mu}_{\perp}=Q^{\mu}_{\perp};
\label{B5}
\end{eqnarray}

\begin{eqnarray}
&&{\hat {\tilde x}}={\hat x}^{\mu}+{1\over 2}\, \epsilon^A_{\nu}(u(p))
\eta_{AB}{ {\partial \epsilon^B_{\rho}(u(p))}\over {\partial p_{\mu}} }\,
{\hat S}^{\nu\rho}=\nonumber \\
&&={\hat x}^{\mu}-{ 1\over {\eta \sqrt{p^2}(p^o+\eta \sqrt{p^2})} }\, [p_{\nu}
{\hat S}^{\nu\mu}+\eta \sqrt{p^2} ({\hat S}^{o\mu}-{\hat S}^{o\nu}
{ {p_{\nu}p^{\mu}}\over {p^2} })]\nonumber \\
&&p^{\mu}\nonumber \\
&&{\hat T}_R=T_R\nonumber \\
&&{\hat \epsilon}_R=\epsilon_R-{{m_1^2-m_2^2}\over {2\epsilon}}\quad
\Rightarrow \chi_{-}=\epsilon {\hat \epsilon}_R\approx 0\nonumber \\
&&{\hat {\vec \rho}}=\vec \rho \nonumber \\
&&{\hat {\vec \pi}}=\vec \pi .
\label{B6}
\end{eqnarray}

Let us remark that Todorov's quasipotential-like Hamiltonian\cite{todo,todor}
corresponds to the following rewriting of the constraint $\chi_{+}$ (modulo
$\chi_{-}\approx 0$):

\begin{eqnarray}
\chi_{+}&\approx& 4[{({\hat Q}^{\mu}_{\perp}+\epsilon_p{{p^{\mu}}\over
{\epsilon}})}^2-m_p^2]=-4[{\vec \pi}^2-b^2(\epsilon^2)]\approx 0
\nonumber \\
&&{}\nonumber \\
&&m_p={{m_1m_2}\over {\epsilon}},\quad\quad \epsilon_p={{\epsilon^2-m_1^2-
m_2^2}\over {2\epsilon}}\nonumber \\
&&b^2(\epsilon^2)={{\epsilon^4+m_1^4+m_2^4-2(m_1^2+m_2^2)\epsilon^2-2m_1^2m_2^2}
\over {4\epsilon^2}}=\epsilon^2_p-m_p^2;
\label{B7}
\end{eqnarray}

\noindent $m_p$ and $\epsilon_p$ are interpreted as the relativistic reduced
mass and energy of a fictitious particle of relative motion.

\vfill\eject

\section{}

Let $\lbrace \Sigma (\tau )\rbrace$ be a one-parameter family of spacelike
hypersurfaces foliating Minkowski spacetime $M^4$. At fixed $\tau$, let
$z^{\mu}(\tau ,\vec \sigma )$ be the coordinates of the points on $\Sigma
(\tau )$ in $M^4$, $\lbrace \vec \sigma \rbrace$ a system of coordinates on
$\Sigma (\tau )$. If $\sigma^{\check A}=(\sigma^{\tau}=\tau ;\vec \sigma
=\lbrace \sigma^{\check r}\rbrace)$ [the notation ${\check A}=(\tau ,
{\check r})$ with ${\check r}=1,2,3$ will be used; note that ${\check A}=
\tau$ and ${\check A}={\check r}=1,2,3$ are Lorentz-scalar indices] and
$\partial_{\check A}=\partial /\partial \sigma^{\check A}$,
one can define the vierbeins

\begin{equation}
z^{\mu}_{\check A}(\tau ,\vec \sigma )=\partial_{\check A}z^{\mu}(\tau ,\vec
\sigma ),\quad\quad
\partial_{\check B}z^{\mu}_{\check A}-\partial_{\check A}z^{\mu}_{\check B}=0,
\label {C1}
\end{equation}

\noindent so that the metric on $\Sigma (\tau )$ is

\begin{eqnarray}
&&g_{{\check A}{\check B}}(\tau ,\vec \sigma )=z^{\mu}_{\check A}(\tau ,\vec
\sigma )\eta_{\mu\nu}z^{\nu}_{\check B}(\tau ,\vec \sigma ),\quad\quad
g_{\tau\tau}(\tau ,\vec \sigma ) > 0\nonumber \\
&&g(\tau ,\vec \sigma )=-det\, ||\, g_{{\check A}{\check B}}(\tau ,\vec
\sigma )\, || ={(det\, ||\, z^{\mu}_{\check A}(\tau ,\vec \sigma )\, ||)}^2
\nonumber \\
&&\gamma (\tau ,\vec \sigma )=-det\, ||\, g_{{\check r}{\check s}}(\tau ,\vec
\sigma )\, ||.
\label{C2}
\end{eqnarray}

If $\gamma^{{\check r}{\check s}}(\tau ,\vec \sigma )$ is the inverse of the
3-metric $g_{{\check r}{\check s}}(\tau ,\vec \sigma )$ [$\gamma^{{\check r}
{\check u}}(\tau ,\vec \sigma )g_{{\check u}{\check s}}(\tau ,\vec
\sigma )=\delta^{\check r}_{\check s}$], the inverse $g^{{\check A}{\check B}}
(\tau ,\vec \sigma )$ of $g_{{\check A}{\check B}}(\tau ,\vec \sigma )$
[$g^{{\check A}{\check C}}(\tau ,\vec \sigma )g_{{\check c}{\check b}}(\tau ,
\vec \sigma )=\delta^{\check A}_{\check B}$] is given by

\begin{eqnarray}
&&g^{\tau\tau}(\tau ,\vec \sigma )={{\gamma (\tau ,\vec \sigma )}\over
{g(\tau ,\vec \sigma )}}\nonumber \\
&&g^{\tau {\check r}}(\tau ,\vec \sigma )=-[{{\gamma}\over g} g_{\tau {\check
u}}\gamma^{{\check u}{\check r}}](\tau ,\vec \sigma )\nonumber \\
&&g^{{\check r}{\check s}}(\tau ,\vec \sigma )=\gamma^{{\check r}{\check s}}
(\tau ,\vec \sigma )+[{{\gamma}\over g}g_{\tau {\check u}}g_{\tau {\check v}}
\gamma^{{\check u}{\check r}}\gamma^{{\check v}{\check s}}](\tau ,\vec \sigma
),
\label{C3}
\end{eqnarray}

\noindent so that $1=g^{\tau {\check C}}(\tau ,\vec \sigma )g_{{\check C}\tau}
(\tau ,\vec \sigma )$ is equivalent to

\begin{equation}
{{g(\tau ,\vec \sigma )}\over {\gamma (\tau ,\vec \sigma )}}=g_{\tau\tau}
(\tau ,\vec \sigma )-\gamma^{{\check r}{\check s}}(\tau ,\vec \sigma )
g_{\tau {\check r}}(\tau ,\vec \sigma )g_{\tau {\check s}}(\tau ,\vec \sigma ).
\label{C4}
\end{equation}

We have

\begin{equation}
z^{\mu}_{\tau}(\tau ,\vec \sigma )=(\sqrt{ {g\over {\gamma}} }l^{\mu}+
g_{\tau {\check r}}\gamma^{{\check r}{\check s}}z^{\mu}_{\check s})(\tau ,
\vec \sigma ),
\label{C5}
\end{equation}

\noindent and

\begin{eqnarray}
\eta^{\mu\nu}&=&z^{\mu}_{\check A}(\tau ,\vec \sigma )g^{{\check A}{\check B}}
(\tau ,\vec \sigma )z^{\nu}_{\check B}(\tau ,\vec \sigma )=\nonumber \\
&=&(l^{\mu}l^{\nu}+z^{\mu}_{\check r}\gamma^{{\check r}{\check s}}
z^{\nu}_{\check s})(\tau ,\vec \sigma ),
\label{C6}
\end{eqnarray}

\noindent where

\begin{eqnarray}
l^{\mu}(\tau ,\vec \sigma )&=&({1\over {\sqrt{\gamma}}
}\epsilon^{\mu}{}_{\alpha
\beta\gamma}z^{\alpha}_{\check 1}z^{\beta}_{\check 2}z^{\gamma}_{\check 3})
(\tau ,\vec \sigma )\nonumber \\
&&l^2(\tau ,\vec \sigma )=1,\quad\quad l_{\mu}(\tau ,\vec \sigma )z^{\mu}
_{\check r}(\tau ,\vec \sigma )=0,
\label{C7}
\end{eqnarray}

\noindent is the unit (future pointing) normal to $\Sigma (\tau )$ at
$z^{\mu}(\tau ,\vec \sigma )$.

For the volume element in Minkowski spacetime we have

\begin{eqnarray}
d^4z&=&z^{\mu}_{\tau}(\tau ,\vec \sigma )d\tau d^3\Sigma_{\mu}=d\tau (z^{\mu}
_{\tau}(\tau ,\vec \sigma )l_{\mu}(\tau ,\vec \sigma ))\sqrt{\gamma
(\tau ,\vec \sigma )}d^3\sigma=\nonumber \\
&=&\sqrt{g(\tau ,\vec \sigma )} d\tau d^3\sigma.
\label{C8}
\end{eqnarray}

Let us remark that according to the geometrical approach of Ref.\cite{ku}, by
using Eq.(\ref{C5}) in the form $z^{\mu}_{\tau}(\tau ,\vec \sigma )=N(\tau ,
\vec \sigma )l^{\mu}(\tau ,\vec \sigma )+N^{\check r}(\tau ,\vec \sigma )
z^{\mu}_{\check r}(\tau ,\vec \sigma )$ [with $N=\sqrt{g/\gamma}$ and
$N^{\check r}=g_{\tau \check s}\gamma^{\check s\check r}$ being the standard
lapse and shift functions, so that $g_{\tau \tau}=N^2+g_{\check r\check s}
N^{\check r}N^{\check s}, g_{\tau \check r}=g_{\check r\check s}N^{\check s},
g^{\tau \tau}=N^{-2}, g^{\tau \check r}=-n^{\check r}/N^2, g^{\check r\check
s}=\gamma^{\check r\check s}+{{N^{\check r}N^{\check s}}\over {N^2}}$],
we should write

\begin{equation}
A_{\tau }(\tau ,\vec \sigma )=
N(\tau ,\vec \sigma )A_l(\tau ,\vec \sigma )+N^{\check r}
(\tau ,\vec \sigma )A_{\check r}(\tau ,\vec \sigma )
\label{C9}
\end{equation}

\noindent and use $A_l
(\tau ,\vec \sigma )$ as the genuine field configuration variable independent
from the motion of the embedded hypersurface.

Then, by denoting $K_{\check r\check s}=K_{\check s\check r}=l_{\mu}\partial
_{\check r}\partial_{\check s}z^{\mu}$ and $\Gamma^{\check u}_{\check r\check
s}=z^{\check u}_{\mu}\partial_{\check r}\partial_{\check s}z^{\mu}={1\over 2}
\gamma^{\check u\check n}(\partial_{\check r}g_{\check s\check n}+\partial
_{\check s}g_{\check r\check n}-\partial_{\check n}g_{\check r\check s})$
the extrinsic curvature and the Christoffel symbols respectively of the
spacelike hypersurface (with metric $g_{\check r\check s}$) embedded in the
flat Minkowski spacetime, we get for the field strengths [``;" and ``$|$"
are the total and ordinary covariant derivatives]

\begin{eqnarray}
F_{\check r\check s}&=&A_{\check s;\check r}-A_{\check r;\check s}=
A_{\check s\, |\, \check r}-K_{\check s\check r}A_l-(A_{\check r\, |\,
\check s}-K_{\check r\check s}A_l)=\nonumber \\
&=&A_{\check s\, |\, \check r}-A_{\check r\, |\, \check s}=\partial_{\check r}
A_{\check s}-\Gamma^{\check n}_{\check r\check s}A_{\check n}-(\partial_{\check
s}A_{\check r}-\Gamma^{\check n}_{\check s\check r}A_{\check n})=
\partial_{\check r}A_{\check s}-\partial_{\check s}A_{\check r},\nonumber \\
F_{\tau \check r}&=&\partial_{\tau}A_{\check r}-\partial_{\check r}(NA_l+
N^{\check u}A_{\check u})=NF_{l\check r}+N^{\check s}F_{\check s\check r},
\nonumber \\
F_{l\check r}&=&A_{l;\check r}-A_{\check r;l}=\partial_{\check r}A_l+K_{\check
r\check u}\gamma^{\check u\check s}A_{\check s}-{1\over N}(\partial_{\tau}-
{\cal L}_{\vec N}A_{\check r}+NK_{\check r\check u}\gamma^{\check u\check s}
A_{\check s}-A_l\partial_{\check r}N)=\nonumber \\
&=&-{1\over N}[\partial_{\tau}A_{\check r}-{\cal L}_{\vec N}A_{\check r}-
\partial_{\check r}(NA_l)],
\label{C10}
\end{eqnarray}

\noindent where ${\cal L}_{\vec N}A_{\check r}=N^{\check s}A_{\check r\, |\,
\check s}+A_{\check s}N^{\check s}{}_{|\, \check r}=N^{\check s}\partial
_{\check s}A_{\check r}+A_{\check s}\partial_{\check r}N^{\check s}$ is the
Lie derivative along $\vec N$.

By using the second line of Eq.(\ref{C6}) to evaluate $-{1\over 4}\sqrt{g}
\eta^{\mu\nu}\eta^{\rho\sigma}F_{\mu\rho}F_{\nu\sigma}=-{1\over 4}N
\sqrt{\gamma}(2\gamma^{\check r\check s}F_{l\check r}F_{l\check s}+\gamma
^{\check r\check s}\gamma^{\check u\check v}F_{\check r\check u}
F_{\check s\check v})$, the Lagrangian (\ref{114}) becomes

\begin{eqnarray}
{\cal L}&(&\tau ,\vec \sigma )={i\over 2}\sum_{i=1}^N\delta^3(\vec \sigma -
{\vec \eta}_i(\tau ))[\theta^{*}_i(\tau ){\dot \theta}_i(\tau )-{\dot \theta}
^{*}_i(\tau )\theta_i(\tau )]-\nonumber \\
&-&\sum_{i=1}^N\delta^3(\vec \sigma -{\vec \eta}_i(\tau ))[\eta_im_i
\sqrt{N^2(\tau ,\vec \sigma )+g_{\check r\check s}(\tau ,\vec \sigma )
(N^{\check r}(\tau ,\vec \sigma )+{\dot \eta}_i^{\check r}(\tau ))
(N^{\check s}(\tau ,\vec \sigma )+{\dot \eta}_i^{\check s}(\tau ))}+\nonumber
\\
&+&e_i\theta^{*}_i(\tau )\theta_i(\tau )(N(\tau ,\vec \sigma )A_l(\tau ,
\vec \sigma )+A_{\check r}(\tau ,\vec \sigma )(N^{\check r}(\tau ,\vec
\sigma )+{\dot \eta}_i^{\check r}(\tau ))]-\nonumber \\
&-&\sqrt{\gamma (\tau ,\vec \sigma )}[{1\over {2N}}\gamma^{\check r\check s}
(\partial_{\tau}A_{\check r}-{\cal L}_{\vec N}A_{\check r}-\partial_{\check r}
(NA_l))(\partial_{\tau}A_{\check s}-{\cal L}_{\vec N}A_{\check s}-\partial
_{\check s}(NA_l))+\nonumber \\
&+&{N\over 4}\gamma^{\check r\check s}\gamma^{\check u\check v}F_{\check r
\check u}F_{\check s\check v}](\tau ,\vec \sigma ),
\label{C11}
\end{eqnarray}

\noindent which is independent from both the extrinsic curvature and the
Christoffel symbols [this is true only in the case of the spin 1 vector
field but not for general tensors fields].
Now the configuration variables are $z^{\mu}(\tau ,\vec
\sigma ), A_l(\tau ,\vec \sigma ), A_{\check r}(\tau ,\vec \sigma )$, and we
perform the constraint analysis with the new associated momenta. However,
since we have $\pi^l(\tau ,\vec \sigma )=N(\tau ,\vec \sigma )\pi^{\tau}(\tau ,
\vec \sigma )=0$ and the other electromagnetic momenta still given by
Eqs.(\ref{115}), in the case of (spin 1) vector fields
both the formulation with $A_l(\tau ,\vec \sigma )$ and that with $A_{\tau}
(\tau ,\vec \sigma )$ give the same results as in Section VI, since both
$A_{\tau}(\tau ,\vec \sigma )$ and $A_l(\tau ,\vec \sigma )$ are gauge
variables.

\vfill\eject

\end{document}